\documentclass[aps,pre,twocolumn,groupedaddress,showpacs,floatfix]{revtex4}
\usepackage{dcolumn}
\usepackage{amsmath}
\usepackage{amssymb}
\usepackage{graphicx}
\usepackage{bm}
\usepackage[T1]{fontenc}
\usepackage{color}
\usepackage{url}
\usepackage[bf]{subfigure}
\usepackage{rotating}

\usepackage{multirow}

\begin{document}

\title{The role of centrality for the identification of influential spreaders in complex networks}

\author{Guilherme Ferraz de Arruda}
\affiliation{Departamento de Matem\'{a}tica Aplicada e Estat\'{i}stica, Instituto de Ci\^{e}ncias Matem\'{a}ticas e de Computa\c{c}\~{a}o,
Universidade de S\~{a}o Paulo - Campus de S\~{a}o Carlos, Caixa Postal 668,
13560-970 S\~{a}o Carlos, SP, Brazil.}

\author{Andr\'e Luiz Barbieri}
\affiliation{Departamento de Matem\'{a}tica Aplicada e Estat\'{i}stica, Instituto de Ci\^{e}ncias Matem\'{a}ticas e de Computa\c{c}\~{a}o,
Universidade de S\~{a}o Paulo - Campus de S\~{a}o Carlos, Caixa Postal 668,
13560-970 S\~{a}o Carlos, SP, Brazil.}

\author{Pablo Mart\'in Rodr\'iguez}
\affiliation{Departamento de Matem\'{a}tica Aplicada e Estat\'{i}stica, Instituto de Ci\^{e}ncias Matem\'{a}ticas e de Computa\c{c}\~{a}o,
Universidade de S\~{a}o Paulo - Campus de S\~{a}o Carlos, Caixa Postal 668,
13560-970 S\~{a}o Carlos, SP, Brazil.}

\author{Yamir Moreno}
\affiliation{Institute for Biocomputation and Physics of Complex Systems (BIFI) \& Department of Theoretical Physics, University of Zaragoza, 50018 Zaragoza, Spain}
\affiliation{Complex Networks and Systems Lagrange Lab, Institute for Scientific Interchange, Turin, Italy}

\author{Luciano da Fontoura Costa}
\affiliation{Instituto de F\'{\i}sica de S\~{a}o Carlos, Universidade de S\~{a}o Paulo, Av. Trabalhador S\~{a}o Carlense 400, Caixa Postal 369, CEP 13560-970, S\~{a}o
Carlos, S\~ao Paulo, Brazil}

\author{Francisco A. Rodrigues}
\email{francisco@icmc.usp.br}
\affiliation{Departamento de Matem\'{a}tica Aplicada e Estat\'{i}stica, Instituto de Ci\^{e}ncias Matem\'{a}ticas e de Computa\c{c}\~{a}o,
Universidade de S\~{a}o Paulo - Campus de S\~{a}o Carlos, Caixa Postal 668,
13560-970 S\~{a}o Carlos, SP, Brazil.}

\begin{abstract} 
The identification of the most influential spreaders in networks is important to control and understand the spreading capabilities of the system as well as to ensure an efficient information diffusion such as in rumor-like dynamics. Recent works have suggested that the identification of influential spreaders is not independent of the dynamics being studied. For instance, the key disease spreaders might not necessarily be so when it comes to analyze social contagion or rumor propagation. Additionally, it has been shown that different metrics (degree, coreness, etc) might identify different influential nodes even for the same dynamical processes with diverse degree of accuracy. In this paper, we investigate how nine centrality measures correlate with the disease and rumor spreading capabilities of the nodes that made up different synthetic and real-world (both spatial and non-spatial) networks. We also propose a generalization of the random walk accessibility as a new centrality measure and derive analytical expressions for the latter measure for simple network configurations. Our results show that for non-spatial networks, the $k$-core and degree centralities are most correlated to epidemic spreading, whereas the average neighborhood degree, the closeness centrality and accessibility are most related to rumor dynamics. On the contrary, for spatial networks, the accessibility measure outperforms the rest of centrality metrics in almost all cases regardless of the kind of dynamics considered. Therefore, an important consequence of our analysis is that previous studies performed in synthetic random networks cannot be generalized to the case of spatial networks.
\end{abstract}


\maketitle

\section{Introduction}

Spreading phenomena are ubiquitous in Nature \cite{Hethcote00:SR, Costa011:AP}. Rumors and viruses spread from person to person, worms contaminate computers worldwide and innovations are diffused from place to place. The advent of new technology and modern transportation means has led to radical changes of classical transmission channels, making in much cases natural and manmade systems more prone to contagion processes. On the other hand, new tools have been developed to study such phenomena, for instance, by explicitly dealing with the topology and dynamics of so-called complex networks, which are nothing else but the backbone on top of which information and diseases propagate \cite{Barrat08:book, Newman010:book}.

Networks are made up by nodes, that represent the elements of the system, and edges, which define the possible interaction patterns among nodes \cite{Costa07:AP, Boccaletti06:PR}. A large body of recent studies have verified that the way in which such nodes are organized plays a fundamental role in spreading processes~\cite{Newman02:PRE, Boccaletti06:PR}. For instance, Pastor-Satorras and Vespignani showed that a disease outbreak takes place when the spreading rate, $\beta$, is larger than the epidemic threshold \cite{Satorras01:PRL}, i.e., if $\beta>\beta_ c = \langle k \rangle / \langle k^2\rangle$, where $\langle k^m \rangle$ is the $m$-th moment of the degree distribution. Therefore, most scale-free networks (those for which the degree distribution follows a power law $P(k)\sim k^{-\gamma}$ with $\gamma<3$) are particularly prone to the spreading of diseases, since $\beta_c \rightarrow 0$ when $N\rightarrow\infty$. Additional network properties, such as assortativity~\cite{Newman02:PRL, Boguna03:PRL} or modular organization~\cite{Liu:EPL} also play a fundamental role in disease spreading.

One of the most interesting challenges in network science is to understand the relation between the structure of the system and its emergent dynamical properties. This is why finding determinant structural factors is important, as a better knowledge would allow controlling the function of the system, which for the scope of this paper, means determining what network properties are more closely related to information and viruses diffusion. In particular, we will focus our attention in one topological feature: centrality. Since the most central nodes can diffuse their influence to the whole network faster than the rest of nodes, it is expected that such agents are the most influential spreaders. Recently, Kitsak et al.~\cite{Kitsak10:NP} found evidences that confirmed this hypothesis for the case of epidemic outbreaks. The authors verified that the most influential spreaders can be forecasted from the $k$-shell decomposition analysis. Such agents are located within the core of the network and do not need to be the most connected. Silva et al.~\cite{Silva012} explored the correlations between heterogeneous spread and central attributes of the vertices that were first seeded with a disease, finding that degree and accessibility are measures mostly related to the efficient spread of the disease. On the other hand, Borge-Holthoefer and Moreno~\cite{Borge12} showed that, for standard rumor models, it is not  possible to identify the most influential spreaders using the same metrics.

Although many works have provided evidences for the presence of influential spreaders in epidemic spreading, the conclusions are not general. Indeed, there is no general consensus on the definition of network ``centrality'', because there are many measures able to quantify the centrality of a node, each one considering specific concepts~\cite{Newman010:book}. For instance, the betweenness and closeness centrality take into account only the shortest distance between pairs of nodes~\cite{Boccaletti06:PR, Newman010:book}, ignoring alternative paths. At the same time, the $k$-core decomposition may eliminate important sets of vertices, which can be connected to the main core through nodes with a small number of links~\cite{Seidman83}. Thus, to overcome such a lack of a universal definition of node centrality, it is necessary to look at additional measures. In this paper, we study the problem of the identification of influential spreaders using eight centrality measures in order to complement previous studies~\cite{Kitsak10:NP, Borge12}. Moreover, we introduce a new metric, the generalized accessibility, as a centrality measure that is based on random walks. We observe that in social and scale-free networks, the accessibility, average neighborhood degree and closeness centrality are the measures most related to rumor spreading. Other measures, such as the $k$-core and degree correlate well only with epidemic spreading in social networks, as found previously in~\cite{Kitsak10:NP, Borge12}.

Another important result is related to the kind of networks studied in this work. Despite the fact that many diffusion processes take place on spatially embedded networks \cite{Barthelemy2011}, previous studies have disregarded spatial networks~\cite{Kitsak10:NP, Borge12, Silva012}. 
These networks have several topological constraints that greatly influence the way connections are established, and thus, one expect an impact in  network centrality metrics and consequently on the spreading dynamics. In this paper, we intend to fill this gap by exploring the role of centrality measures in predicting the spreading capabilities of nodes of spatial networks. Specifically, we consider both real networks (road networks of four countries) and artificial spatial networks with exponential and power-law degree distributions and find that correlations between spreading capacity and centrality measures is spatial networks differs significantly from those observed in non-spatial networks. 

This paper is organized as follows. Sec.~\ref{Sec:cent} presents the centrality measures considered in our investigations. The generalized random walk accessibility is introduced in Sec.~\ref{Sec:Acc}. The analytical expressions for complete graphs, stars and rings are also evaluated in this section. Concepts of epidemic and rumor spreading are discussed in Sec.~\ref{Sec:Spreading} and the databases are described in Sec.~\ref{Sec:data}. The analysis of spatial networks is outlined in Sec.\ref{Sec:spatial}, where it is shown that the accessibility is strongly correlated to the node capacity for rumor and epidemic spreading. Sec.~\ref{Sec:nonspatial} presents the analysis of non-spatial networks, which complements the investigations in~\cite{Kitsak10:NP, Borge12}. Our final conclusions are developed in Sec.~\ref{Sec:conclusion}. 


\section{Centrality measures} 
\label{Sec:cent}

As mentioned before, one can in principle consider several metrics to define the centrality of a node \cite{Newman010:book}. For completeness, here we provide the basic definitions of those that will be used in the rest of the paper. For more details, we refer the reader to the literature cited.

\paragraph{Basic centrality measures.} The most basic definition of centrality takes into account the number of connections of a node $i$, called node degree, $k_i$. In this case, the most central node has the largest number of connections. Alternatively, the centrality of a vertex can be defined in terms of the degree of its second neighbors, since strongly connected vertices can surround a central node. In this case, the average degree of the nearest neighbors of $i$ is defined as
\begin{equation}
r_i = \frac{1}{k_i}\sum_{j \in \nu(i)}k_j,
\end{equation}
where $\nu(i)$ is the set of nodes connected to $i$. It has been verified that the average neighborhood degree is related to epidemic spreading in networks~\cite{Barrat08:book}. 

\paragraph{Eigenvector centrality.} It considers that the centrality of each node is the sum of the centrality values of the nodes that it is connected to. The eigenvector centrality is defined by the eigenvector associated to the largest eigenvalue of the adjacency matrix $A$. Formally,
\begin{equation}
x_i = \kappa^{-1} \sum_j A_{ij} x_j,
\end{equation}
or in the matrix form $A \textbf{x} = \kappa \textbf{x}$, where $\textbf{x}$ is the right leading eigenvector~\cite{Newman010:book} and $\kappa$ is the largest eigenvalue.

\paragraph{Distance-based centrality metrics.} Centrality can also be established in terms of the shortest distances between pairs of nodes, since the more central a node is, the lower its total distance to all other nodes is. The closeness centrality of $i$ is defined as~\cite{Newman010:book}
\begin{equation}
C_i = \frac{N}{\sum_{j=1, j \neq i}^N d_{ij}},
\end{equation}
where $d_{ij}$ is the shortest distance between nodes $i$ and $j$, and $N$ is the number of nodes in the network. 

Alternatively, the effective load of a node can also be considered as a centrality measure. Betweenness centrality quantifies the load as the number of times a node acts as a bridge along the shortest path between two other nodes~\cite{Girvan02:PNAS}. Thus, for a node $i$,
\begin{equation}
B_i = \sum_{(a,b)} \frac{\sigma(a,i,b)}{\sigma(a,b)},
\label{betweenness}
\end{equation}
where $\sigma(a,i,b)$ is the number of shortest paths connecting vertices $a$ and $b$ that pass through vertex $i$ and $\sigma(a,b)$ is the total number of shortest paths between $a$ and $b$. The sum is over all pairs $(a,b)$ of distinct vertices. In this case, a central node should be crossed by many paths and shows the highest value of $B_i$.

\paragraph{Clustering.} The clustering coefficient quantifies the occurrence of triangles in the networks. It is defined as~\cite{Boccaletti06:PR}
\begin{equation}
cc(i) = \frac{3N_{\triangle}(i)}{N_3(i)},
\label{eq:cci}
\end{equation}
where $N_{\triangle}(i)$ is the number of triangles involving the node $i$ and $N_3(i)$ is the number of triples centered around $i$. $cc(i)$ can be also understood as a centrality measure in the sense that if two nodes are connected only via the node $i$, this node can control the information flow~\cite{Newman010:book}. Thus, the  clustering coefficient could be thought off as a local version of the betweenness centrality. Note that $cc(i)$ takes smaller values for more central nodes, in opposite to the other centrality measures.

\paragraph{Coreness.} The $k$-shell decomposition partitions a network into sub-structures and assigns an integer index to each node $i$, $k_c(i)$, in such a way that $k_c(i)=k$ if $i$ belongs to the $k$-core, but it is not in the $(k+1)$-core~\cite{Seidman83}. Nodes with low values of $k_c$ are located at the periphery of the network. This measure was adopted recently to detect influential spreaders in networks~\cite{Kitsak10:NP}. The most central nodes should have the highest values of coreness, whereas high-degree nodes localized in the periphery of networks should display small values of coreness~\cite{Kitsak10:NP}. Therefore, only hubs at the main core of networks present the highest values of $k_c$.

\paragraph{Random-walk based centrality measures.} The number of visits that a given node receives when an agent travels through the network without a preferential route can also be taken into account to quantify the node centrality. In this case, a possible measure is the Google PageRank~\cite{Brin98theanatomy}. PageRank is calculated as 
\begin{equation}
\pi^T = \pi^T G,
\end{equation}
$G$ is the Google matrix, i.e.,
\begin{equation}
G = \kappa \left(P + \frac{ae^T}{N} \right) + \frac{(1-\kappa)}{N} e e^T,
\end{equation}
and $a$ is the binary vector called dangling node vector ($a_i$ is equal to one if $i$ is a dangling node and 0 otherwise), $e$ is a vector of ones of length $N$ and $P$ is the transition probability matrix of the respective network ($P(i,j) = \frac{1}{\sum_j A_{ij}}$, where $A_{ij}$ are the elements of the adjacency matrix). The original version of the algorithm considers $\kappa = 0.85$~\cite{Brin98theanatomy}. The PageRank of a node $i$, $\pi_i$, is given by the $i$-th entry of the dominant eigenvector $\pi$ of $G$, given that $\sum_i \pi_i = 1$. $\pi_i$ can be understood as the probability of arriving at the node $i$ after a large number of steps following a random walk navigation through the network.


\section{Generalized random walk accessibility}\label{Sec:Acc}

The accessibility is related to the diversity of access of individual nodes through random walks~\cite{Travenccolo2008:PLA}. This measure has been considered for identification of the border of complex networks~\cite{Travencolo2009:NJP}. Let $P^{(h)}(i,j)$ be the probability of reaching node $j$ by performing random walks of length $h$ departing from $i$. The accessibility of the node $i$ for a given distance $h$ is defined by the exponential of the Shannon entropy~\cite{Travenccolo2008:PLA}, i.e.,
\begin{equation}
\alpha_h(i) = \exp\left({-\sum_{j}P^{(h)}(i,j)\log P^{(h)}(i,j)}\right),
\label{Eq:acch}
\end{equation}
where $1 \leq \alpha_h(i) \leq N$. The maximum value corresponds to the case in which all nodes are reached with the same probability $1/N$. Note that this metric was defined in a multilevel fashion, depending on the parameter $h$ that defines the scale of the dynamics~\cite{Travenccolo2008:PLA, Travencolo2009:NJP}. In addition, though here we will be constrained to random walks, virtually any other type of dynamics yielding transition probabilities between adjacent nodes can be considered in the accessibility, which makes this measurement adaptable to the dynamics of each problem being studied.

\begin{figure} [!t]
\centerline{\includegraphics[width=0.9\columnwidth]{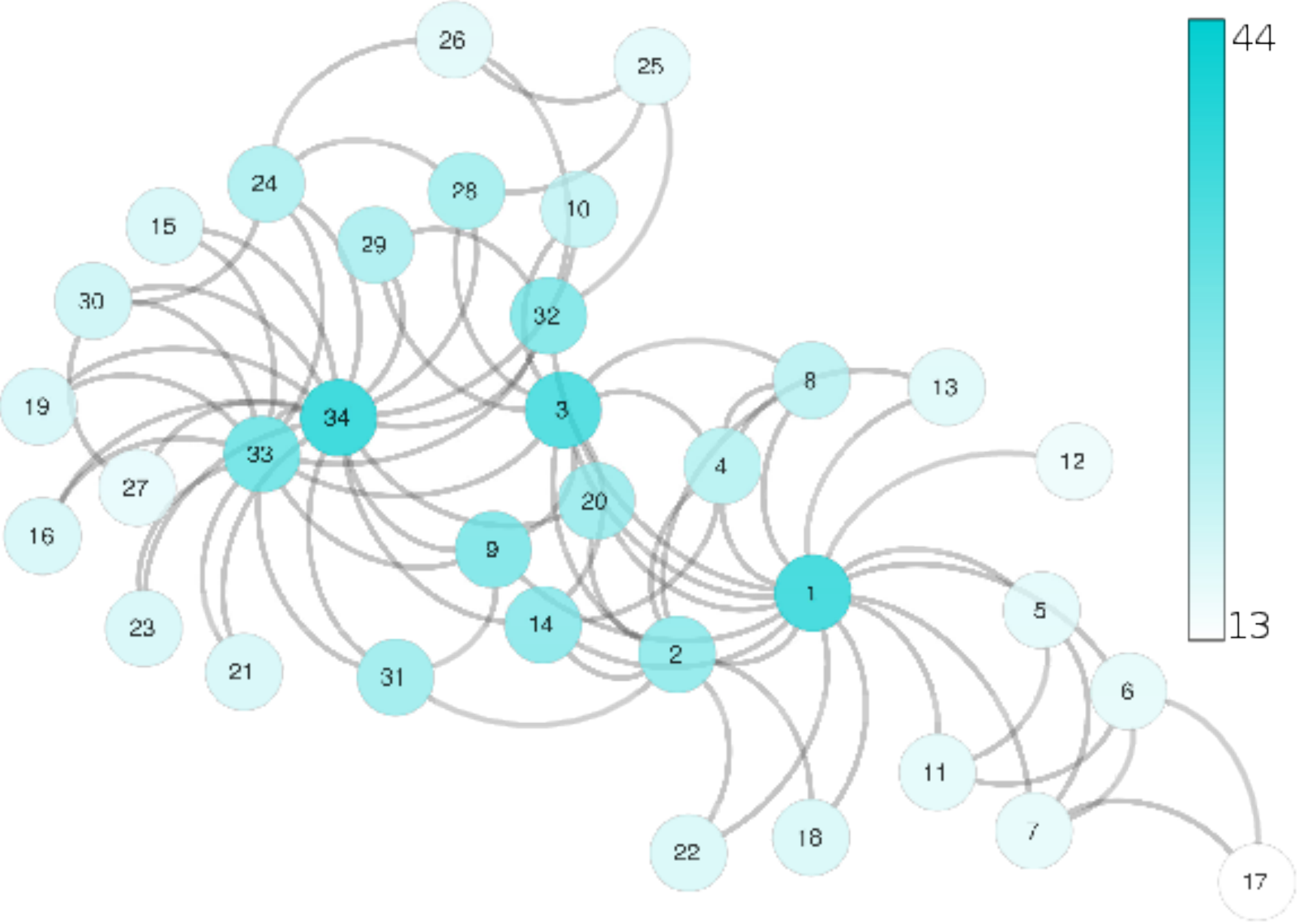}}
\caption{Illustration of the concept of the accessibility (values calculated from Eq.\ref{Eq:acc}) in the Zachary Karate-club network~\cite{Zachary77}. Nodes at the center of the network present the highest accessibility.}
\label{fig:ex_acc}
\end{figure}

In order to generalize the accessibility, here we introduce a new version of this metric, which is based on the matrix exponential operation~\cite{Bhatia1997:book}. This matrix enables the calculus of the probability of transition considering walks of all lengths between any pair of vertices. In this way, if $P$ is the transition matrix, the exponential of $P$ is defined as
\begin{equation}\label{Eq:W}
\textbf{W} = \sum_{k=0}^{\infty} \frac{1}{k!} {P}^k = e^{P}.
\end{equation}
The matrix $\textbf{W}$ is based on a modified random walk, which penalizes longer paths. To construct such stochastic process we consider an usual random walk $(X_n)_{n\geq 0}$, where $X_n$ represents the node visited by the agent at time $n$. We take a collection of independent and identically distributed uniform random variables in the interval $(0,1)$, i.e. $\{U_1, U_2,\ldots\}$, which represents a kind of ``fitness'' associated to each step of the walk. Also, we assume independence between the collection of uniform random variables and the random walk. This modified random walk, which we call accessibility random walk (ARW) in the rest of the paper, considers walks through the network such that all associated fitnesses along a trajectory are in ascending order. We say that node $j$ is visited by the ARW, at time $n$, if $X_n =j$ and $U_1 < U_2 < U_3 < \cdots < U_n$. We denote by $(\tilde{X}_n)_{n\geq 0}$ the new process and note that $\{\tilde{X}_n = j\}$ implies $\{X_n = j\}$, but the opposite is not necessarily true. A quantity of interest is the number of visits that a given node $j$ receives when an agent travels through the network according to the ARW. This quantity can be written as $\sum_{n=1}^{\infty}I_{\left\{\tilde{X}_n =j\right\}}$, where $I_{A}$ is the indicator function of the event $A$. We are interested in the mean of this value, by assuming that the agent starts from node $i$, i.e. $\sum_{n=1}^{\infty}E(I_{\left\{\tilde{X}_n =j\right\}}|\tilde{X}_0 =i)$. In order to compute this value we observe that the term of the sum is the probabilitiy $P(\tilde{X}_n =j|\tilde{X}_0 =i)$ which, by our definition, is equal to $P(\{X_n=j\}\cap\{U_1< U_2 < U_3 < \cdots < U_n\}|X_0 = i)$. This probability is exactly $(1/n!)P^{(n)}(i,j)$, where $P^{(n)}(i,j)$ is the probability of transition from $i$ to $j$ through walks of length $n$. Therefore, the matrix $\textbf{W}$ considered in Eq.~\eqref{Eq:W} is a matrix of mean values associated to the ARW. The element $\textbf{W}(i,j)$ provides the mean number of visits that node $j$ receives when the agent starts at node $i$ following and follows ARW. 

The probability of transition between any pair of vertices through ARW is given by
\begin{equation}\label{Eq:P}
\textbf{P} = \frac{\textbf{W}}{e}.
\end{equation}
Note that the matrix $\textbf{W}$ weights all walks by the inverse of the factorial of their lengths. Therefore, this definition penalizes longer walks, i.e., the shortest walks receive more weight than the longest ones. We define the generalized expression for the accessibility as
\begin{equation}\label{Eq:acc}
\alpha(i) = \exp\left(-\sum_j \textbf{P}(i,j)\log \textbf{P}(i,j)\right),
\end{equation}
which we call generalized random walk accessibility. Figure~\ref{fig:ex_acc} illustrates this measure.

We note that the exponential matrix is also considered in the definition of the communicability \cite{Estrada08, Estrada2011}. The difference is that the accessibility is based on the concept of diversity~\cite{Hill1973, Jost06} whereas the communicability is associated to the communication between any pair of vertices~\cite{Estrada2011}. Moreover, the former is related to the probability transition matrix, whereas the latter on the adjacency matrix. In this way, there is no trivial relation between these two measures in irregular graphs.

Let us provide in what follows some exact expression for the metric just introduced. Although the graphs considered below are not representatives of real world networks, we believe that the analysis helps understanding what can be learned from the new metric. In addition, there are structures that already capture some important features of real networks, such as the star graph, which is an extreme example of an heterogenous configuration but that have provided insightful hints about the dynamics under study in other cases \cite{Gomez11:PRL, Peron12:PRE}.


\subsection{Accessibility in star graphs}

For a star graph, the probability of transition between the central node $i$ and any of the $k$ leaves considering an ARW is given by (see Eqs.~\ref{Eq:W} and~\ref{Eq:P})
\begin{equation}
\textbf{P}(i,j) =\frac{1}{ek}\sum_{n=0}^\infty \frac{1}{(2n + 1)!} = \frac{1}{ek}\sinh(1), \quad i\neq j
\end{equation}
and between the leaves and central node $i$,
\begin{equation}
\textbf{P}(j,i) = \frac{\sinh(1)}{e}.
\end{equation}
In addition, 
\begin{equation}
\textbf{P}(i,i) = \frac{\cosh(1)}{e}.
\end{equation}
The probability of transition between leaves $j$ and $l$ is given by
\begin{equation}
\textbf{P}(j,l) = \frac{1}{ek}(\cosh(1)-1), 
\end{equation}
and for $l=j$,
\begin{equation}
\textbf{P}(j,j) = \frac{1}{e} + \frac{1}{ek}(\cosh(1)-1), 
\end{equation}
Therefore, the general form of the exponential matrix, considering the node number one as the hub of the star graph, is given as
\begin{widetext}
\begin{equation} \label{eq:P_star}
\textbf{P} = \frac{1}{e}
\begin{bmatrix} 
\cosh(1) & \frac{1}{k}\sinh(1) &\cdots &\cdots & \frac{1}{k}\sinh(1)\\
\sinh(1) & 1 + \frac{1}{k}(\cosh(1)-1)& \frac{1}{k}(\cosh(1)-1) & \cdots & \frac{1}{k}(\cosh(1)-1)\\
\vdots & \frac{1}{k}(\cosh(1)-1) & \ddots &  & \vdots\\
\vdots & \vdots & & \ddots & \frac{1}{k}(\cosh(1)-1) \\
\sinh(1) & \frac{1}{k}(\cosh(1)-1) &\cdots & \frac{1}{k}(\cosh(1)-1) &1 + \frac{1}{k}(\cosh(1)-1)\\
\end{bmatrix}.
\end{equation}
\end{widetext}
In this way, since $k = N-1$, the accessibility of the hub $i$ is 
\begin{equation}\label{Eq:hub}
\alpha(i) =\exp\left\{-x\log(x) - y\log\left(\frac{y}{N-1} \right) \right\},
\end{equation}
where $x = \frac{\cosh(1)}{e}$ and $y = \frac{\sinh(1)}{e}$. For any leaf $j$ connected with $i$,
\begin{equation}\label{Eq:leaf}
\begin{split}
\alpha(j) =\exp\{-x\log(x) +  (N-2)y\log(y) + \\
(1/e + y)\log(1/e + y) \},
\end{split}
\end{equation}
where $x = \frac{\sinh(1)}{e}$, $y = \frac{(\cosh(1)-1)}{e(N-1)}$. 

\begin{figure} [!t]
\centerline{\includegraphics[width=1\columnwidth]{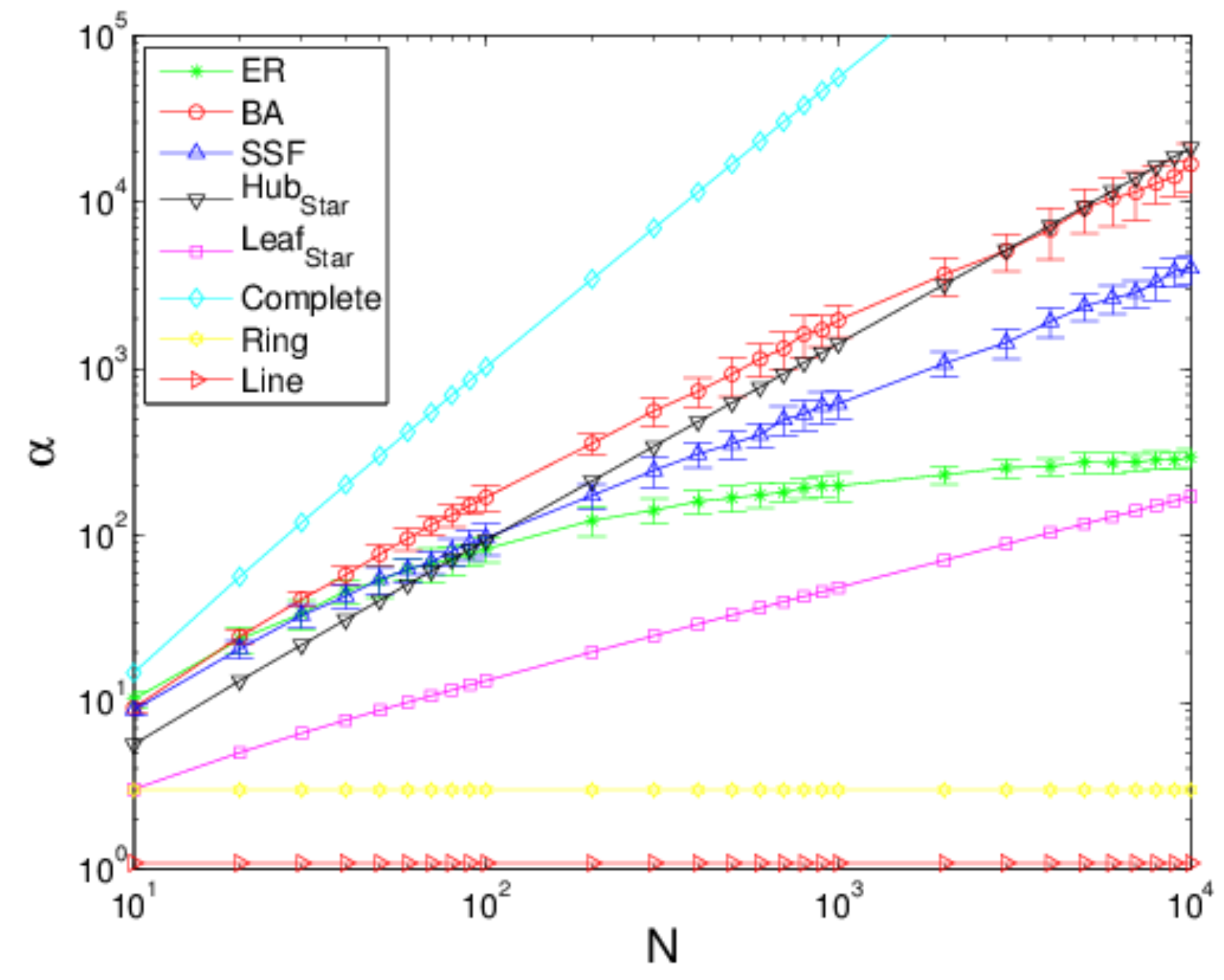}}
\caption{Accessibility calculated in star (from Eqs.~\ref{Eq:hub} and~\ref{Eq:leaf}), complete, ring and line graphs (extreme nodes) compared to the maximum value
obtained in Erd\"os-R\'enyi (ER) random graphs, scale-free networks of Barab\'asi-Albert (BA), and spatial scale-free networks (SSF). $N$ is the network size. For complex
networks, each point is an average over 50 networks with $\langle k \rangle \approx 4$.}
\label{fig:acc_star_hub}
\end{figure}

We show in Fig.~\ref{fig:acc_star_hub} the results obtained for the accessibility on top of different networks and configurations. As it can be seen, Eq.~\ref{Eq:hub} can be considered to be a good predictor of the accessibility of the hubs in scale-free networks. However, as expected from the fact that the star graph does not capture any topological aspect of homogeneous networks, the star-graph approximation is not accurate for random Erd\"os-R\'enyi networks.

\subsubsection{Eigendecomposition analysis}

The exact values of accessibility in star graphs can also be calculated by the eigen-decomposition analysis of $P$. The exponential matrix, Eq.~\ref{Eq:W}, can be obtained as
\begin{equation}\label{eq:expm_spec}
 \textbf{W} = e^{P} = \mathcal{V} \mathcal{D} \mathcal{V}^{-1}.
\end{equation}
where $\mathcal{V}$ is a matrix whose columns are the eigenvectors of the matrix $\textbf{W}$ and $\mathcal{D}$ is a matrix whose diagonal presents the exponential of each eigenvalue of $P$,
\begin{equation}\label{eq:eig_def}
 \left( P - \lambda \textbf{I} \right) v = 0
\end{equation}
where $P$ is the transition matrix, $\lambda$ is its eigenvalue and $v$ is the associated eigenvector. 

In this way, for the star graph, the transition matrix is sparse and its characteristic polynomial, $\text{det}(P - \lambda I) = 0$, is calculated by the Laplace rule as
\begin{equation}
\begin{split}
 \text{det}(P - \lambda I) = \left( -\lambda \right)^N - \left( -\lambda \right)^{(N-2)} = \\
 \left( -\lambda \right)^{(N-2)} \left( \left( -\lambda \right)^2 - 1 \right) = 0,
\end{split}
\end{equation}
whose solutions are $\lambda_1 = -1$, $\lambda_2 = 1$ and $\lambda_i=0$, $\forall~i=2,3,\ldots,N$. Therefore, using the definition of an eigenvalue and eigenvector
problem, it is possible to obtain the following equations for the eigenvectors. For $\lambda_1 = -1$,
\begin{equation}
\begin{cases}
  v_{11} = &-\frac{1}{N-1}\sum_{j=2}^N v_{1j} \\
  v_{1j} =& -v_{11}  \hspace{0.5cm}  j = 2, 3,\ldots,N;
\end{cases}
\end{equation}
where $v_{pj}$ is the $j$-th element of the eigenvector $v_p$ associated with the eigenvalue $\lambda_p$. For $\lambda = 1$,
\begin{equation}
\begin{cases}
 v_{21} =& \frac{1}{N-1}\sum_{j=2}^N v_{2j} \\
 v_{2j} =& v_{21}  \hspace{0.5cm}  j = 2, 3, \ldots,N ;
\end{cases}
\end{equation}
finally, for $\lambda_p = 0$ where $p = 3, \ldots,N$, which has multiplicity $(N-2)$, 
\begin{equation}
\begin{cases}
  0 =& \frac{1}{N-1}\sum_{j=2}^N v_{pj} \\
  v_{p1} =& 0\\
\end{cases}
\end{equation}
which yields the matrix
\begin{equation} \label{eq:Mat_Vstar}
\mathcal{V} = 
\begin{bmatrix} 
 -1     &  1         & 0      & \cdots & \cdots & 0    \\
 1      &  1         & -1     & \cdots & \cdots & -1   \\
 \vdots &  \vdots         & 1      &  0     & \cdots &    0  \\
 \vdots &  \vdots    & 0      & \ddots &  &    \vdots  \\
 \vdots &  \vdots    & \vdots &        & \ddots &    0  \\
 1      &  1         & 0      & \cdots &  0 & 1    \\
\end{bmatrix},
\end{equation}
whose inverse is
\begin{equation} \label{eq:Mat_Vstar_inv}
\mathcal{V}^{-1} = 
\begin{bmatrix} 
 \frac{-1}{2}  & \frac{1}{2(N-1)}         & \frac{1}{2(N-1)}     & \cdots & \frac{1}{2(N-1)} \\
 \frac{1}{2}  & \frac{1}{2(N-1)}        & \frac{1}{2(N-1)}     & \cdots & \frac{1}{2(N-1)} \\
 0 &  \frac{-1}{N-1}         & \frac{N-2}{N-1} &        & \frac{-1}{N-1}  \\
 \vdots &  \vdots    &        & \ddots & \vdots \\
 0      &  \frac{-1}{N-1}         & \cdots &        & \frac{N-2}{N-1}  \\
\end{bmatrix}.
\end{equation}
Note that we used non-unit vectors to construct the matrices $\mathcal{V}$. This is not necessary since $\mathcal{D}$ is also multiplied by $\mathcal{V}^{-1}$ and the
non-unit norms are compensated. Substituting in matrices~\ref{eq:Mat_Vstar} and~\ref{eq:Mat_Vstar_inv} in Eq.~\ref{eq:expm_spec}, after some algebra, we recover Eq.~\ref{eq:P_star}. The
accessibilty of hubs and leafes are calculated by Eqs.~\ref{Eq:hub} and~\ref{Eq:leaf}, respectively.

\subsection{Accessibility in ring graphs}

The generalized random walk accessibility can also be calculated exactly in rings, that are a special case of K-regular graphs, where $K = 2$. The
probability transition matrix has the form
\begin{equation} \label{eq:Mat_P_ring}
P = 
\begin{bmatrix} 
 0           & \frac{1}{2} &  0     & \cdots & 0      & \frac{1}{2}    \\
 \frac{1}{2} &  \ddots     & \ddots &        &        & 0     \\
 0           &  \ddots     & \ddots & \ddots &        & \vdots  \\
 \vdots      &             & \ddots & \ddots & \ddots & \vdots \\
 0           &             &        & \ddots & \ddots & \frac{1}{2} \\
 \frac{1}{2} &  0          & \cdots & 0   & \frac{1}{2} & 0  \\
\end{bmatrix}.
\end{equation}
Such matrix has a well known spectra and is widely used in finite difference methods~\cite{LeVeque}. As exposed in~\cite{LeVeque}, the eigenvalues of $P$ are
\begin{equation} \label{eq:eigval_ring}
 \lambda_p = \frac{1}{2} \left( \exp\left(  \frac{2 \pi i p}{N} \right) + \exp\left(  \frac{- 2 \pi i p}{N} \right) \right) = \cos \left( \frac{2 \pi p}{N} \right),
\end{equation}
where $i = \sqrt{-1}$ and the associated elements of the eigenvector can be expressed as
\begin{equation} \label{eq:eigvec_ring}
 u_{pj} = \frac{\exp\left( \frac{2 \pi i p j}{N} \right)}{\sqrt{N}},
\end{equation}
where $\sqrt{N}$ is just a normalization factor. This set of eigenvectors diagonalizes the matrix $P$ as $P = U \Lambda U^H$, where $\Lambda$ is the diagonal matrix with
the eigenvalues of $P$ (Eq.~\ref{eq:eigval_ring}), $U$ is the matrix whose columns are the eigenvector of $P$ and $U^H$ is the conjugate transpose of $U$. We can write
the closed expression for $\textbf{P}$ as
\begin{equation}
 \textbf{P}(j,k) = \frac{1}{e}\sum_p \exp \left(\lambda_p\right) u_{pj} u_{pk}^*,
\end{equation}
where $u_{pj}^*$ is the conjugate transpose of $u_{pj}$. Note that we used the complex domain to solve the problem, however the solution is on the real domain. Using Eqs.~\ref{eq:eigval_ring} and~\ref{eq:eigvec_ring}, we obtain
\begin{equation}
\begin{split}
\textbf{P}(j,k) = 
\frac{1}{eN} \sum_{p=1}^N \exp \left(\cos \left( \frac{2 \pi p}{N} \right) \right) \exp \left( \frac{2 \pi i p (j - k)}{N} \right),
\end{split}
\end{equation}
which is a closed form for the evaluation of $\textbf{P}$ in ring graphs. 
Furthermore we can use some graph spectra properties to separate the first eigenvalue from the summation 
\begin{equation}
\begin{split}
 \textbf{P}(j,k) =\frac{1}{e} \left( \frac{k_j}{2M} \right) \exp(1) + \\
 \frac{1}{N} \sum_{p=2}^N \exp \left(\cos \left( \frac{2 \pi p}{N} \right) + \left( \frac{2 \pi i p (j - k)}{N}
\right) \right).
\end{split}
\end{equation}
Figure~\ref{fig:acc_star_hub} shows the comparison between network models and the analytical solutions for the regular structures. Note that the solution for the ring does not depend on the network size. The results for the line graph, which again do not depend on the network size, are also presented in this figure. We also remark that the extremes of the line present the lowest values of accessibilility, whereas the nodes in the center have the highest values.
\subsection{Accessibility in complete graphs}

The generalized random walk accessibility can also be calculated exactly for a complete graph, in which every pair of nodes is connected without self connections. In this way, the probability of transition between any pair of nodes is $P(i,j) = \frac{1}{N-1}$ and the exponential matrix (see Eq.~\ref{Eq:W}) is given by
\begin{equation}
\begin{split}
 \textbf{P}(i,j) = \frac{1}{eN} \sum_{n = 0}^{\infty} \frac{(N-1)^n + (-1)^n(N-1)}{(N-1)^n n!} = \\
\frac{\exp(1) + (N-1)\exp\left({\frac{-1}{N-1}}\right)}{eN}, \quad i\neq j.
\end{split}
\end{equation}
The main diagonal of $\textbf{P}$, which considers the paths starting and ending at the same node, is expressed as
\begin{equation}
\textbf{P}(i,i) = \frac{1}{eN} \sum_{n = 0}^{\infty} \frac{(N-1)^n + (-1)^{n+1}}{(N-1)^n n!} = 
\frac{\exp(1) -\exp\left({\frac{-1}{N-1}}\right)}{eN}.
\end{equation}

Therefore, the general form of the exponential matrix is given as
\begin{widetext}
\begin{equation} \label{eq:P_complete}
\textbf{P} = \frac{1}{e}
\begin{bmatrix} 
\frac{\exp(1) + (N-1)\exp\left({\frac{-1}{N-1}}\right)}{N} & \frac{\exp(1) -\exp\left({\frac{-1}{N-1}}\right)}{N} &\cdots & \frac{\exp(1)
-\exp\left({\frac{-1}{N-1}}\right)}{N}\\
\frac{\exp(1) -\exp\left({\frac{-1}{N-1}}\right)}{N} & \ddots & & \vdots\\
\vdots & & \ddots & \vdots\\
\frac{\exp(1) -\exp\left({\frac{-1}{N-1}}\right)}{N} & \cdots & \cdots & \frac{\exp(1) + (N-1)\exp\left({\frac{-1}{N-1}}\right)}{N}\\
\end{bmatrix}.
\end{equation}
\end{widetext}
The accessibility of each node is
\begin{equation}\label{Eq:acccp}
\begin{split}
\alpha(i) = \exp\{ -\textbf{P}(i,i) \log(\textbf{P}(i,i)) +\\
- (N-1)\textbf{P}(i,j) \log(\textbf{P}(i,j))\} = \\
 e \left( a^{-(N-1)a/e}b^{-b/e} \right),
\end{split}
\end{equation}
where
\begin{equation}
 a = \frac{1}{eN} \left( \exp(1) -\exp\left({\frac{-1}{N-1}}\right) \right)
\end{equation}
and
\begin{equation}
 b = \frac{1}{eN} \left( \exp(1) + (N-1)\exp\left({\frac{-1}{N-1}}\right) \right).
\end{equation}

In the complete graph all nodes present the same value of accessibility and, since a random walker needs just one step to reach any other node, this value is the upper bound of the maximum value of accessibility for a network with $N$ nodes. Figure~\ref{fig:acc_star_hub} shows the variation of the accessibility in complete graphs as a function of the network size.

\subsubsection{Eigendecomposition analysis}

The exact values of accessibility in complete graphs can also be obtained by the eigen-decomposition analysis, the graph spectra and its eigenvectors, as performed for
the star graph. In this way, we get the following system (from Eq.~\ref{eq:eig_def})
\begin{equation}
\begin{bmatrix} 
-\lambda & \frac{1}{N-1} &\cdots & \frac{1}{N-1}\\
\frac{1}{N-1} & \ddots & & \vdots\\
\vdots &  & \ddots & \vdots\\
\frac{1}{N-1} & \cdots & \cdots & -\lambda\\
\end{bmatrix} v = 0,
\end{equation}
which yields
\begin{equation} \label{eq:eig_P_complete}
 \sum_{i, i \neq j}^N v_{pi} \left( \frac{1}{N-1} \right) - \lambda_p v_{pj} = 0, \hspace{0.5cm} \forall i,j; \hspace{0.2cm}  i \neq j,
\end{equation}
where $v_{pj}$ is the $j$-th element of the eigenvector $v_p$ associated with the eigenvalue $\lambda_p$.

The eigenvalues of $P$ for a complete graph is the spectrum of the adjacency matrix multiplied by $\frac{1}{N-1}$, i.e., $\lambda_1 = 1$, $\lambda_2 = \lambda_3 = \ldots =
\lambda_N = \frac{1}{N-1}$~\cite{Mieghem:2011}. Therefore, for $\lambda_1 = 1$ we have that
\begin{equation} 
 \left( N-1 \right) v_{1j} = \sum_{i, i \neq j}^N v_{1i}, \hspace{0.5cm} \forall i,j; \hspace{0.2cm}  i \neq j.
\end{equation}
The solution is $v_{1i} = v_{1j}$. On the other hand, for $\lambda_p = \frac{-1}{N-1}$, where $p = 2, 3, \ldots, N$,
\begin{equation}
 \sum_{i, i \neq j}^N v_{pi} \left( \frac{1}{N-1} \right) + \frac{1}{N-1} v_{pj} = 0,
\end{equation} 
i.e.,
\begin{equation}
\left( \frac{1}{N-1} \right) \sum_{i}^N v_{pi} = 0, \hspace{0.5cm}
\forall i,j; \hspace{0.2cm}  i \neq j
\end{equation}
The respective solution is $\sum_{i}^N v_{pi} = 0$. Note that both solutions are not unique, whereas the Eq.~\ref{eq:expm_spec} has a unique solution. Without loss
generality, we assume
\begin{equation} \label{eq:Mat_V}
\mathcal{V} = 
\begin{bmatrix} 
 1      & -1         & -1     & \cdots & -1    \\
 1      &  1         & 0      & \cdots & 0     \\
 \vdots &  0         & \ddots &        & \vdots  \\
 \vdots &  \vdots    &        & \ddots & \vdots \\
 1      &  0         & \cdots &        & 1  \\
\end{bmatrix},
\end{equation}
whose inverse is
\begin{equation} \label{eq:Mat_V_inv}
\mathcal{V}^{-1} = 
\begin{bmatrix} 
 \frac{1}{N}      & \frac{1}{N}         & \frac{1}{N}     & \cdots & \frac{1}{N} \\
 \frac{-1}{N}      &  \frac{(N-1)}{N}        & \frac{-1}{N}     & \cdots & \frac{-1}{N} \\
 \vdots &  \frac{-1}{N}         & \ddots &        & \vdots  \\
 \vdots &  \vdots    &        & \ddots & \vdots \\
 \frac{-1}{N}      &  \frac{-1}{N}         & \cdots &        & \frac{(N-1)}{N}  \\
\end{bmatrix}.
\end{equation}
Substituting matrices $\mathcal{V} $ and $\mathcal{V}^{-1}$ in Eq.~\ref{eq:expm_spec} and using the information about the eigenvalues, we obtain the matrix in Eq.~\ref{eq:P_complete} and the same expression for the accessibility (Eq.~\ref{Eq:acccp}). 

As a practical comment about the matrix exponential, it is important to mention that it should be computed by the Pad\'e approximation~\cite{Golub1996, Higham05} and not by the truncated Taylor series or by Eqs.~\ref{eq:expm_spec}. The former method is more precise and has a lower computational cost. However, Eq.~\ref{eq:expm_spec} is important for theoretical analysis, since it transforms the calculus of accessibility into a eigenvector and eigenvalue problem, which is well studied in the literature.


\section{Epidemic and rumor spreading} \label{Sec:Spreading}

Many mathematical models have been developed to study epidemic spreading in complex networks~\cite{ Keeling05, Keeling08}. A particularly important model is the susceptible-infectious-recovered (SIR), in which each node can be in one of three states: (i) susceptible, (ii) infected, or (iii) recovered. Susceptible nodes are healthy and can catch the disease, whereas infected individuals are the ones actually transmitting the disease. Finally, individuals in the recovered state are immune to the disease and, therefore, play no role on the dynamics. The transitions between the first two states, i.e., from healthy to infected subjects, occurs via contacts between individuals. At each time step, the infectious nodes spread the disease to their susceptible neighbors with probability $\beta$ and an infected node becomes recovered with probability $\mu$. This is a spontaneous process and does not depend on any contact. The epidemic spreading process terminates when there is no infected node in the network and the disease cannot propagate anymore.

Rumor dynamics are in some aspects similar to epidemic spreading~\cite{Daley01, Castellano09:RMP}. Rumor diffusion is simulated considering that nodes are spreaders, ignorants, or stiflers. Spreaders are those individuals that know the rumor and want to spread it to ignorants, whereas stiflers are those that know the rumor but are not interested on the information anymore. The main difference between rumor and epidemic spreading is that spreader turns into a stifler by a process that involves contacts, whereas infected nodes become recovered by a spontaneous process. The fraction of ignorants ($\psi(t)$), spreaders ($\phi(t)$), and stiflers ($s(t)$) at time $t$  are defined such that $\psi(t) + \phi(t) + s(t) = 1$. The process starts with one spreader and $N-1$ ignorants, where $N$ is the number of nodes in the network. At each time step, spreaders try out to spread the rumor to their ignorant neighbors at a rate $\lambda$. On the other hand, if a spreader contacts another spreader or a stifler, such spreader becomes a stifler at rate $\delta$. This process corresponds to the model proposed by Maki and Thompson (MT model)~\cite{Castellano09:RMP}. In the version proposed by Daley and Kendall (DK model), two interacting spreaders become stiflers at rate $\lambda$~\cite{Castellano09:RMP}. Moreover, Monte Carlo simulations of a rumor spreading dynamics can be performed in two different ways. In a contact process (CP), only one random neighbor of a spreader is contacted at each time step. In the truncated process (TP) the neighbors of a spreader are contacted in a random way until all of them are contacted or the spreader turns into a stifler. The rumor dynamics terminates when there is no spreader in the network and the rumor cannot propagate anymore. 

Here, we consider that the spreading dynamics begin in a single seed node, whereas the remaining nodes are in the susceptible (or ignorant) state. In the SIR model, the spreading potential of each vertex is quantified in terms of the total prevalence of the epidemic process. The spreading capacity of $i$ is the fraction of recovered vertices at the end of the process given that the dynamics started in $i$, i.e., $M(i) = r(t \rightarrow \infty)$. Similarly, the spreading capacity of a node $i$ in rumor dynamics is quantified by the percentage of stiflers at the end of the process given that the spreading started at $i$, i.e., $M(i) = s(t \rightarrow \infty)$.


\section{Database} \label{Sec:data}

We performed numerical simulations of epidemic and rumor spreading processes on top of real-world and artificial networks. Table~\ref{tab:structure} presents some network properties of the road maps and networks generated by the spatial models.

\begin{table*}[t] 
\begin{center}
\caption{Structural properties of the complex networks.}
\begin{tabular}{|c|l|c|c|c|c|c|c|c|c|c|c|}
\hline
 & Network       &$N$    & $\langle k\rangle$ & $\langle cc_i \rangle$ & $\langle B_i \rangle$ & $\langle C_i \rangle$ & $\langle r_i \rangle$ & $\langle \pi \rangle$ &
 
$\langle\alpha\rangle$ & $\langle x_i \rangle$ & $\langle k_c \rangle$ \\

\hline
\multirow{8}{*}{\hspace{0.2cm}\begin{rotate}{90}
\hbox{Spatial}
\end{rotate}\hspace{0.2cm}} & Japan         & 2130 & 3.792 & 0.24 & 3.731$\times 10^{4}$ & 0.03 & 4.290 & 4.695$\times 10^{-4}$ & 6.95 & 2.892$\times 10^{-3}$ & 2.523 \\
& England       & 4460 & 3.415 & 0.14 & 8.163$\times 10^{4}$ & 0.03 & 3.557 & 2.242$\times 10^{-4}$ &  6.65 & 1.401$\times 10^{-3}$ & 2.062 \\
& United States & 6443 & 3.098 & 0.09 & 1.605$\times 10^{5}$ & 0.02 & 3.302 & 1.552$\times 10^{-4}$ & 6.178 & 9.328$\times 10^{-4}$ & 2.038 \\
& Germany       & 3555 & 3.068 & 0.08 & 5.944$\times 10^{4}$ & 0.03 & 3.173 & 2.813$\times 10^{-4}$ & 6.243 & 2.668$\times 10^{-3}$ & 1.988 \\
& SpatialSF     & 5000 & 3.998 & 0.04 & 1.226$\times 10^{4}$ & 0.17 & 9.291 & 2.000$\times 10^{-4}$ & 9.793 & 6.001$\times 10^{-3}$ & 2.000 \\
& Waxman        & 4883 & 4.078 & 0.14 & 4.598$\times 10^{4}$ & 0.05 & 4.863 & 2.048$\times 10^{-4}$ &  8.071 & 1.433$\times 10^{-3}$ & 2.570 \\

\hline
\multirow{5}{*}{\vspace{-1.5cm}\hspace{0.2cm}\begin{rotate}{90}
\hbox{Non-spatial}
\end{rotate}\hspace{0.2cm}} 
& advogato         & 5054 & 15.58 & 0.25 & 5.748$\times 10^{3}$ & 0.31 & 9.962$\times 10^{1}$ & 1.979$\times 10^{-4}$ & 28.92 & 6.819$\times 10^{-3}$ & 8.137 \\
& e-mail           & 1133 & 9.622                & 0.22 & 1.475$\times 10^{3}$ & 0.28 & 1.790$\times 10^{1}$ & 8.826$\times 10^{-4}$ & 17.88 & 1.764$\times 10^{-2}$ & 5.349 \\
& Political blogs & 1222 & 27.36 & 0.32 & 1.061$\times 10^{3}$ & 0.37 & 1.001$\times 10^{2}$ & 8.183$\times 10^{-4}$ & 33.08 & 1.681$\times 10^{-2}$ & 14.82 \\
& Google+ & 23613  & 3.319 & 0.17 & 3.580$\times 10^{4}$ & 0.25 & 7.270$\times 10^{2}$ & 4.235$\times 10^{-5}$ & 15.13 & 2.301$\times 10^{-3}$ & 1.669 \\
& BA      & 10000  & 3.999 & 5.76$\times 10^{-3}$ & 2.005$\times 10^{4}$ & 0.20 & 1.706$\times 10^{1}$ & 1.000$\times 10^{-4}$ & 10.57 & 3.108$\times 10^{-3}$ & 2.000\\
\hline
\end{tabular}
\label{tab:structure}
\end{center} 
\end{table*}

\subsection{Network models} \label{Sec:models}

Barab\'{a}si and Albert proposed a model which considers growth and preferential attachment rules~\cite{albert1999emergence}. In this case, a network is generated starting with a set of $m_0$ connected vertices.  After that, new vertices with $m$ edges are included in the network. The probability of the new vertex $i$ to connect with a vertex $j$ in the network is proportional to the number of connections of $j$, i.e.,
\begin{equation}
p(i,j) = \frac{k_j}{\sum_u k_u}.
\end{equation}
The most connected vertices have greater probability of receiving new vertices. In this way, networks generated by this model present a power-law degree distribution, $P(k) = k^{-\gamma}$, where $\gamma = 3$ in the thermodynamic  limit ($N\rightarrow \infty$)~\cite{albert1999emergence}, $N$ being the number of nodes. 

We also take into account two spatial models. The model proposed by Waxman~\cite{waxman1988routing} considers that nodes are uniformly distributed into a square of unitary area and each pair of nodes is connected according to a probability, that depends on their distances, i.e.,
\begin{equation}
p(i,j) = \eta \exp(-\eta d_{ij}), 
\end{equation}
where $\eta$ is a parameter that controls the average degree and $d_{ij}$ is the Euclidean distance between nodes $i$ and $j$. Such model generates networks with an exponential degree distribution, which means that the probability of a node having a degree different than $\langle k \rangle$ decays exponentially. 

The model introduced by Barth{\'e}lemy~\cite{Barthelemy2003:EPL}, on the other hand, produces scale-free networks embedded in space. Considering a regular $d$ dimensional
lattice with length $L$, the algorithm has three main steps. Initially, $n_0$ initial active nodes are selected at random. Next, an inactive node $i$ is randomly selected, and connected to an active node $j$ with probability 
\begin{equation}
p(i,j) \propto \frac{k_j+1}{\exp(d_{ij}/r_c)},
\end{equation}
where $k_j$ is the number of connections of node $j$, $r_c$ is a finite scale parameter and $d_{ij}$ is the Euclidean distance between nodes $i$ and $j$. Finally, the node $i$ becomes active and the second and third steps are repeated until all nodes are active. For each node, the second and third steps are repeated $m$ times in order to set the average connectivity as $\langle k \rangle = 2m$~\cite{Barthelemy2003:EPL}. The parameter $r_c$ controls the clustering coefficient~\cite{wattssmall} and
assortativity~\cite{Newman02:PRL} of the network. Here we considered $r_c = 0.05$, $L = 1$ and $d = 2$. These values are similar to those used in the original paper~\cite{Barthelemy2003:EPL}. 

\subsection{Road networks}
\label{sec:real}

The road networks have been extracted from the maps available as portable format (\emph{pdf}) at the United Nations website~\footnote{http://www.un.org}. Initially, the maps have been pre-processed in order to eliminate irrelevant information and keep only the main roads. After that, the skeletonization procedure has extracted the so called skeleton of the image~\cite{Costa2000:book}. The node identification has been performed by applying a 8-connected hit-or-miss convolution filter~\cite{Dougherty:book}. Finally, a label propagation procedure has been implemented from each node. When two pair of labels $i$ and $j$ find each other, a connection is established between them. Here, we have considered the networks extracted from maps of Germany, Japan, England and United States.

\subsection{Social networks}

The social networks considered here are: (i) the email contact network obtained from messages exchanged between users within the Universitat Rovira i Virgili~\cite{PhysRevE.68.065103}; (ii) the political blogs network, composed of hyperlinks between web blogs obtained over the period of two months preceding the U.S. Presidential Election of 2004~\cite{Adamic05} ; (iii) the advogato network, which is an online community dedicated to free software development launched in 1999~\cite{konect:massa09, konect:2014:advogato} and (iv) the Google+ network, which is composed by users connected according to their circles of friendships~\cite{konect:McAuley2012, konect:2014:ego-gplus}. Avogato, political blogs and Google+ networks are directed networks. Morevoer, advogato is also a weighted network. However, here we consider only the unweighted and undirected versions of these networks. In addition, our analysis uses only the nodes in the giant component.


\section{Spatial networks} \label{Sec:spatial}

As outlined in Section~\ref{Sec:cent}, we have studied different centrality metrics: the degree ($k$), clustering coefficient ($cc$), betweenness centrality ($B$), average neighborhood degree ($r$), PageRank ($\pi$), eigenvector centrality ($x$), k-core index ($k_c$), closeness centrality ($C$) and accessibility ($\alpha$). We have considered only the unweighted and undirected versions of these measures. Table~\ref{tab:structure} presents the average values obtained for the road maps and networks generated by the Waxman and scale-free spatial models. Spatial networks are sparse, have large characteristic path lengths and non-zero clustering coefficients. In addition, scale-free spatial networks have the smallest average geodesic distance due to the presence of hubs. 

We have conducted numerical simulations of the SIR (epidemic) and MT (rumor) models to inspect correlations between nodes' centrality (as given by the different metrics above) and the final dynamical outcome of the system, the latter being measured by the density of removed and stiflers after the dynamics has come to an end, respectively. Such correlations have been determined by the Spearman rank correlation coefficient, which is defined as the Pearson correlation coefficient between the ranked variables~\cite{Wolfe73}. The reason of our choice is that the Spearman coefficient quantifies monotonic relationships, whereas the Pearson correlation measures linear relationships. As shown below, these correlations can be monotonic, but not necessarily linear.

\begin{figure}[t]
\begin{center}
\subfigure[]{\includegraphics[width=0.45\columnwidth]{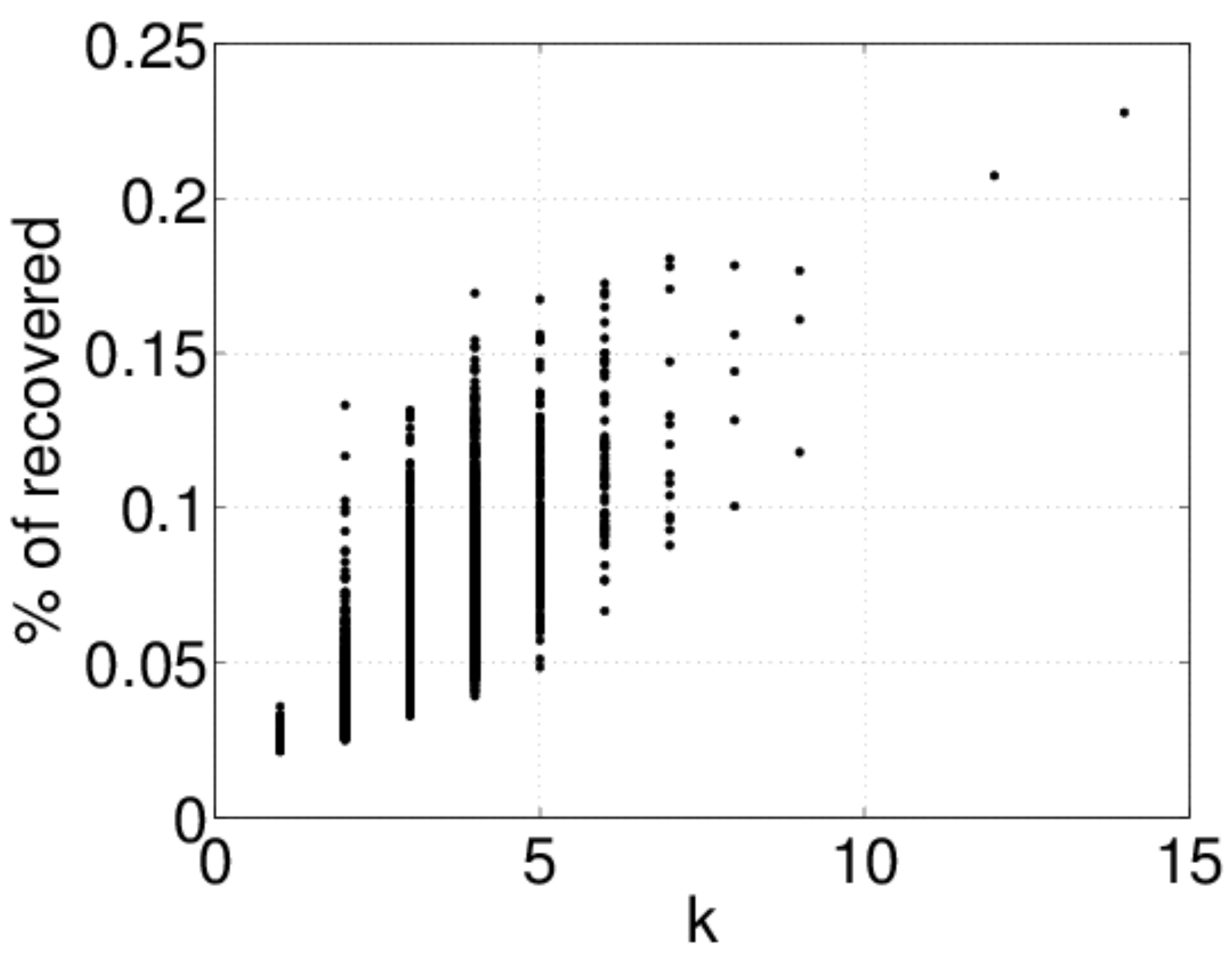}}
\subfigure[]{\includegraphics[width=0.45\columnwidth]{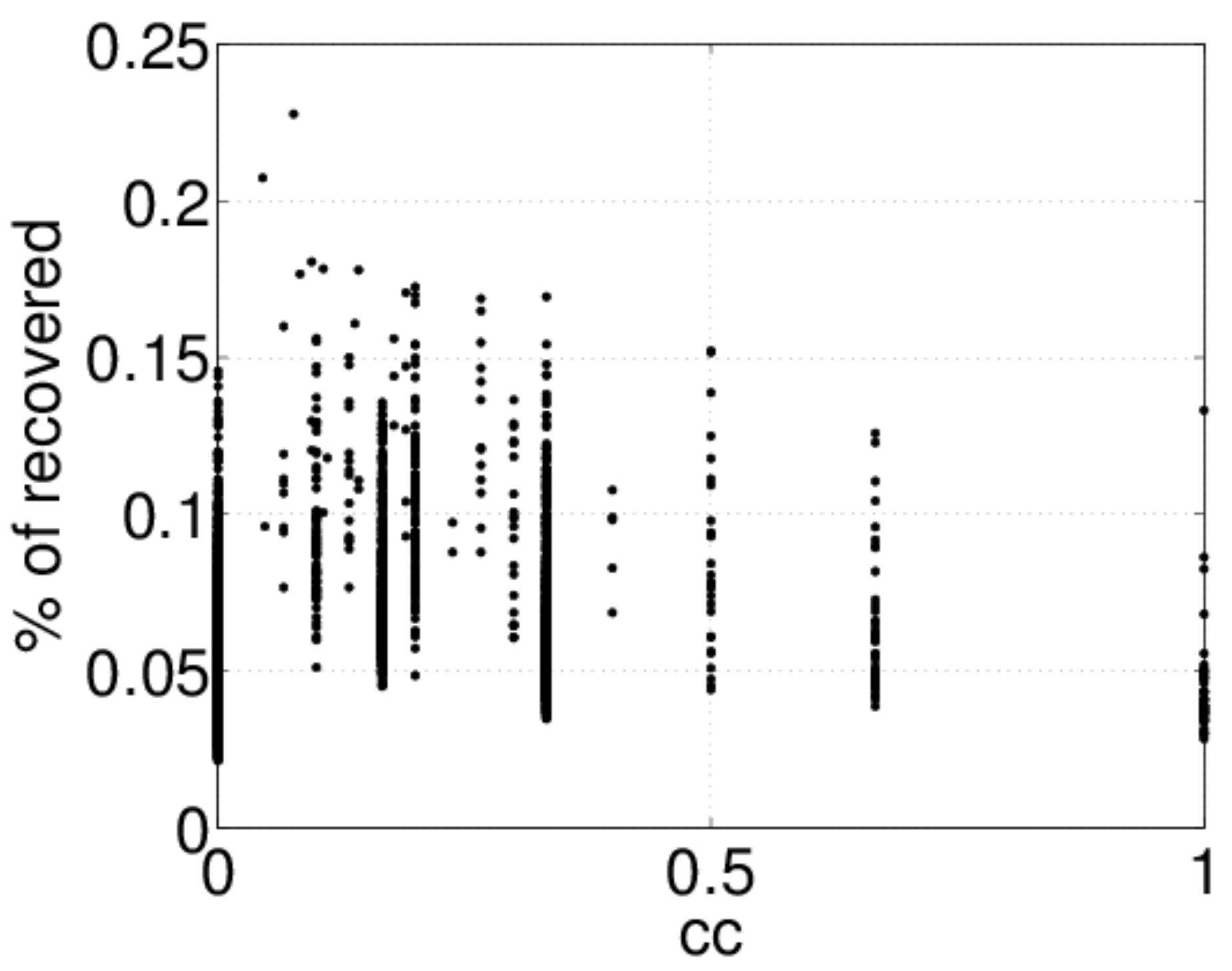}}
\subfigure[]{\includegraphics[width=0.45\columnwidth]{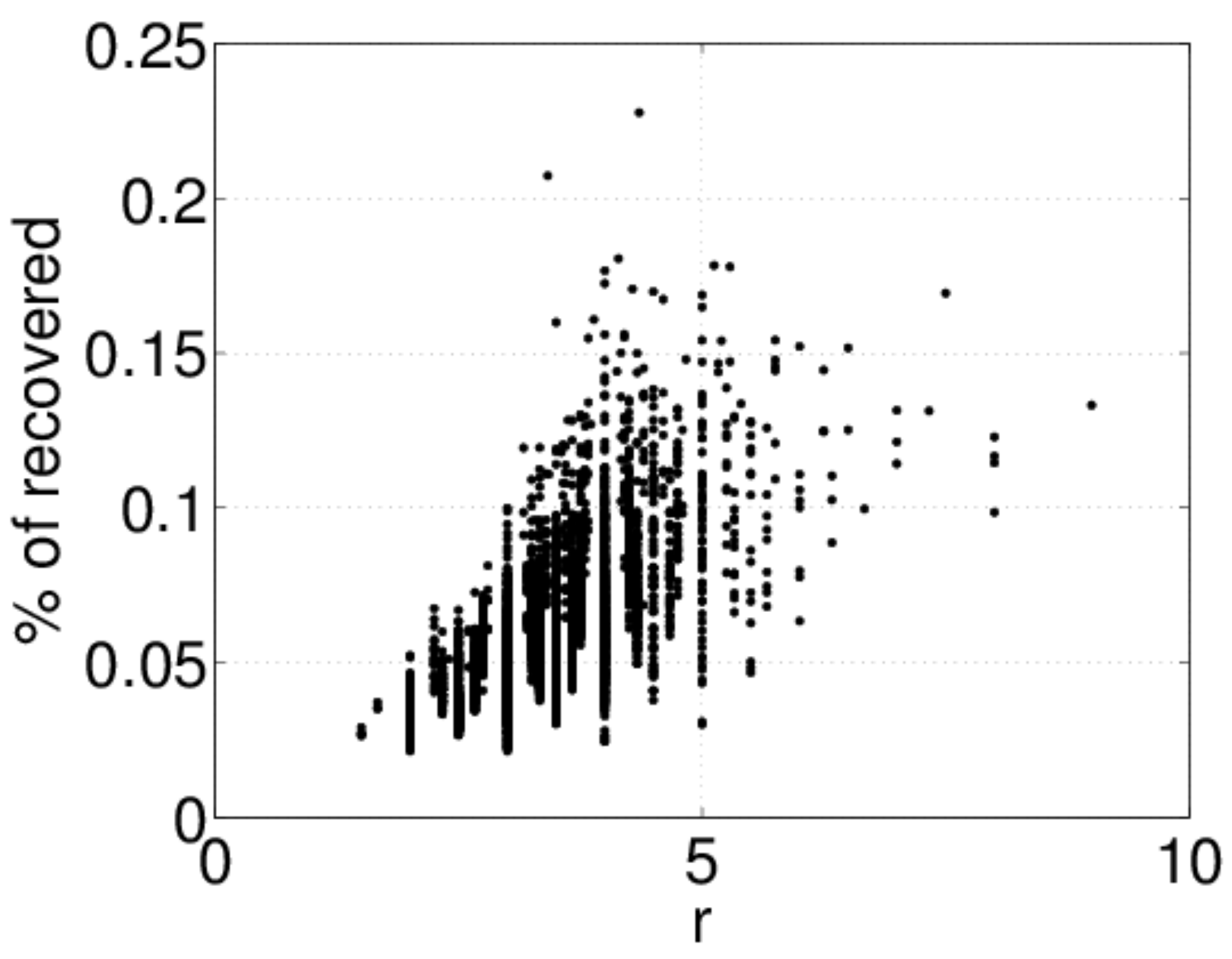}}
\subfigure[]{\includegraphics[width=0.45\columnwidth]{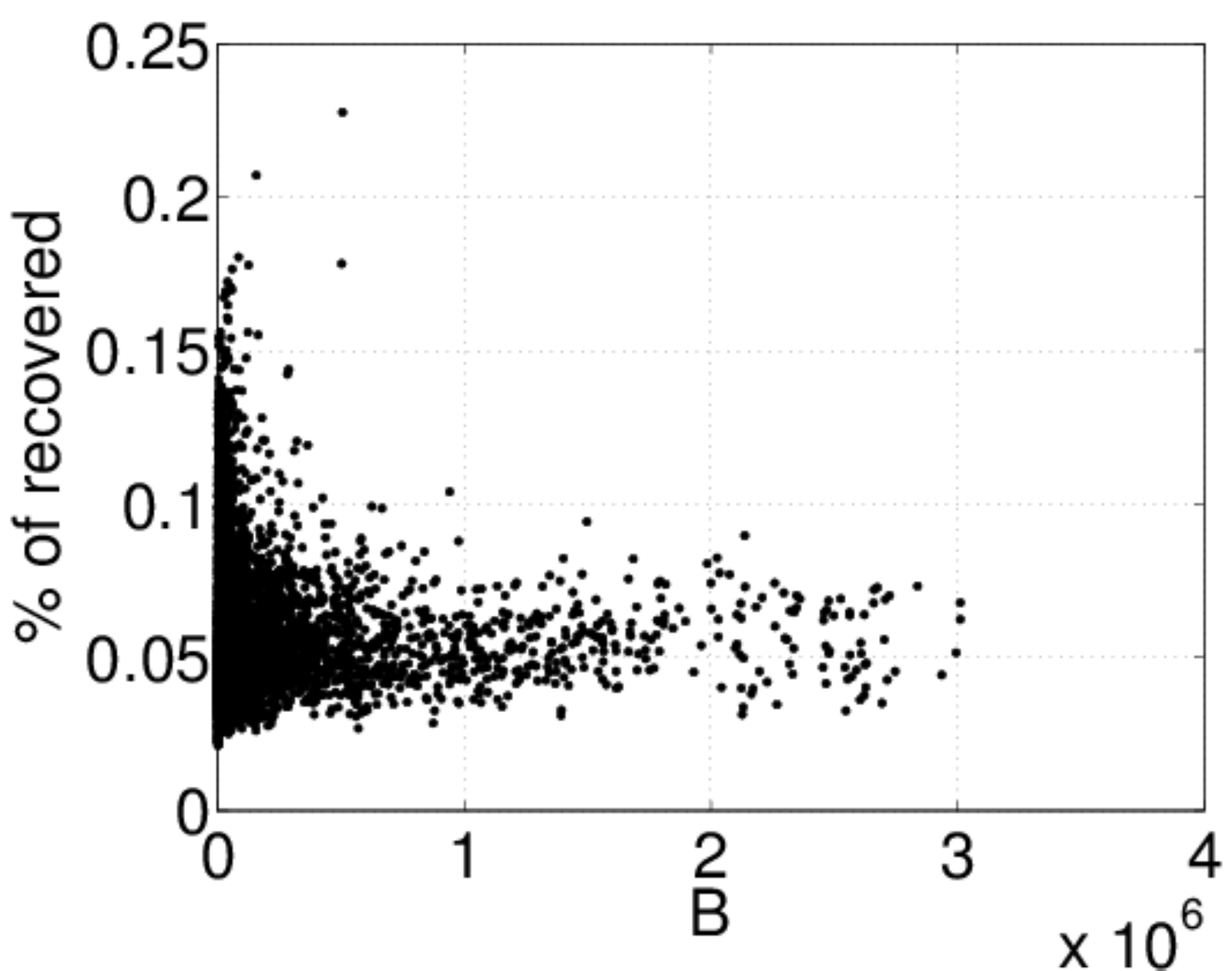}}
\subfigure[]{\includegraphics[width=0.45\columnwidth]{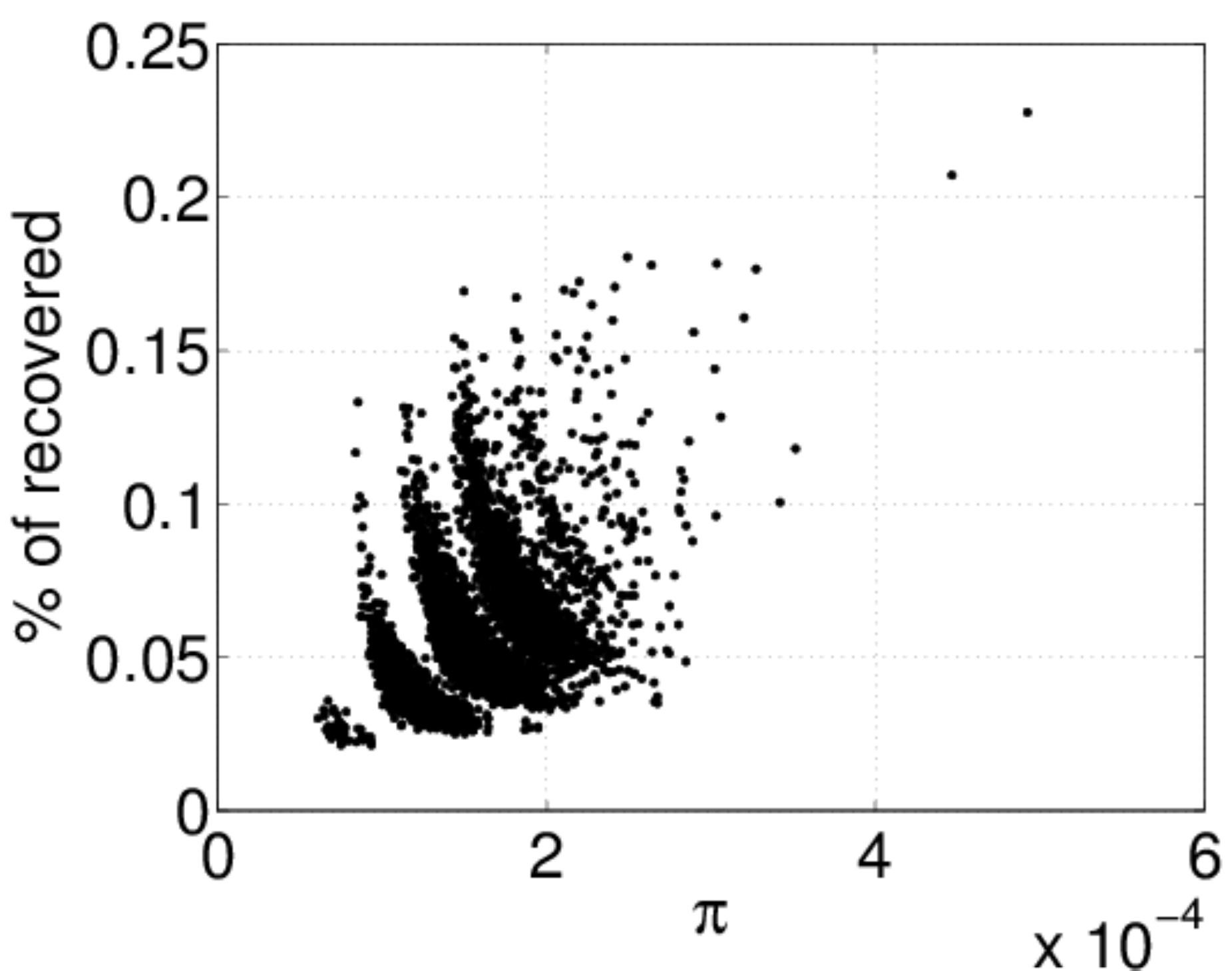}}
\subfigure[]{\includegraphics[width=0.45\columnwidth]{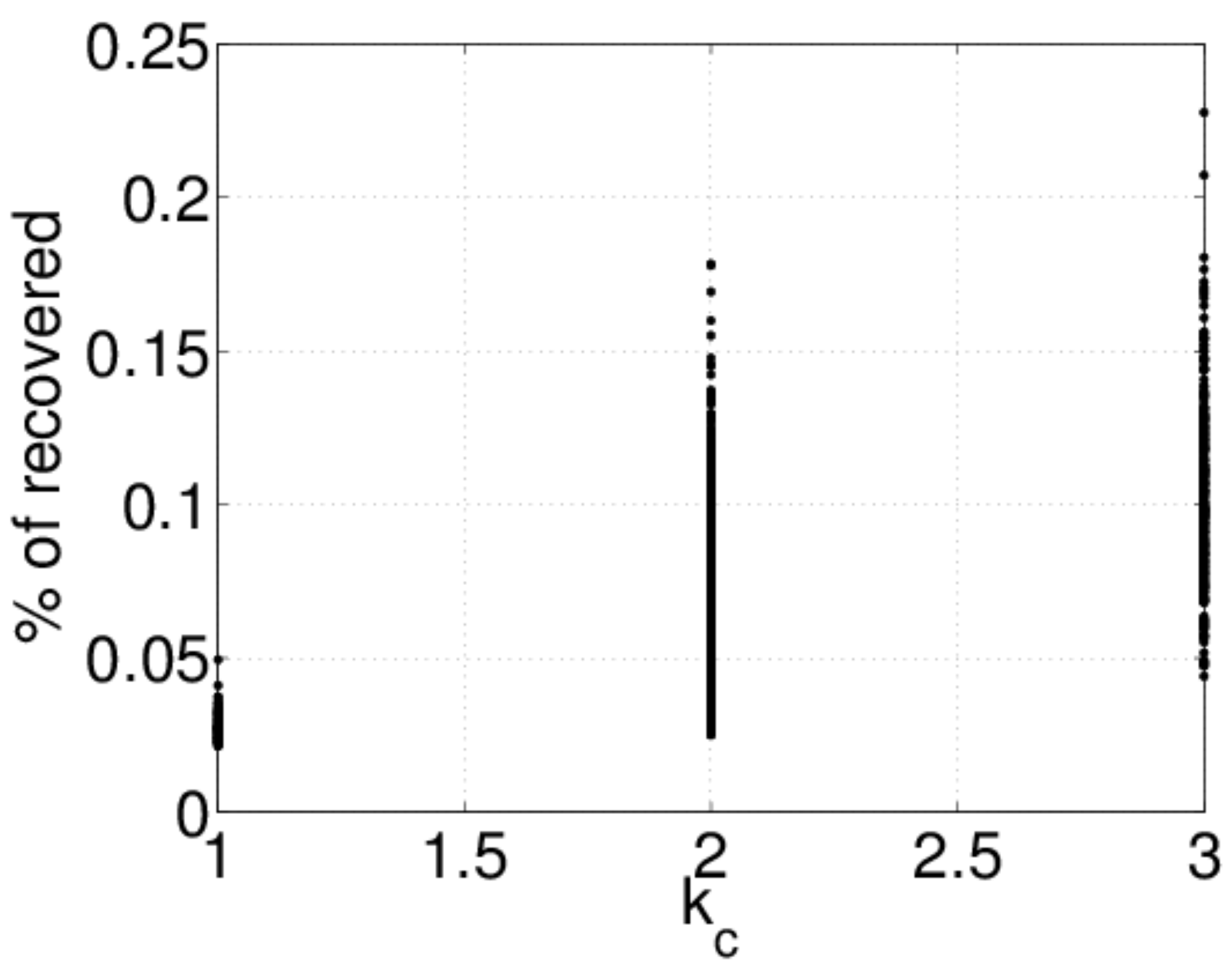}}
\subfigure[]{\includegraphics[width=0.45\columnwidth]{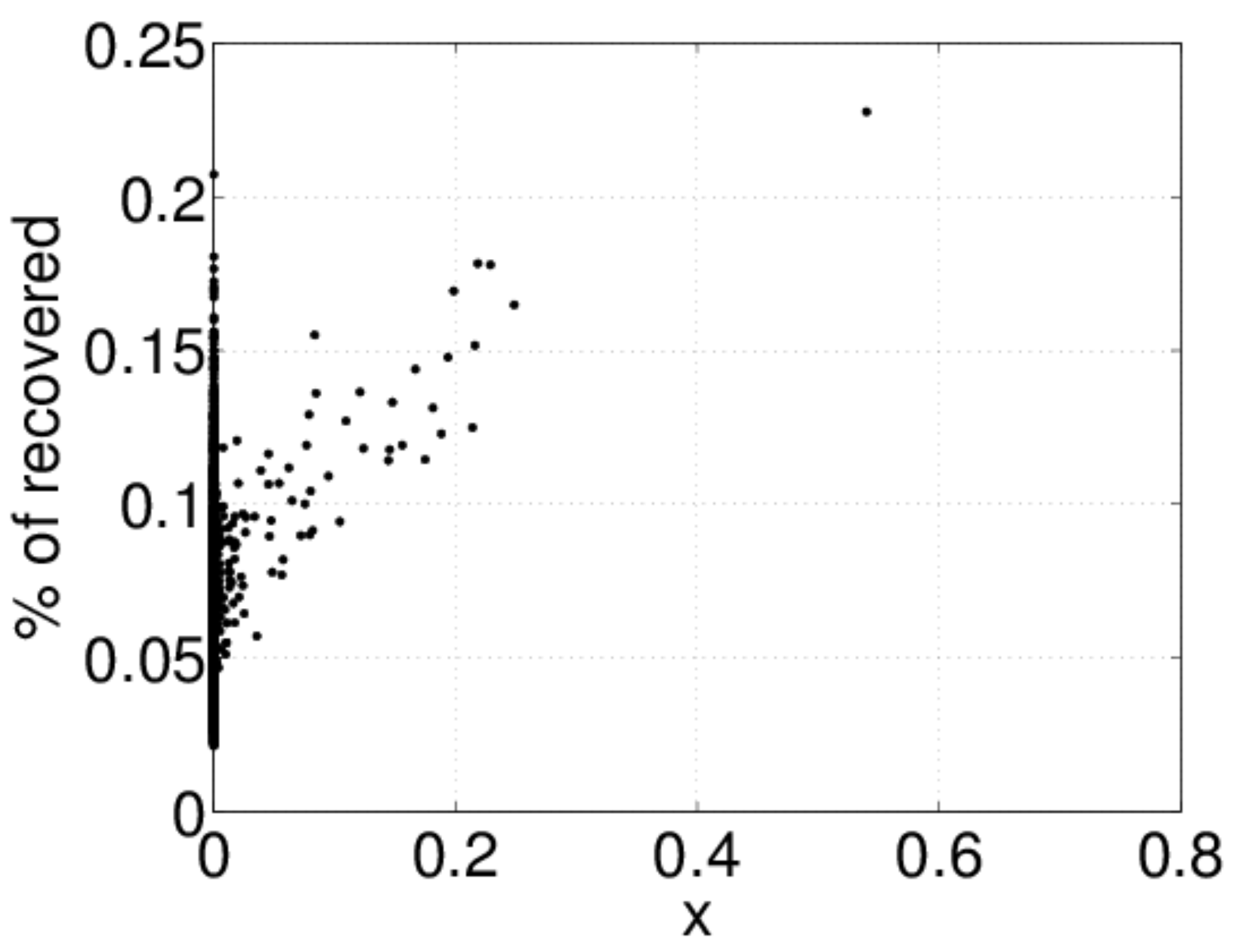}}
\subfigure[]{\includegraphics[width=0.45\columnwidth]{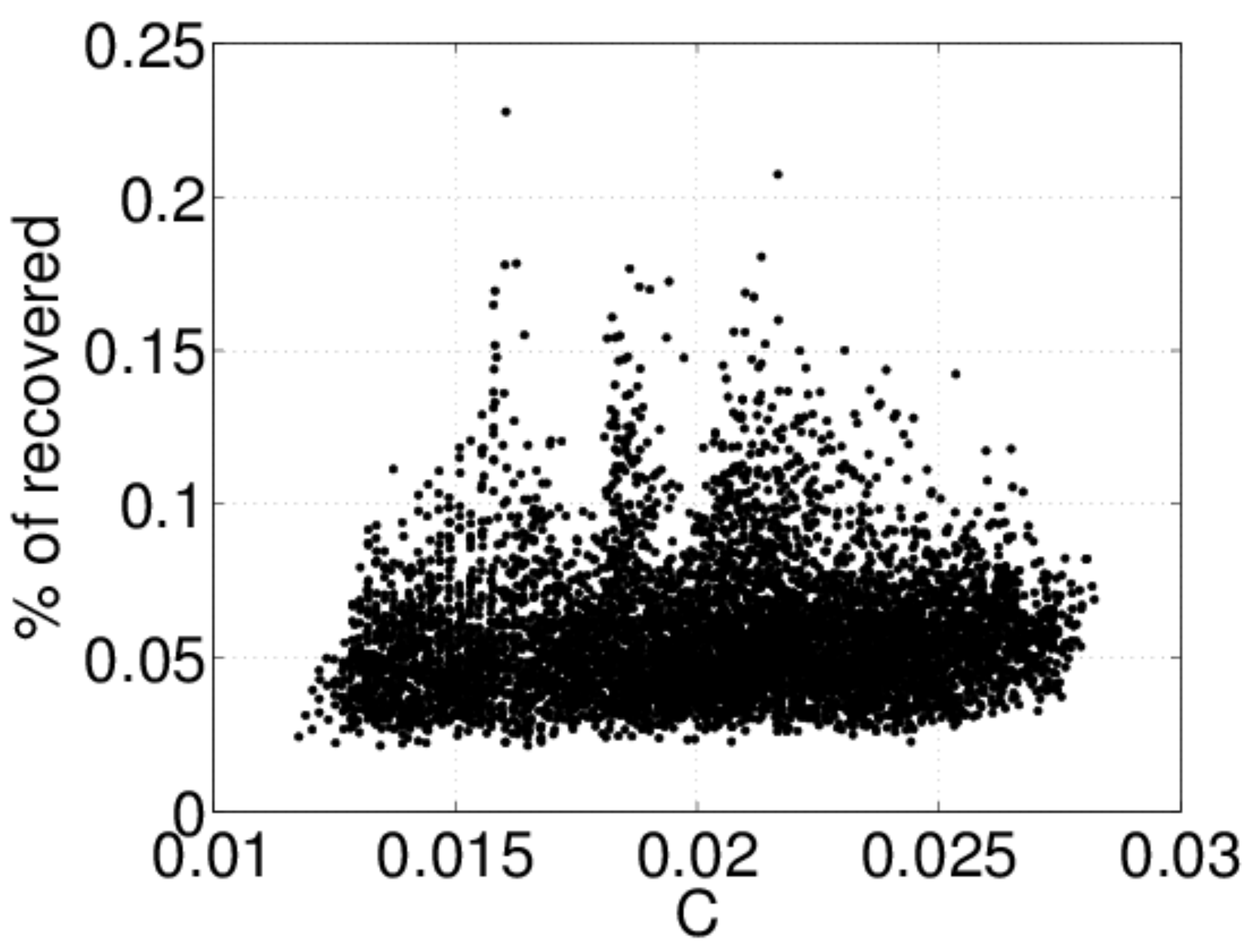}}
\caption{The percentage of recovered individuals in the SIR epidemic spreading model ($\beta = 0.3, \mu = 1.0$) according to the local measures for the US road network: (a) degree; (b) clustering coefficient; (c) average degree of the nearest neighbors; (d) betweenness centrality; (e) PageRank; (f) k-core index; (g) eigenvector centrality and (h) closeness centrality.}
\label{Fig:Corr_SIR_RP}
\end{center}
\end{figure}

\begin{figure}[t]
\begin{center}
\subfigure[]{\includegraphics[width=0.45\columnwidth]{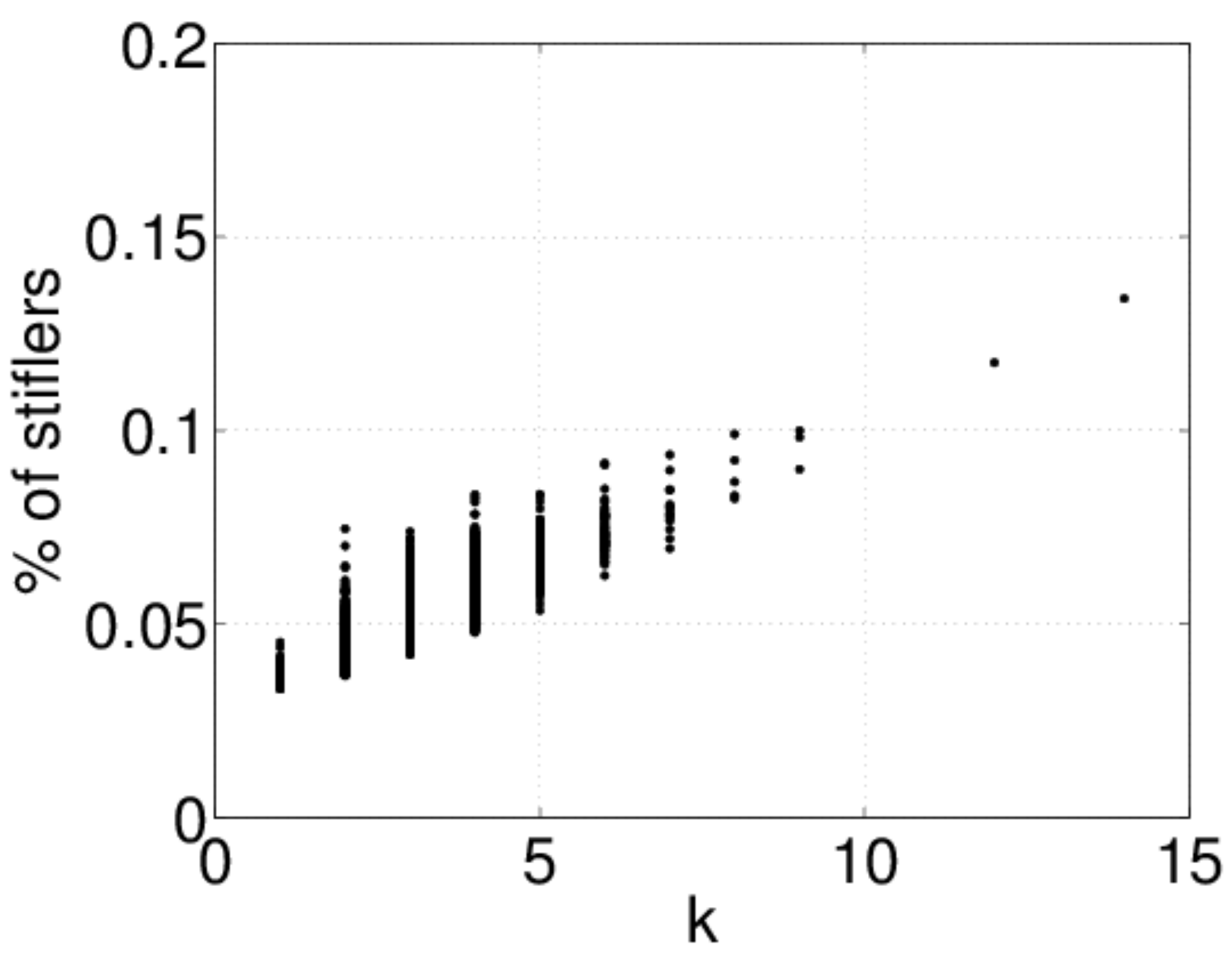}}
\subfigure[]{\includegraphics[width=0.45\columnwidth]{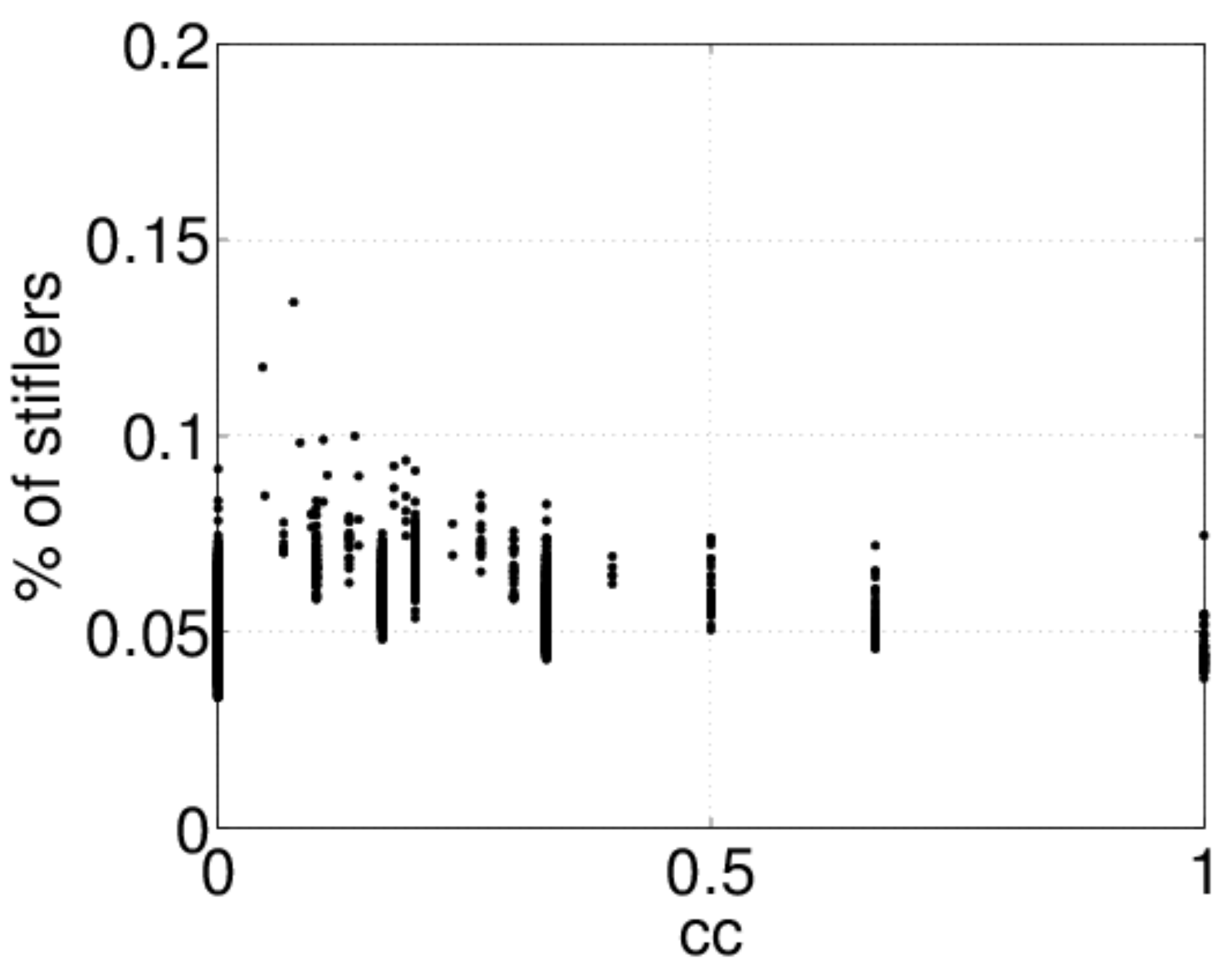}}
\subfigure[]{\includegraphics[width=0.45\columnwidth]{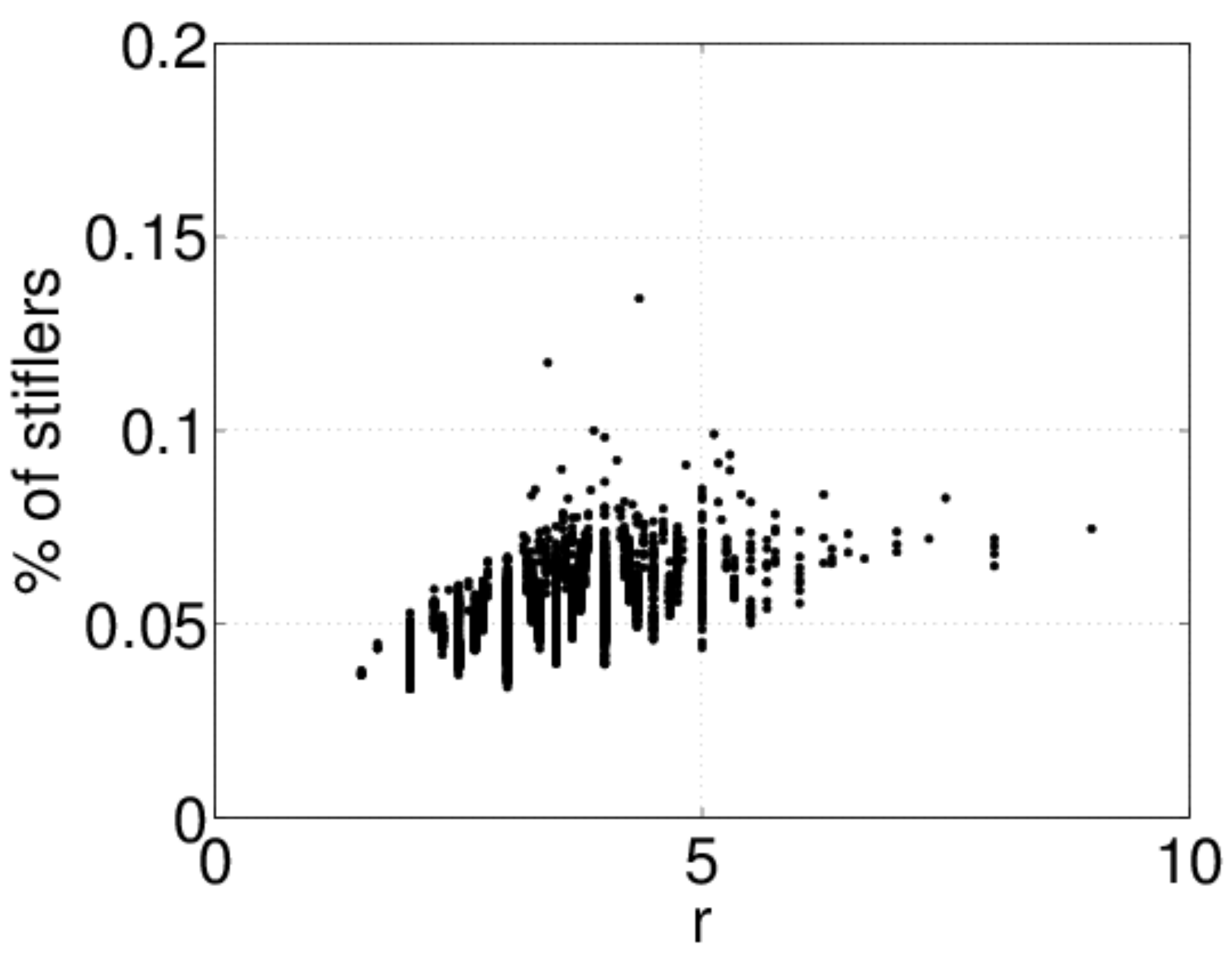}}
\subfigure[]{\includegraphics[width=0.45\columnwidth]{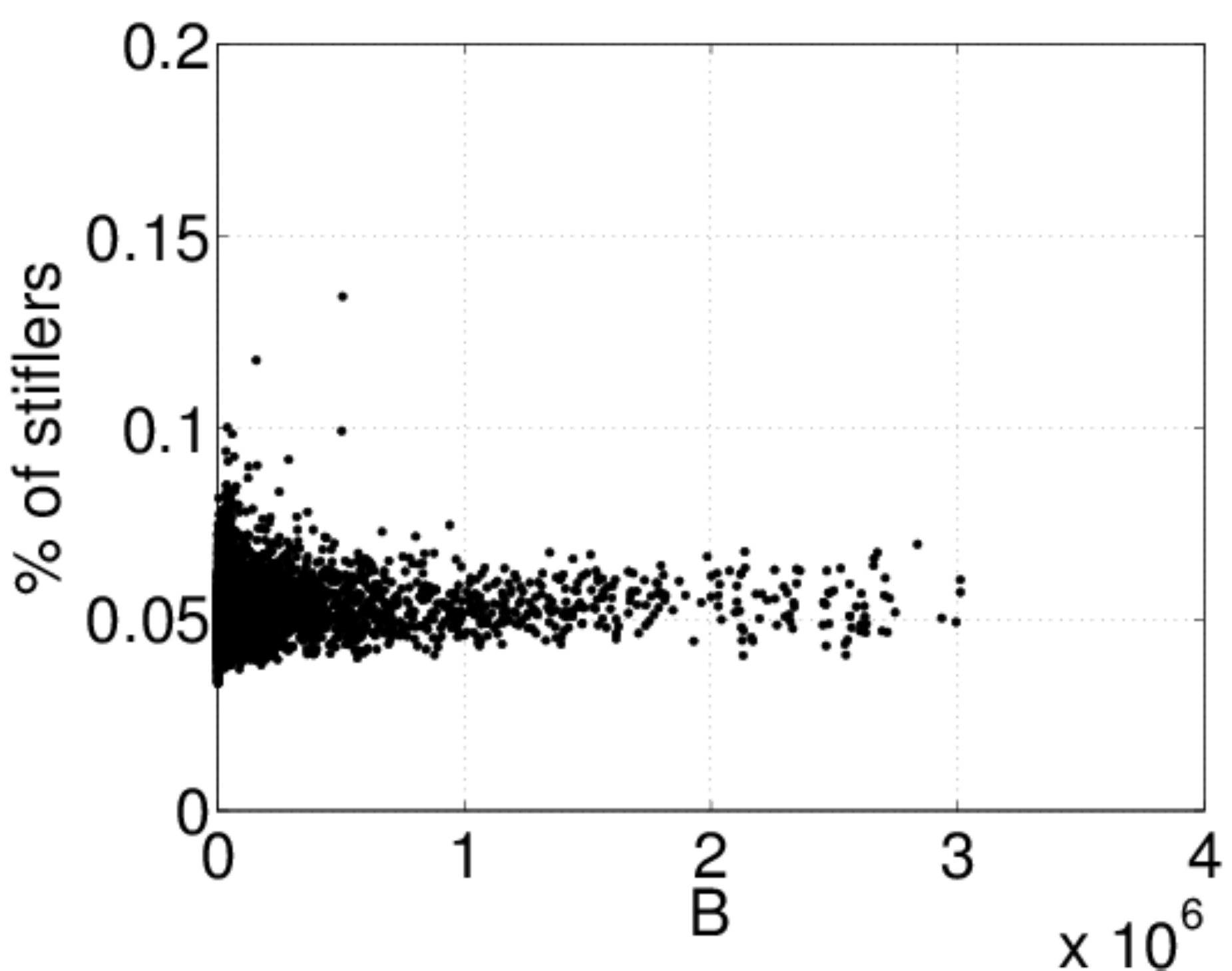}}
\subfigure[]{\includegraphics[width=0.45\columnwidth]{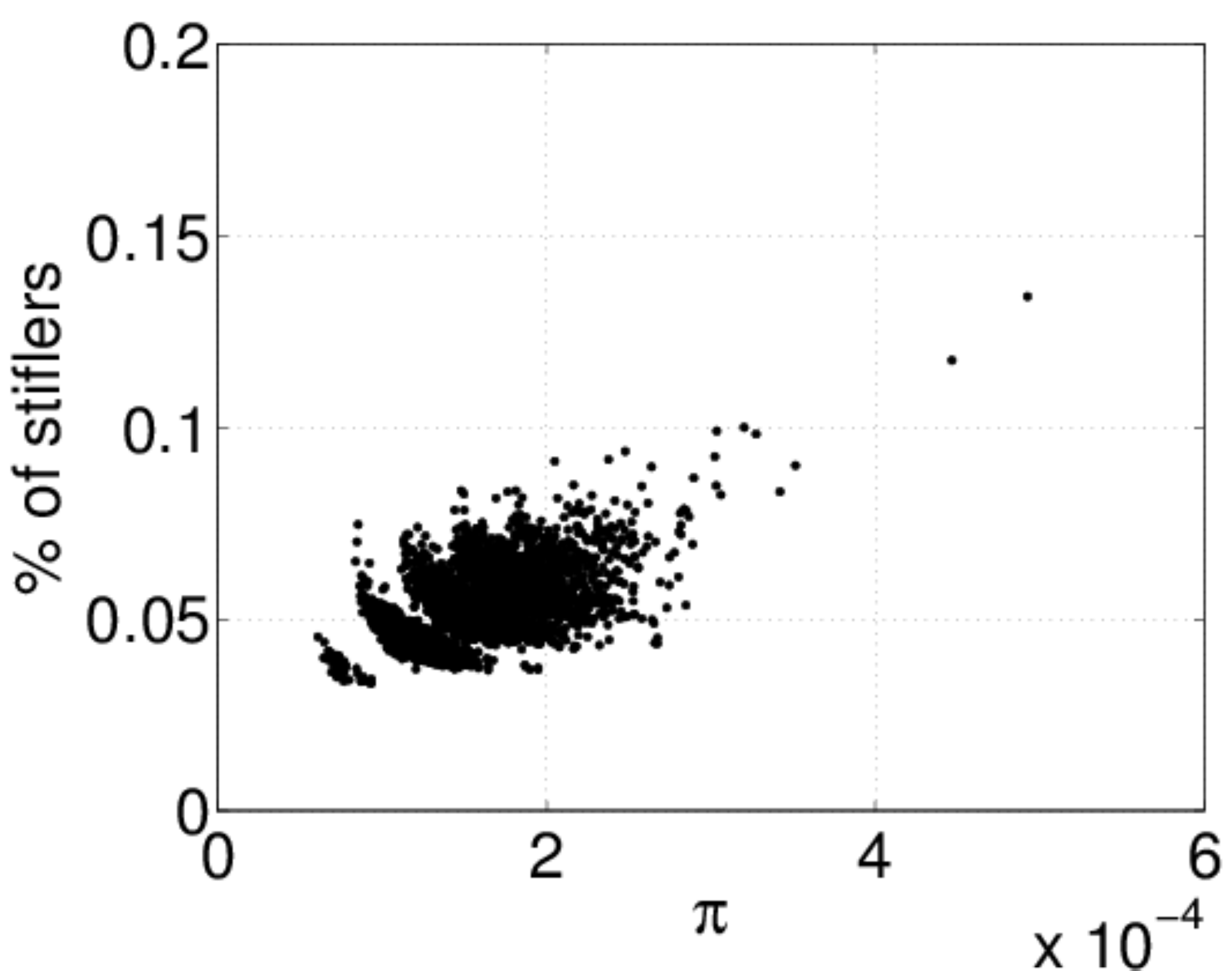}}
\subfigure[]{\includegraphics[width=0.45\columnwidth]{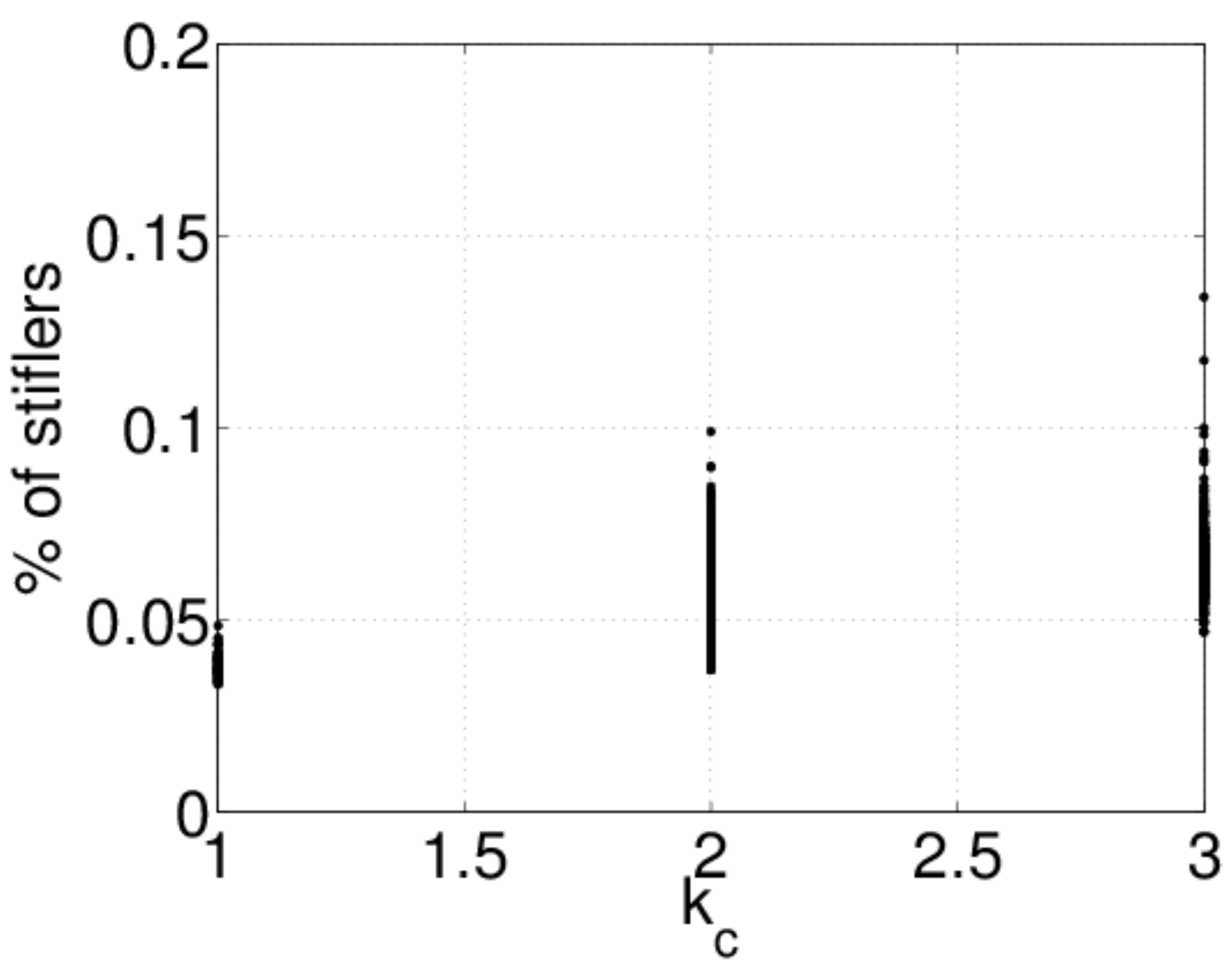}}
\subfigure[]{\includegraphics[width=0.45\columnwidth]{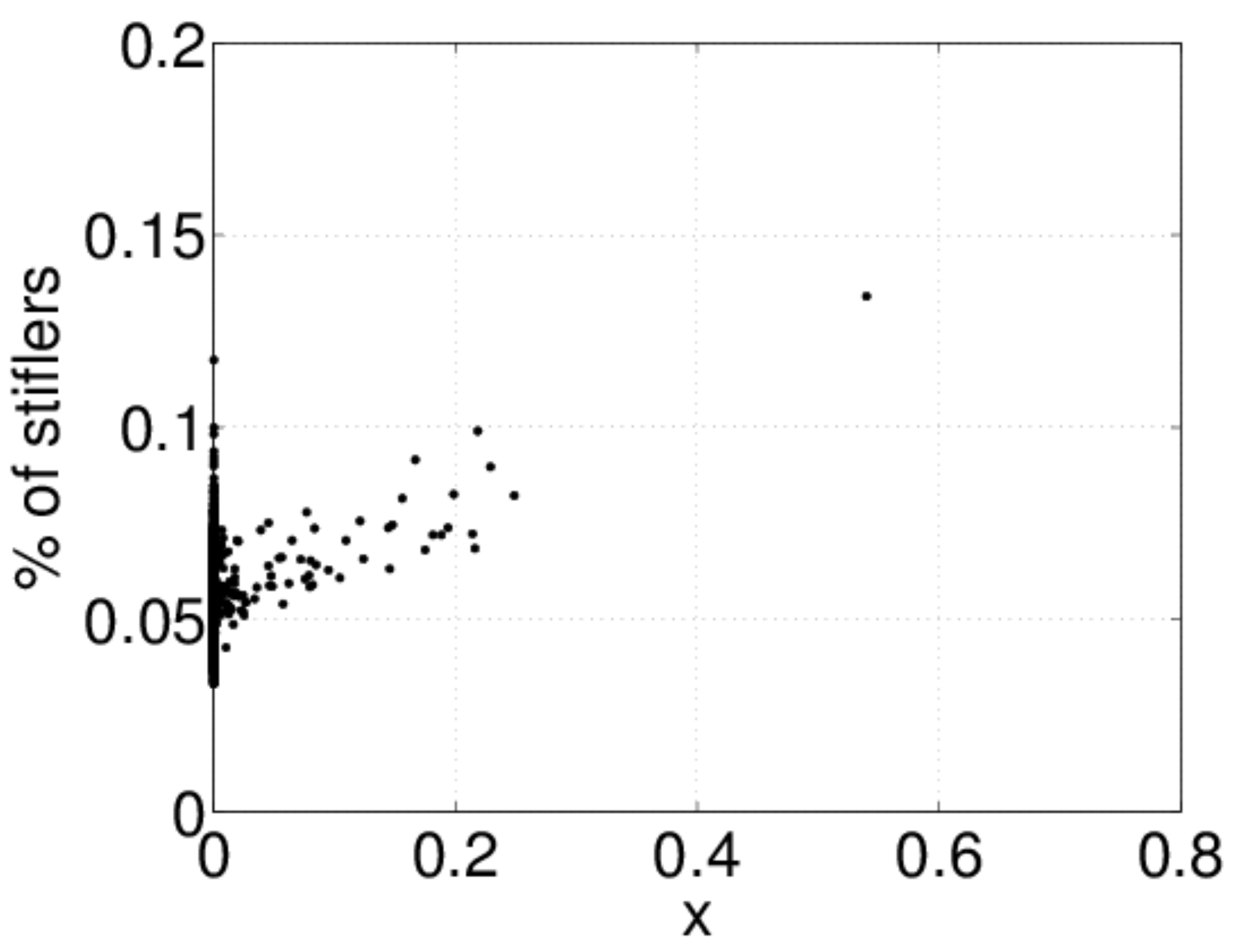}}
\subfigure[]{\includegraphics[width=0.45\columnwidth]{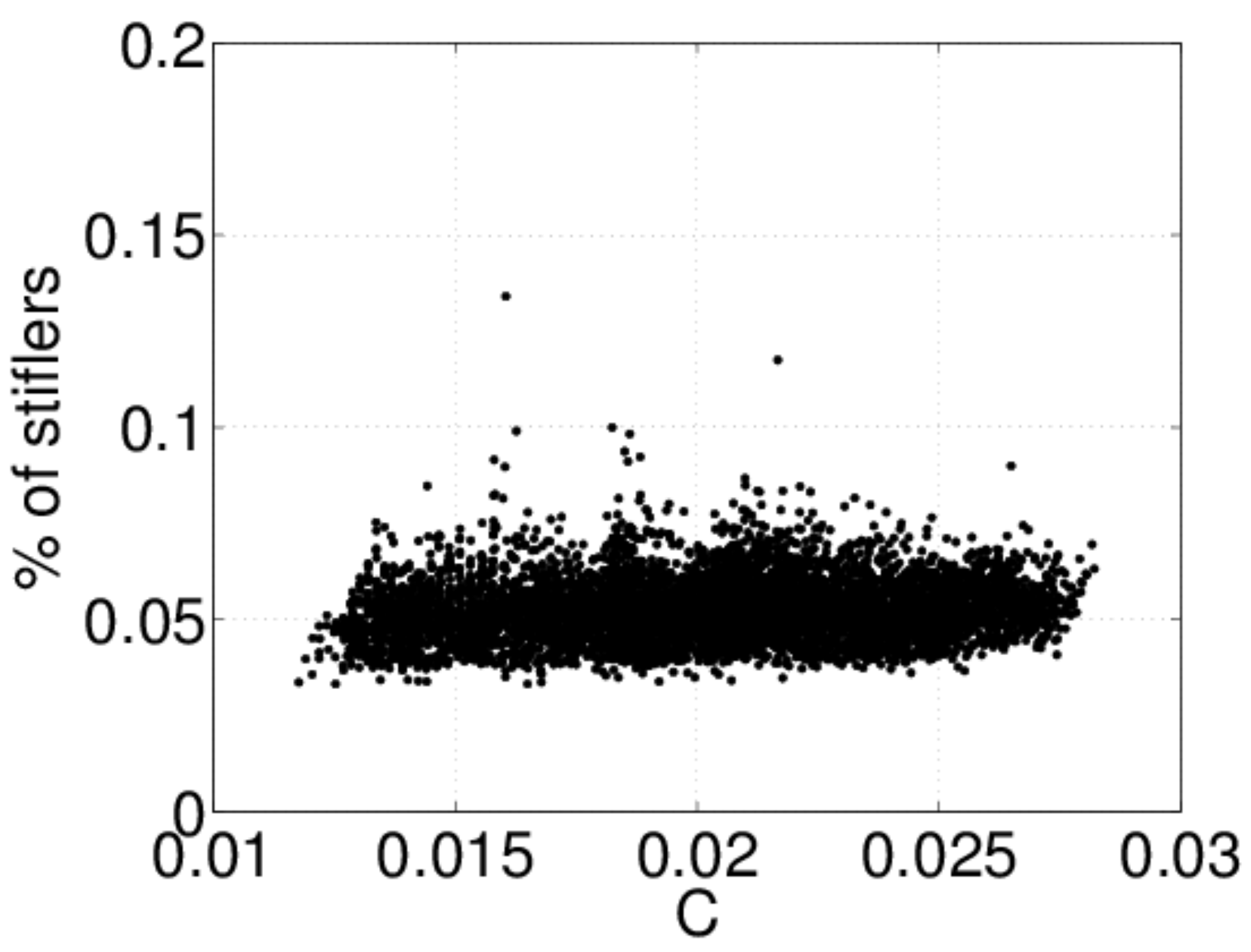}}
\caption{The percentage of stiflers on the MT (TP) rumor model ($\lambda = 0.3, \delta = 1.0$) according to the local measures for the United States network: (a) degree; (b) clustering coefficient; (c) average degree of the nearest neighbors; (d) betweenness centrality; (e) PageRank; (f) k-core index; (g) eigenvector centrality and (h) closeness centrality.}
\label{Fig:Corr_MK_TP}
\end{center}
\end{figure}

\begin{figure}[t]
\begin{center}
\subfigure[]{\includegraphics[width=0.45\columnwidth]{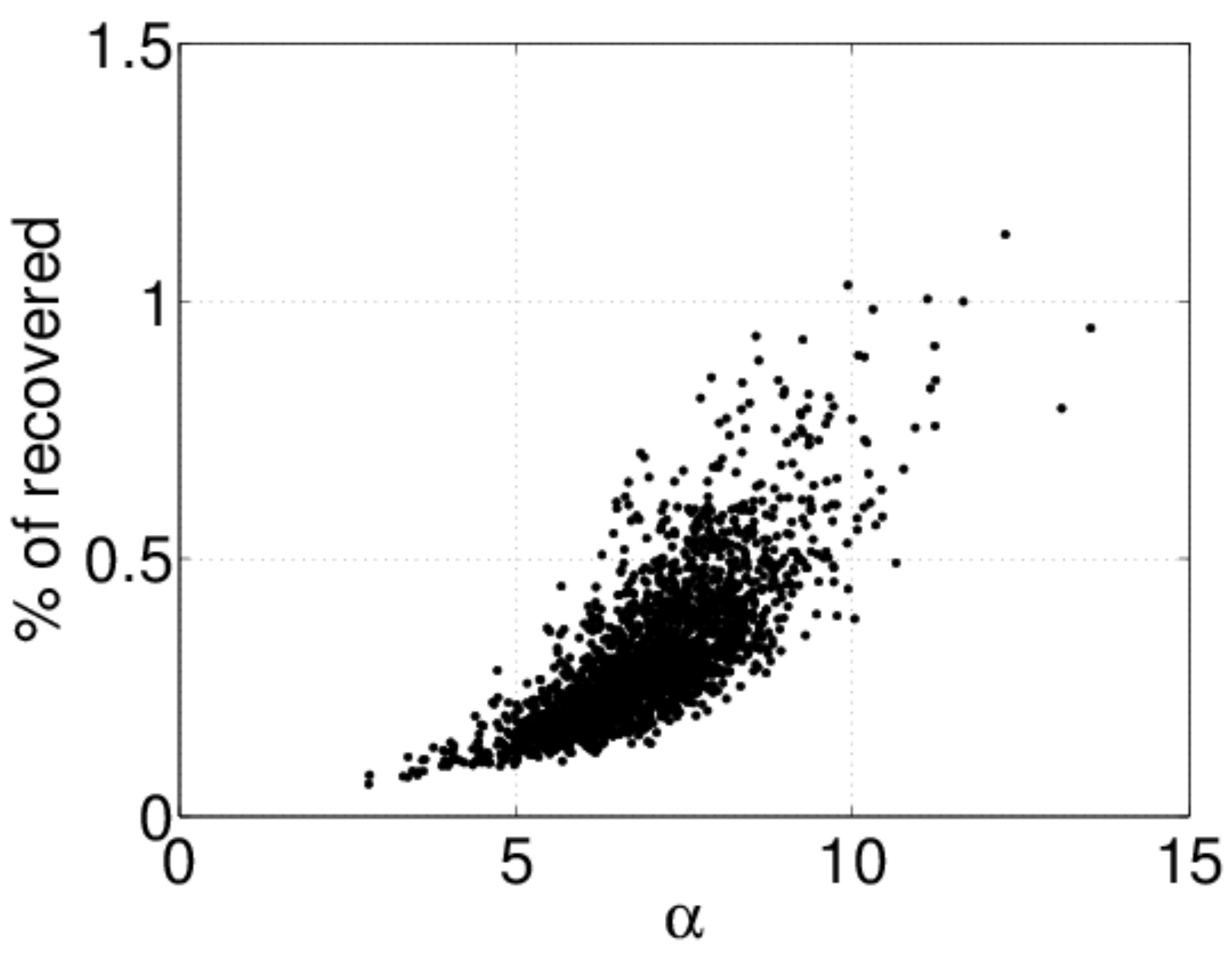}}
\subfigure[]{\includegraphics[width=0.45\columnwidth]{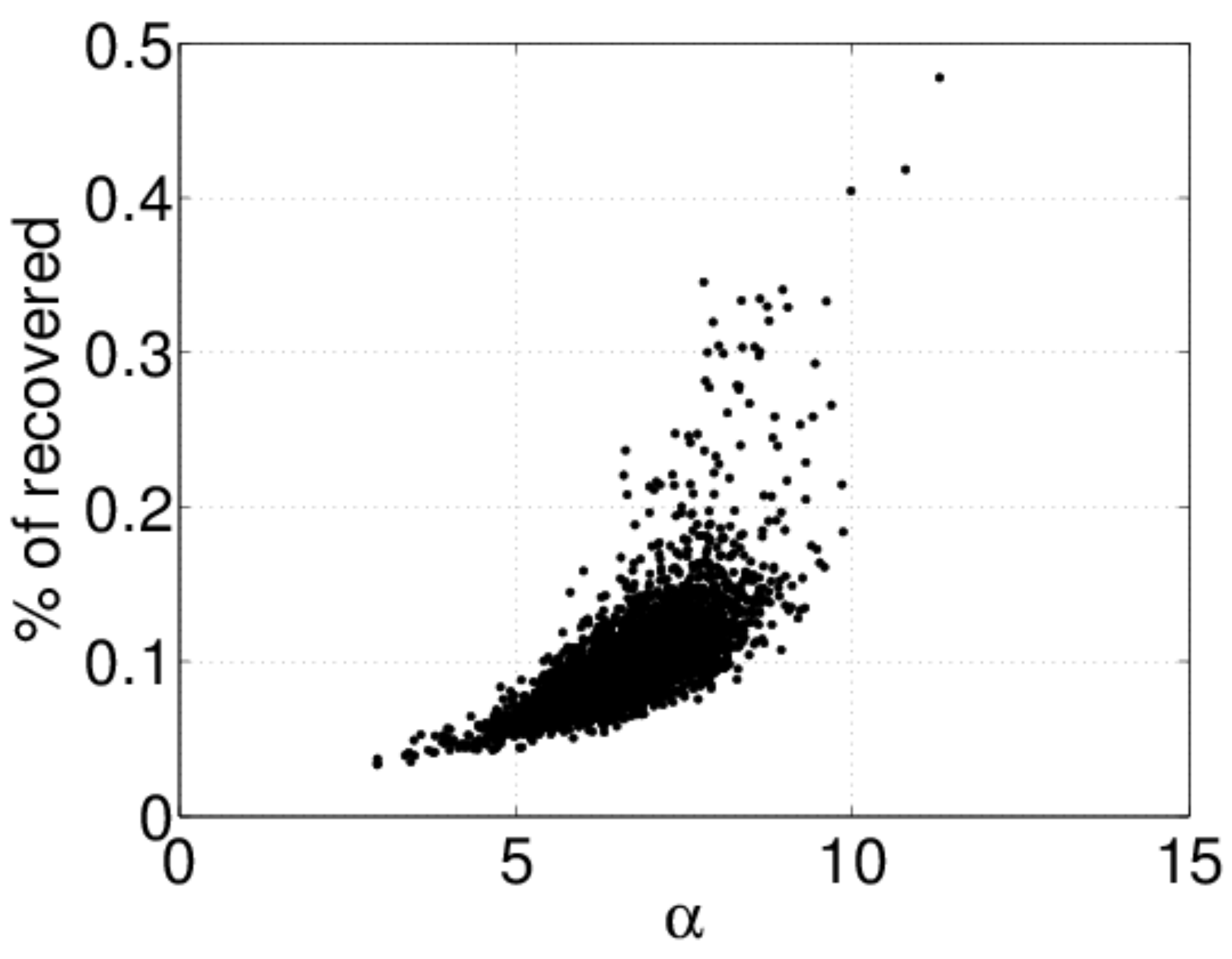}}
\subfigure[]{\includegraphics[width=0.45\columnwidth]{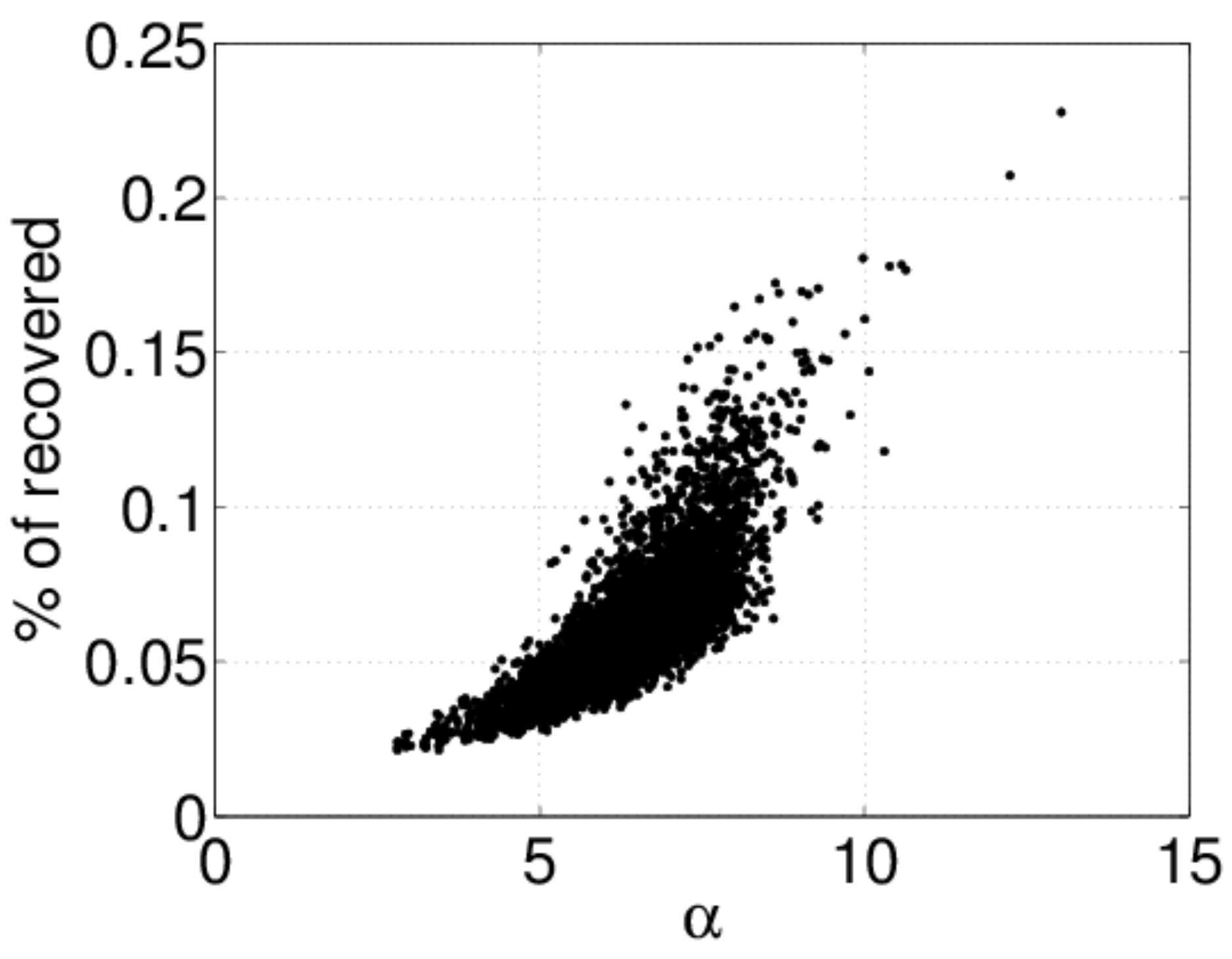}}
\subfigure[]{\includegraphics[width=0.45\columnwidth]{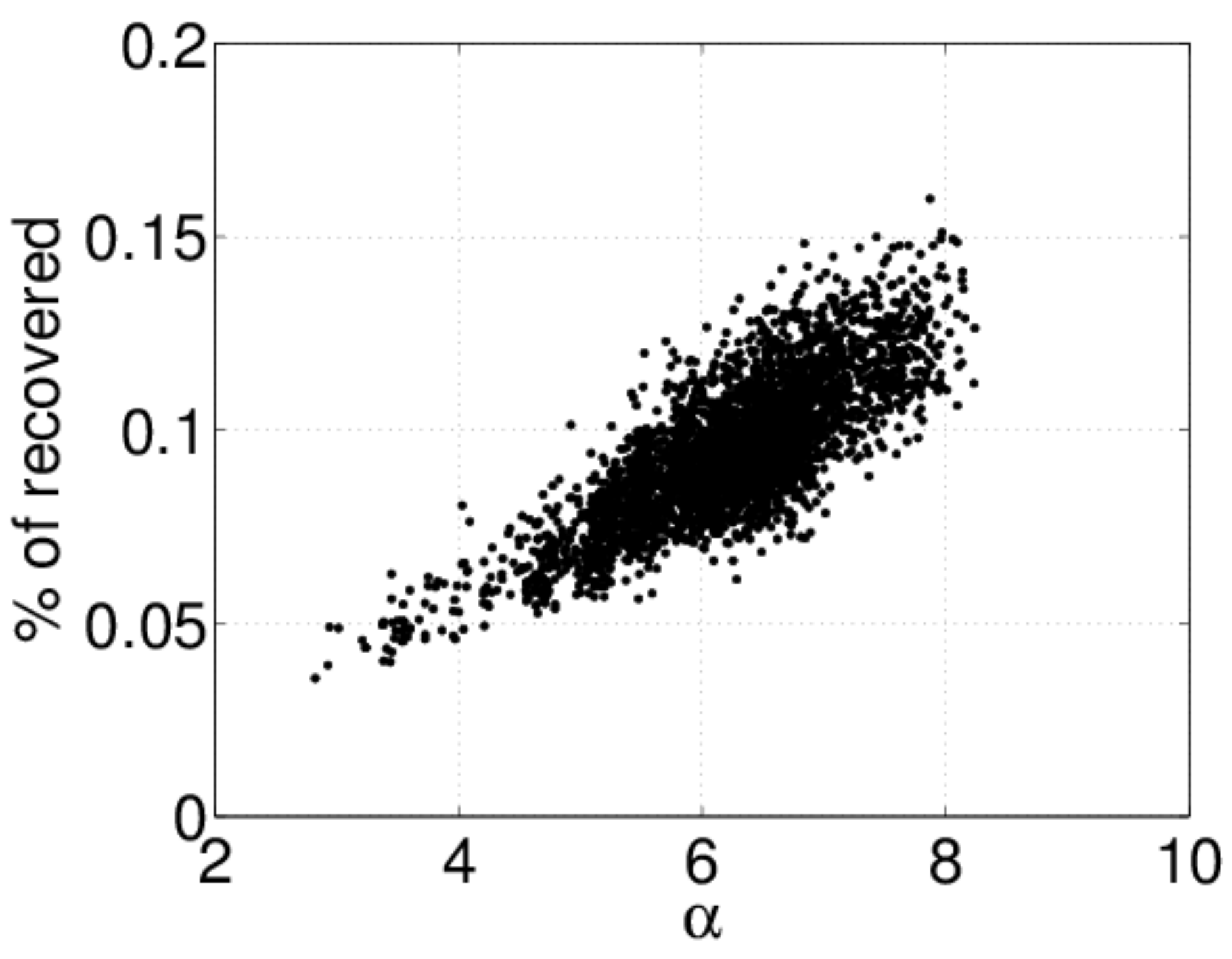}}
\caption{The percentage of recovered individuals on the SIR epidemic spreading model ($\beta = 0.3, \mu = 1.0$) according to the accessibility measure for the road networks of (a) Japan, (b) England, (c) United States and (d) Germany.}
\label{Fig:Corr_alpha_SIR_RP}
\end{center}

\begin{center}
\subfigure[]{\includegraphics[width=0.45\columnwidth]{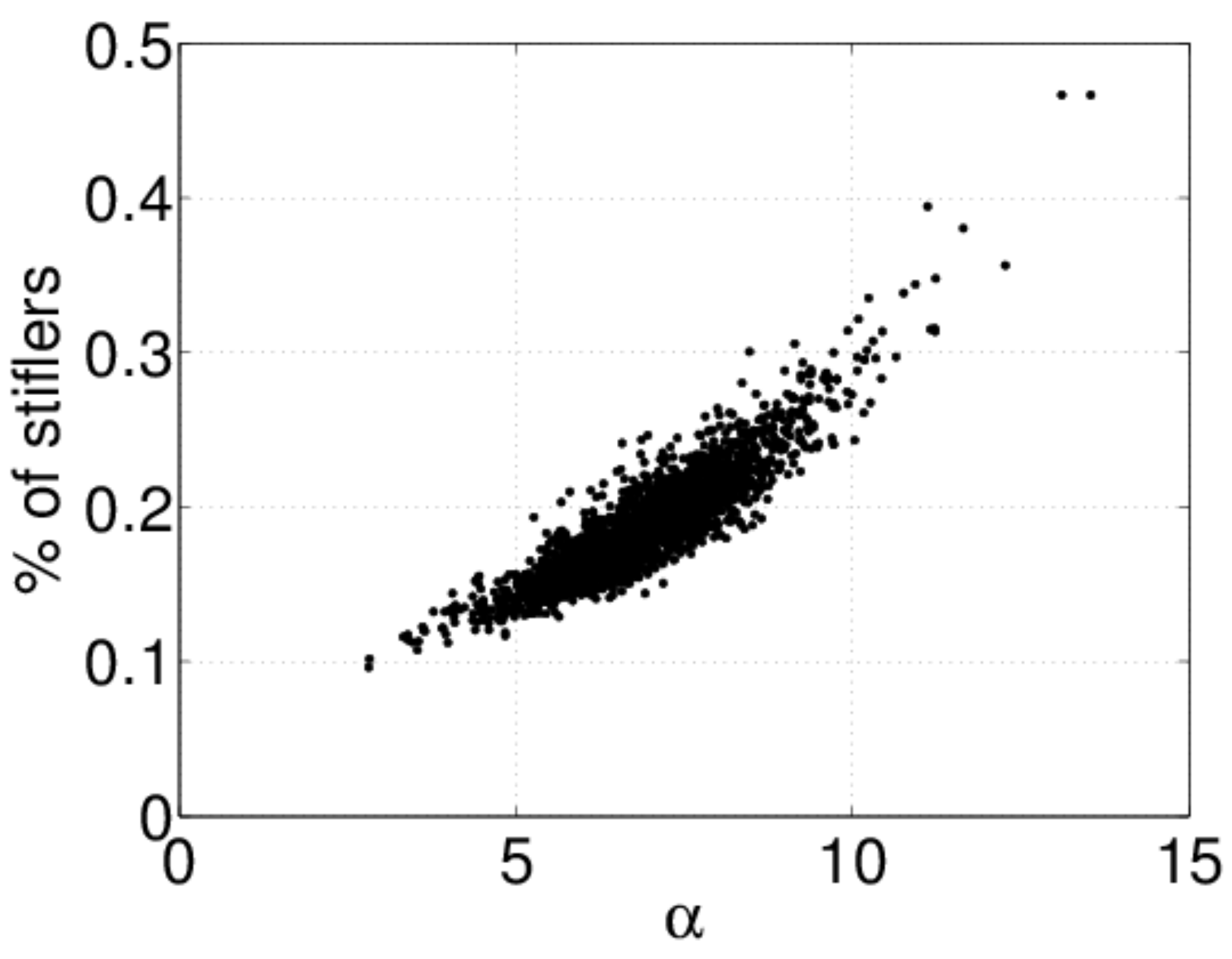}}
\subfigure[]{\includegraphics[width=0.45\columnwidth]{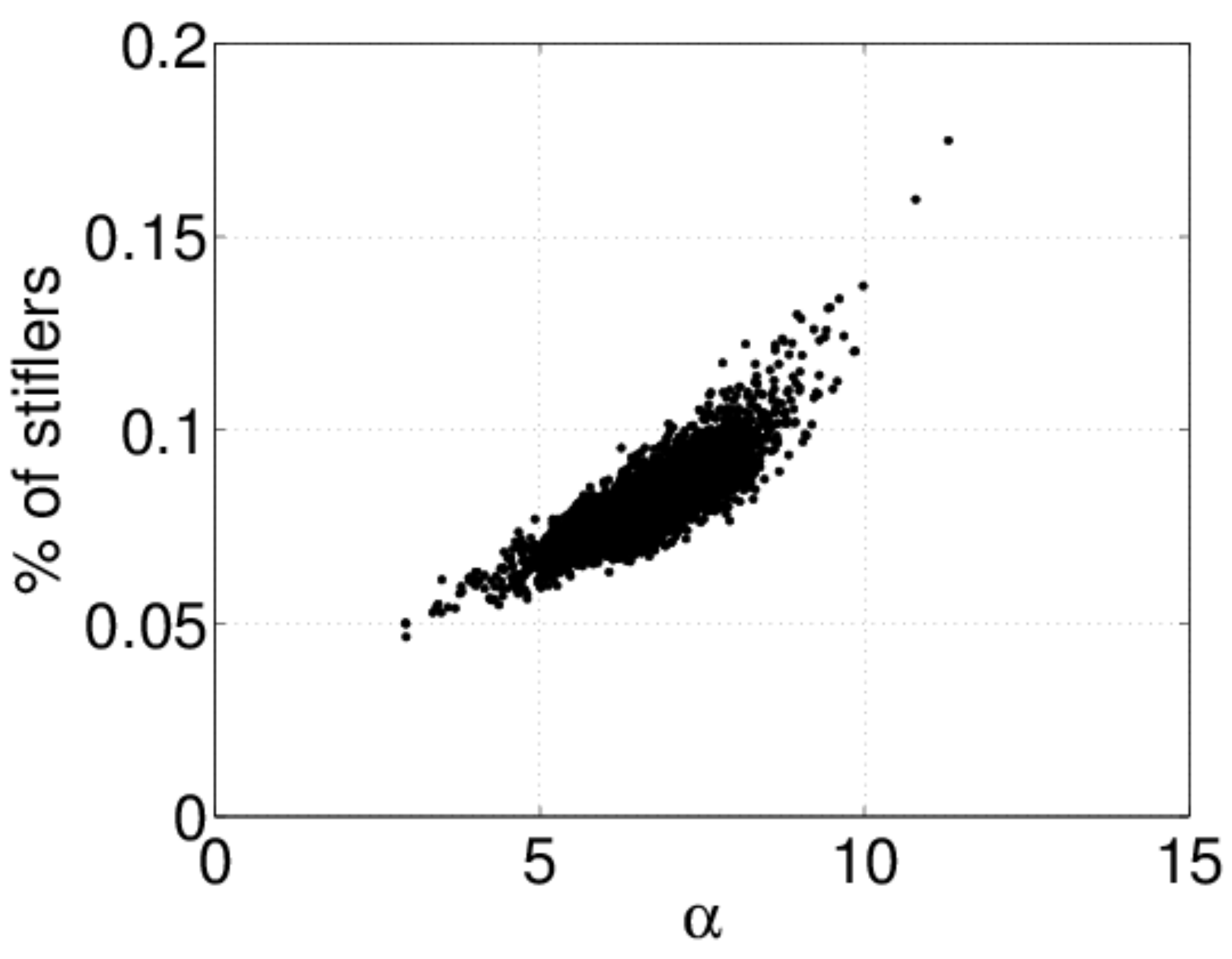}}
\subfigure[]{\includegraphics[width=0.45\columnwidth]{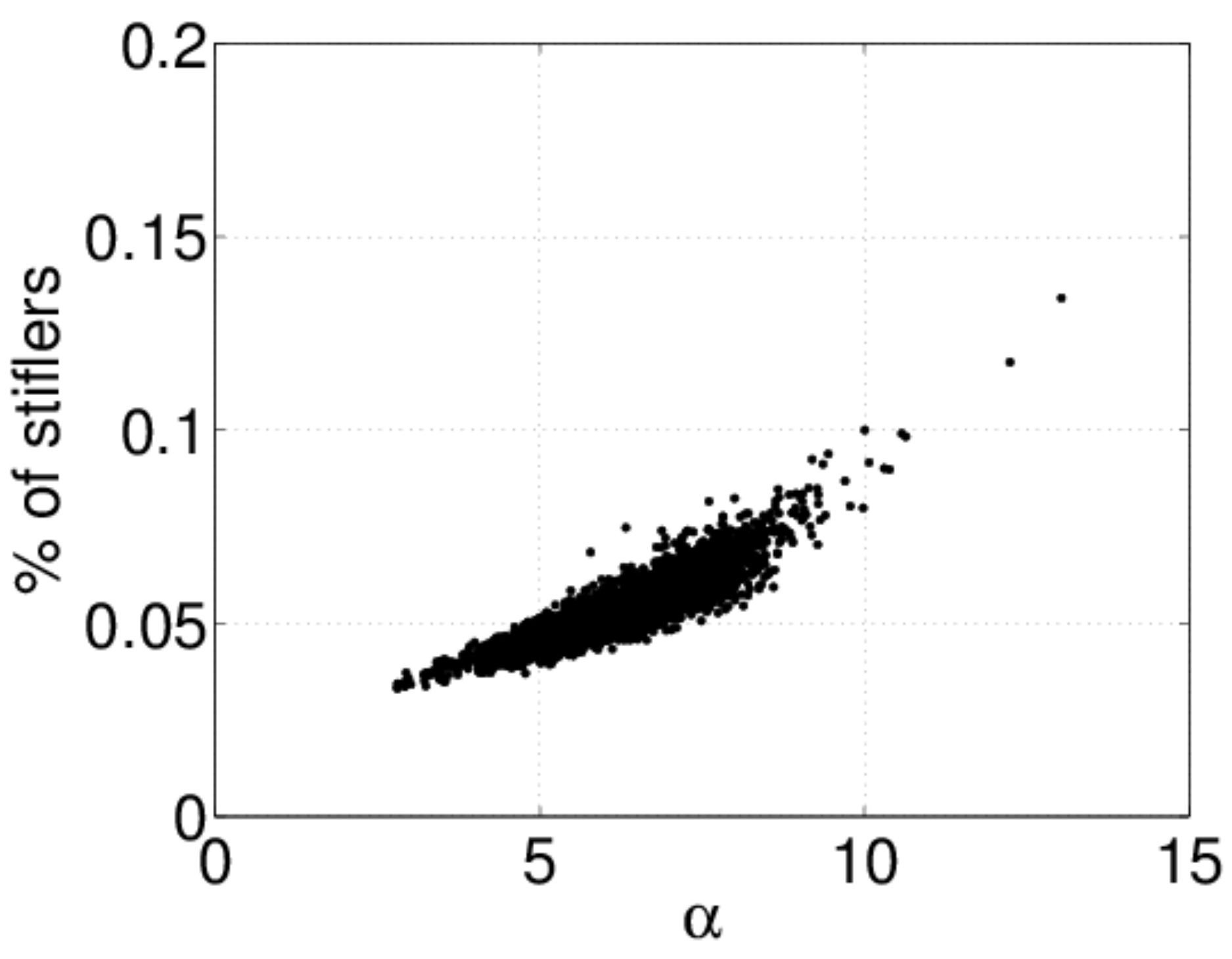}}
\subfigure[]{\includegraphics[width=0.45\columnwidth]{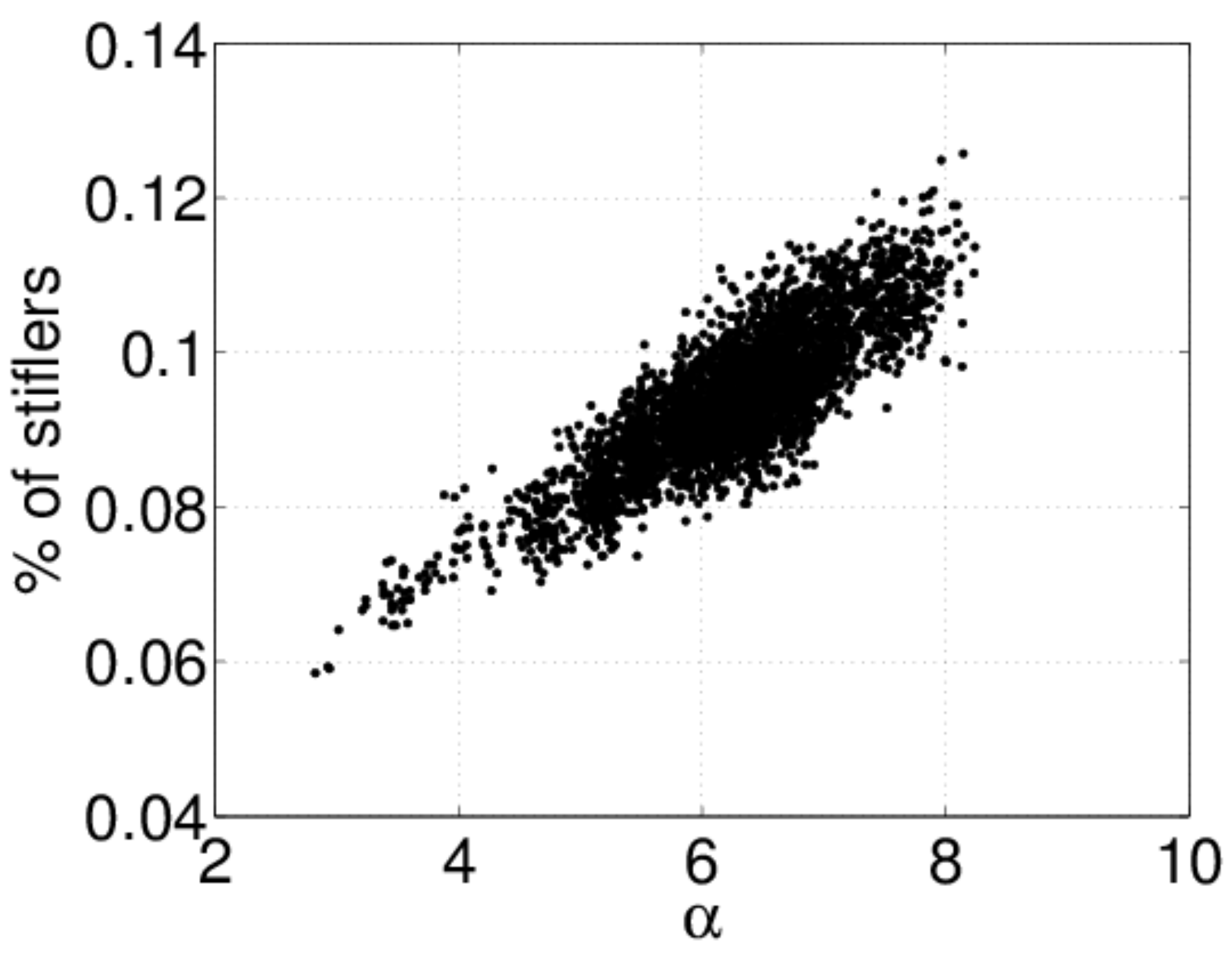}}
\caption{The percentage of stiflers on the MT (TP) rumor model ($\lambda = 0.3, \delta = 1.0$) according to the accessibility measure for the road networks of (a) Japan, (b) England, (c) United States and (d) Germany.}
\label{Fig:Corr_alpha_MK_TP}
\end{center}
\end{figure}

Figures~\ref{Fig:Corr_SIR_RP} and~\ref{Fig:Corr_MK_TP} show the scatter plots for the epidemic and rumor dynamics in the US road network, respectively. The strongest correlation correspond to the degree centrality, while for other metrics, correlations are weak and positive, though not zero. On the contrary, the clustering coefficient leads to a negative correlation because the more central a node is, the smaller its clustering coefficient is. On the other hand, Figures~\ref{Fig:Corr_alpha_SIR_RP} and~\ref{Fig:Corr_alpha_MK_TP} show that the correlations between the generalized random walk accessibility and the potential of rumor and epidemic spreading processes are almost linear and positive for all road networks analyzed.

Furthermore, Table~\ref{tab:Corr} shows that for both spreading processes, the highest correlations between a centrality measure and the impact of the disease or rumor correspond to the case of the generalized accessibility centrality, which values often higher than 0.7. Interesting, the $k$-core centrality yields small correlation values, contrary to what has been observed in~\cite{Kitsak10:NP}, which considered networks not embedded in space. However, this result agrees with~\cite{Borge12}, in the case of rumor dynamics. The node degree is highly correlated with the final fraction of recovered nodes, but less if we look the results corresponding to the final fraction of stiflers, mainly for the case of a MT model simulated using a contact process setting, again as found in~\cite{Borge12}. Moreover, PageRank, closeness and betweenness centrality metrics do not show significant correlations with disease and rumor spreading capabilities $-$except when the parameter $\delta$ in rumor models is small, in which case the closeness gives high correlation. It is also worth noticing that the eigenvector centrality shows high correlation only for the spatial scale-free network model. 

\begin{table*}[t] 
\begin{center}
\caption{Correlation between centrality measures and the final fraction of recovered individuals (SIR model) or the final fraction of stiflers (MT model for the contact (CP) or truncated (TP) cases). The measures are the degree ($k$), clustering coefficient ($cc$), betweenness centrality ($B$),  average neighborhood degree ($r$), PageRank ($\pi$), eigenvector centrality ($x$), k-core index ($k_c$), closeness centrality ($C$) and accessibility ($\alpha$). The highest correlations are in bold. }
\begin{tabular}{|l|c|c|c|c|c|c|c|c|c|c|c|}
\hline
Process & Rates & Network & $k$ & $cc$ & $B$ & $r$ & $\pi$ & $x$ & $k_c$ & $C$ & $\alpha$ \\
\hline
\multirow{12}{*}{SIR} & \multirow{6}{*}{$\beta = 0.8, \mu = 1.0$} & Japan & 0.40 & 0.11 & 0.24 & 0.26 & 0.27 & 0.27 &
0.35 & \textbf{0.47} & \textbf{0.47} \\
 &  & England & 0.55 & 0.10 & 0.26 & 0.38 & 0.30 & -0.04 & 0.21 & 0.27 & \textbf{0.58} \\
 &  & USA & 0.60 & 0.25 & 0.19 & 0.53 & 0.28 & 0.49 & 0.26 & 0.41 & \textbf{0.73} \\
 &  & Germany & 0.54 & 0.05 & 0.42 & 0.35 & 0.20 & 0.22 & 0.19 & 0.34 & \textbf{0.63} \\
 &  & SpatialSF & \textbf{0.81} & 0.25 & 0.65 & -0.04 & 0.75 & 0.35 & -- & 0.32 & 0.60 \\
 &  & Waxman & \textbf{0.74} & 0.34 & 0.47 & 0.29 & 0.62 & 0.05 & 0.70 & 0.19 & \textbf{0.73} \\
\cline{2-12}
 & \multirow{6}{*}{$\beta = 0.3, \mu = 1.0$} & Japan & 0.65 & 0.30 & 0.31 & 0.65 & 0.37 & 0.36 & 0.65 & 0.41 & \textbf{0.79}
\\
 &  & England & 0.68 & 0.13 & 0.27 & 0.61 & 0.31 & -0.07 & 0.38 & 0.21 & \textbf{0.76} \\
 &  & USA & 0.77 & 0.38 & 0.10 & 0.68 & 0.38 & 0.59 & 0.37 & 0.14 & \textbf{0.86} \\
 &  & Germany & 0.69 & 0.08 & 0.46 & 0.42 & 0.31 & 0.22 & 0.19 & 0.25 & \textbf{0.74} \\
 &  & SpatialSF & 0.70 & 0.39 & 0.70 & 0.46 & 0.49 & 0.72 & -- & 0.66 & \textbf{0.91} \\
 &  & Waxman & 0.68 & 0.30 & 0.42 & 0.58 & 0.45 & 0.04 & 0.72 & 0.30 & \textbf{0.81} \\
\hline
\multirow{24}{*}{MT TP} & \multirow{6}{*}{$\lambda = 0.8, \delta = 1.0$} & Japan & 0.50 & -0.08 & 0.43 & 0.48 & 0.29 & 0.33 & 0.45 &
0.42 & \textbf{0.81} \\
 &  & England & 0.54 & -0.20 & 0.43 & 0.49 & 0.22 & -0.03 & 0.34 & 0.32 &
\textbf{0.85} \\
 &  & USA & 0.67 & 0.11 & 0.22 & 0.62 & 0.30 & 0.55 & 0.33 & 0.24 & \textbf{0.90} \\
 &  & Germany & 0.57 & -0.24 & 0.62 & 0.41 & 0.20 & 0.24 & 0.18 & 0.33 & \textbf{0.88} \\
 &  & SpatialSF & 0.63 & 0.29 & 0.71 & 0.55 & 0.40 & 0.75 & -- & 0.70 & \textbf{0.94} \\
&   & Waxman & 0.56 & 0.06 & 0.52 & 0.46 & 0.37 & 0.04 & 0.58 & 0.45 & \textbf{0.76} \\
\cline{2-12}
 & \multirow{6}{*}{$\lambda = 0.8, \delta = 0.3$} & Japan & 0.17 & -0.02 & 0.22 & 0.23 & 0.04 & 0.26
& 0.21 & \textbf{0.66} & 0.35 \\
 &  & England & 0.32 & -0.07 & 0.31 & 0.38 & 0.05 & 0.05 & 0.26 & \textbf{0.60} & 0.53 \\
 &  & USA & 0.26 & 0.00 & 0.25 & 0.29 & 0.07 & 0.08 & 0.12 & \textbf{0.83} & 0.43 \\
 &  & Germany & 0.29 & -0.12 & 0.46 & 0.28 & 0.01 & 0.45 & 0.17 & \textbf{0.67} & 0.52 \\
&   & SpatialSF & 0.40 & 0.16 & 0.43 & 0.28 & 0.27 & 0.41 & -- & 0.37 & \textbf{0.53} \\
&   & Waxman & 0.61 & 0.13 & 0.51 & 0.38 & 0.47 & 0.06 & 0.62 & 0.31 & \textbf{0.74}\\
\cline{2-12}
 & \multirow{6}{*}{$\lambda = 0.3, \delta = 1.0$} & Japan & 0.77 & 0.22 & 0.43 & 0.61 & 0.54 & 0.25 & 0.59 & 0.28 & \textbf{0.88}
\\
 &  & England & 0.77 & 0.03 & 0.34 & 0.53 & 0.47 & -0.07 & 0.30 & 0.16 & \textbf{0.83} \\
 &  & USA & 0.84 & 0.32 & 0.19 & 0.63 & 0.50 & 0.56 & 0.35 & 0.13 & \textbf{0.91} \\
 &  & Germany & 0.73 & 0.01 & 0.45 & 0.40 & 0.39 & 0.17 & 0.19 & 0.20 & \textbf{0.79} \\
  &   & SpatialSF & 0.34 & 0.32 & 0.53 & 0.71 & 0.12 & \textbf{0.89} & -- & 0.84 & 0.77\\
  &   & Waxman & 0.84 & 0.30 & 0.56 & 0.59 & 0.64 & 0.06 & 0.77 & 0.25 & \textbf{0.94}\\
\cline{2-12}
 & \multirow{6}{*}{$\lambda = 0.3, \delta = 0.3$} & Japan & 0.37 & 0.00 & 0.32 & 0.50 & 0.13 & 0.35 & 0.42 & 0.49 &
\textbf{0.68} \\
 &  & England & 0.42 & -0.09 & 0.34 & 0.52 & 0.07 & 0.01 & 0.37 & 0.38 & \textbf{0.71} \\
 &  & USA & 0.54 & 0.12 & 0.15 & 0.64 & 0.16 & 0.54 & 0.33 & 0.28 & \textbf{0.80} \\
 &  & Germany & 0.42 & -0.20 & 0.54 & 0.41 & 0.06 & 0.29 & 0.18 & 0.39 & \textbf{0.73} \\
  &   & SpatialSF & 0.42 & 0.31 & 0.53 & 0.62 & 0.19 & 0.71 & -- & 0.65 & \textbf{0.84}\\
  &   & Waxman & 0.44 & 0.08 & 0.41 & 0.46 & 0.25 & 0.09 & 0.51 & 0.55 & \textbf{0.64}\\
\hline
\multirow{24}{*}{MT CP} & \multirow{6}{*}{$\lambda = 0.8, \delta = 1.0$} & Japan & 0.42 & 0.05 & 0.30 & 0.57 & 0.16 & 0.32 & 0.48
& 0.42 & \textbf{0.73} \\
 &  & England & 0.43 & -0.10 & 0.32 & 0.56 & 0.08 & -0.05 & 0.36 & 0.32 & \textbf{0.73} \\
 &  & USA & 0.57 & 0.16 & 0.13 & 0.682& 0.17 & 0.55 & 0.34 & 0.24 & \textbf{0.82} \\
 &  & Germany & 0.45 & -0.18 & 0.52 & 0.44 & 0.07 & 0.26 & 0.18 & 0.35 & \textbf{0.75} \\
  &   & SpatialSF & 0.26 & 0.27 & 0.42 & 0.69 & 0.02 & \textbf{0.75} & -- & 0.71 & 0.73\\
  &   & Waxman & 0.50 & 0.13 & 0.39 & 0.57 & 0.27 & 0.04 & 0.59 & 0.38 & \textbf{0.72}\\
\cline{2-12}
 & \multirow{6}{*}{$\lambda = 0.8, \delta = 0.3$} & Japan & 0.17 & 0.01 & 0.18 & 0.28 & 0.01 & 0.27 & 0.26 & \textbf{0.67} &
0.36 \\
 &  & England & 0.24 & 0.00 & 0.24 & 0.37 & -0.03 & 0.16 & 0.29 & \textbf{0.70} & 0.43 \\
 &  & USA & 0.29 & 0.04 & 0.20 & 0.36 & 0.04 & 0.20 & 0.16 & \textbf{0.74} & 0.47 \\
 &  & Germany & 0.20 & -0.07 & 0.41 & 0.24 & -0.05 & 0.51 & 0.16 & \textbf{0.81} & 0.40 \\
  &   & SpatialSF & 0.28 & 0.22 & 0.36 & 0.46 & 0.10 & 0.48 & -- & 0.44 & \textbf{0.61}\\
  &   & Waxman & 0.54 & 0.16 & 0.44 & 0.55 & 0.34 & 0.10 & 0.64 & 0.44 & \textbf{0.76}\\
\cline{2-12}
 & \multirow{6}{*}{$\lambda = 0.3, \delta = 1.0$} & Japan & 0.60 & 0.18 & 0.36 & 0.66 & 0.34 & 0.28 & 0.56 & 0.31 & \textbf{0.82} \\
 &  & England & 0.58 & -0.02 & 0.31 & 0.60 & 0.25 & -0.10 & 0.31 & 0.18 & \textbf{0.77} \\
 &  & USA & 0.68 & 0.26 & 0.12 & 0.71 & 0.30 & 0.56 & 0.35 & 0.14 & \textbf{0.85} \\
 &  & Germany & 0.55 & -0.10 & 0.45 & 0.47 & 0.20 & 0.17 & 0.18 & 0.21 & \textbf{0.74} \\
  &   & SpatialSF & 0.23 & 0.27 & 0.42 & 0.71 & 0.01 & \textbf{0.81} & -- & 0.77 & 0.70\\
  &   & Waxman & 0.68 & 0.24 & 0.45 & 0.67 & 0.45 & 0.06 & 0.72 & 0.26 & \textbf{0.87}\\
\cline{2-12}
 & \multirow{6}{*}{$\lambda = 0.3, \delta = 0.3$} & Japan & 0.37 & 0.03 & 0.27 & 0.54 & 0.12 & 0.33 & 0.45 & 0.46 &
\textbf{0.68} \\
 &  & England & 0.40 & -0.08 & 0.32 & 0.53 & 0.06 & -0.01 & 0.36 & 0.36 & \textbf{0.69} \\
 &  & USA & 0.54 & 0.14 & 0.13 & 0.65 & 0.14 & 0.53 & 0.33 & 0.27 & \textbf{0.79} \\
 &  & Germany & 0.40 & -0.20 & 0.52 & 0.43 & 0.03 & 0.27 & 0.18 & 0.39 & \textbf{0.72} \\
  &   & SpatialSF & 0.27 & 0.27 & 0.43 & 0.67 & 0.03 & 0.71 & -- & 0.66 & \textbf{0.75}\\
  &   & Waxman & 0.44 & 0.10 & 0.38 & 0.51 & 0.23 & 0.05 & 0.53 & 0.46 & \textbf{0.65}\\
\hline
\end{tabular}
\label{tab:Corr}
\end{center}
\end{table*}

\subsection{Road networks}

\begin{figure*}[t]
\subfigure[]{\includegraphics[width=0.4\linewidth]{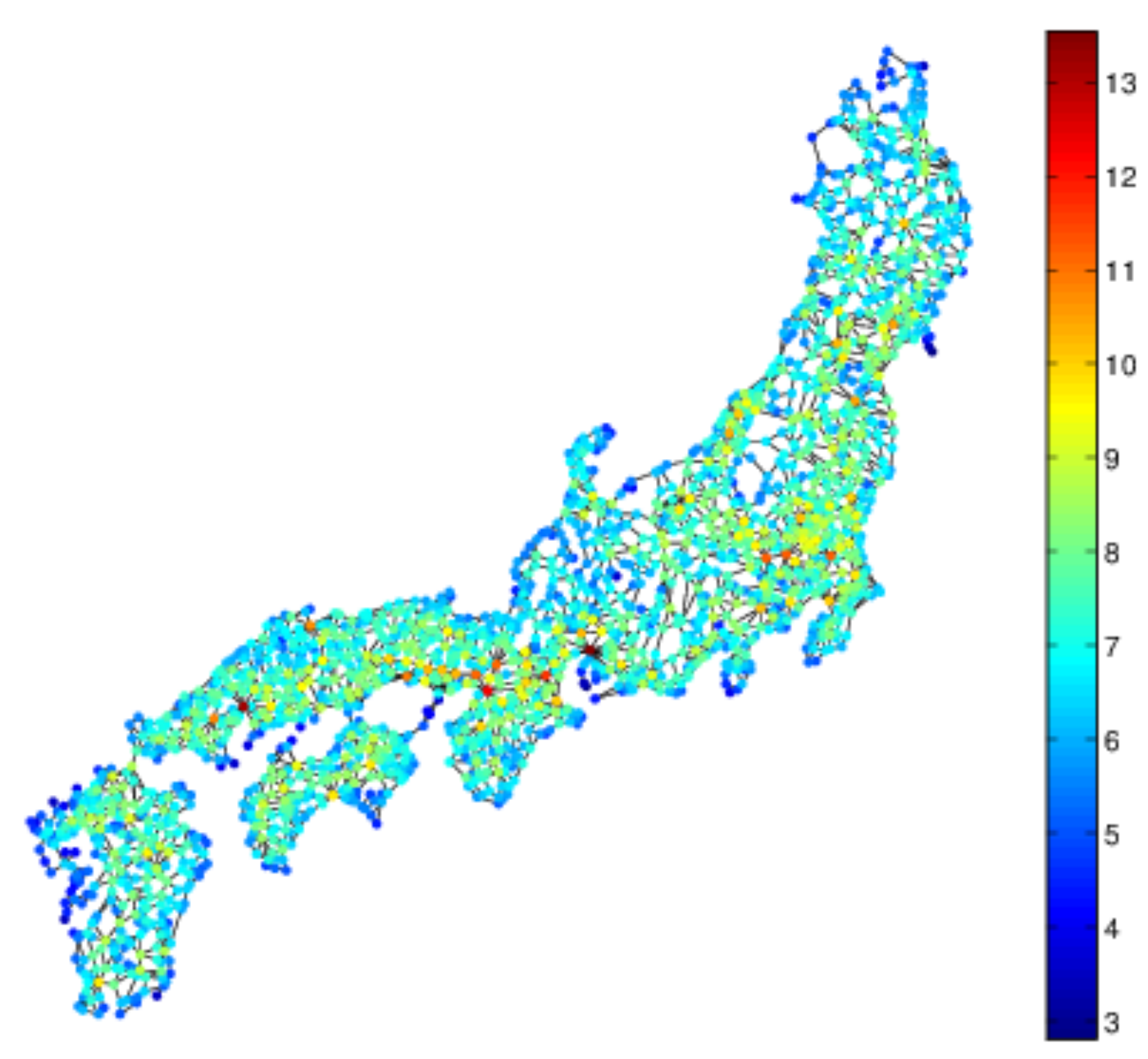}}
\subfigure[]{\includegraphics[width=0.35\linewidth]{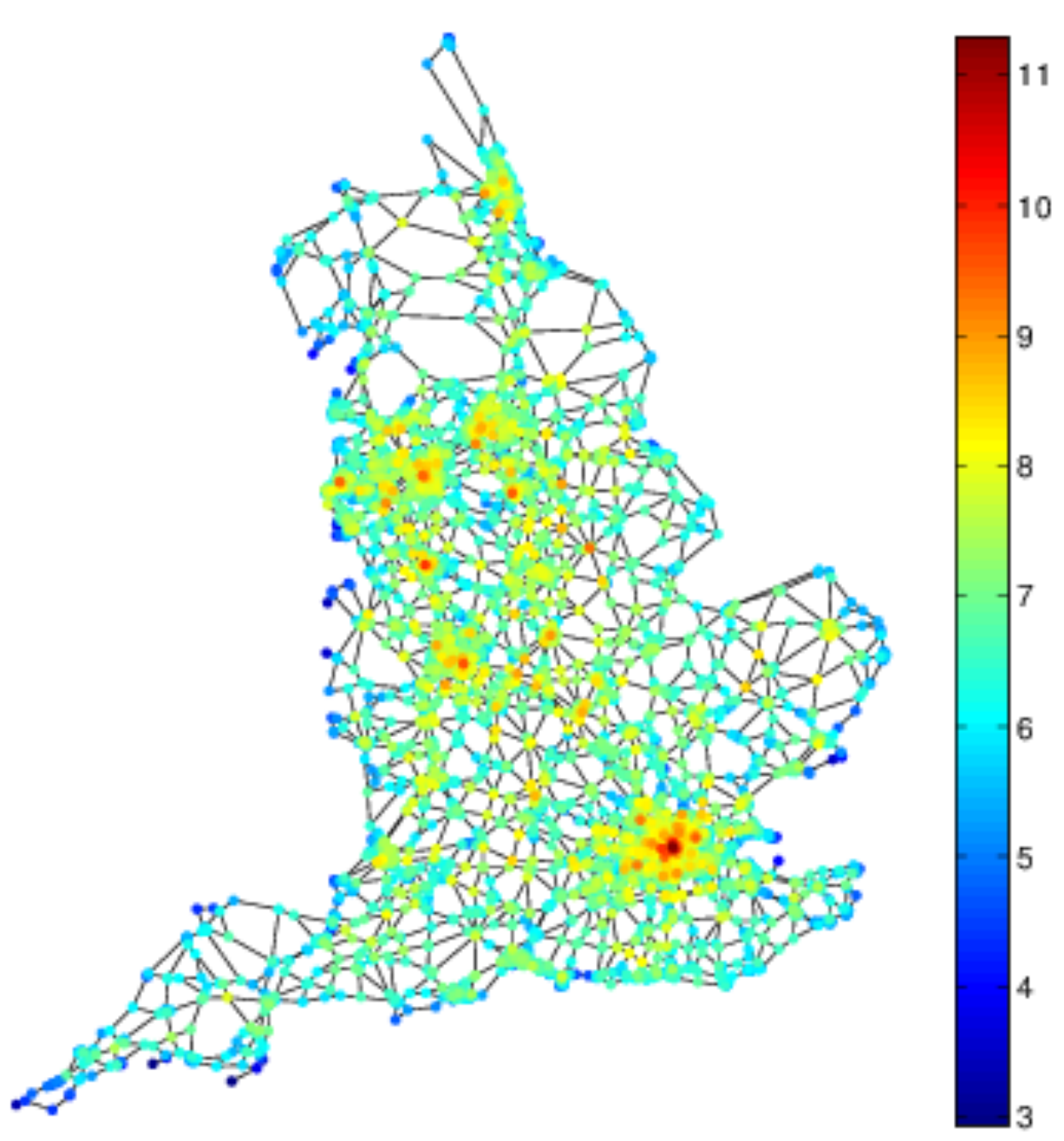}}
\subfigure[]{\includegraphics[width=0.55\linewidth]{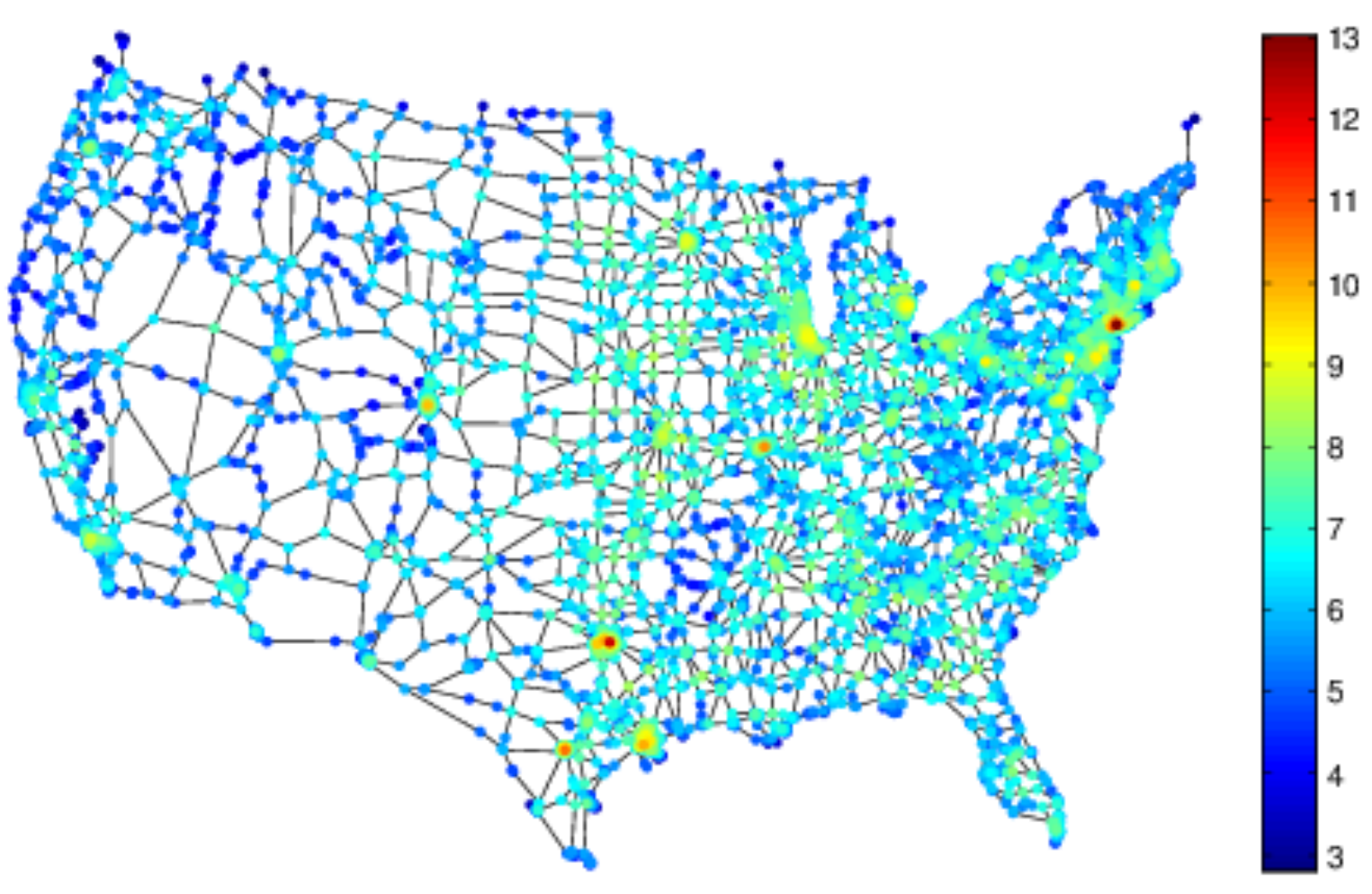}}
\subfigure[]{\includegraphics[width=0.33\linewidth]{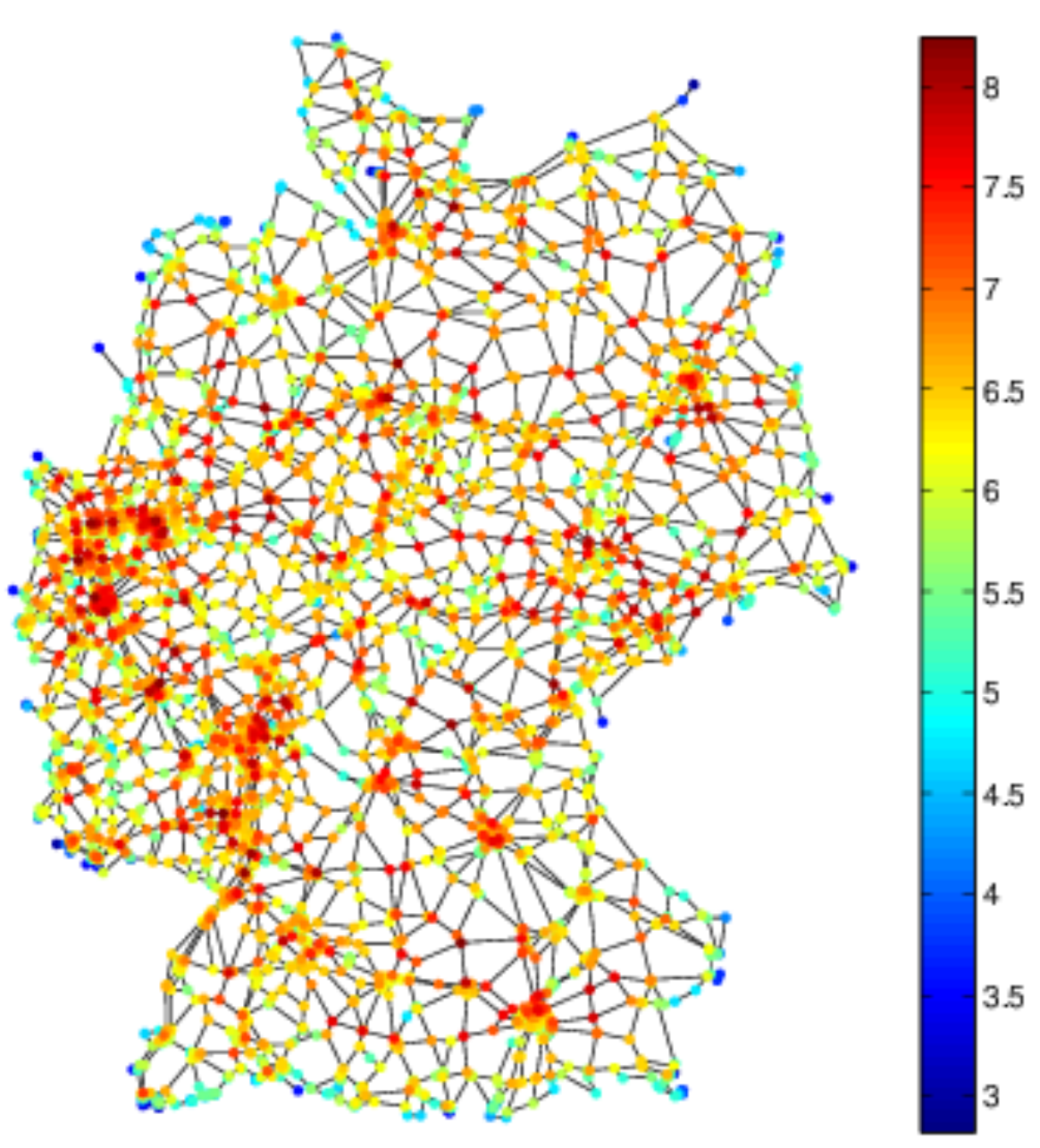}}
\caption{Network visualisation of the real road networks of (a) Japan, (b) England, (c) United States and (d) Germany. The colors represent the values of the accessibility.}
\label{fig:MapsAcc}
\end{figure*}

Focusing on real networks, Figure~\ref{fig:MapsAcc} shows results obtained for the generalized random walk accessibility of each node for the road networks of Japan, England, United States and Germany. In Japan, the most influential spreaders are the cities of Nagoya, Osaka and Hiroshima. Tokyo is highly connected, but does not have the same spreading capability of these cities, since it is a peripheral hub. London, Liverpool and Manchester have the highest values of accessibility in England, while in the US, the cities with the highest accessibility are New York, Houston, Dallas and Chicago $-$interesting enough, these cities are also air transportation hubs$-$. Finally, Berlin, M\"unchen and D\"usseldorf have the highest accessibility in Germany. Note that nodes at the border of the countries present the smallest values of accessibility. Therefore, this measure can be considered for identification of border of networks, as previously pointed out for the original definition of accessibility in~\cite{Travencolo2009:NJP}. 

\begin{figure}[t]
\includegraphics[width=0.9\linewidth]{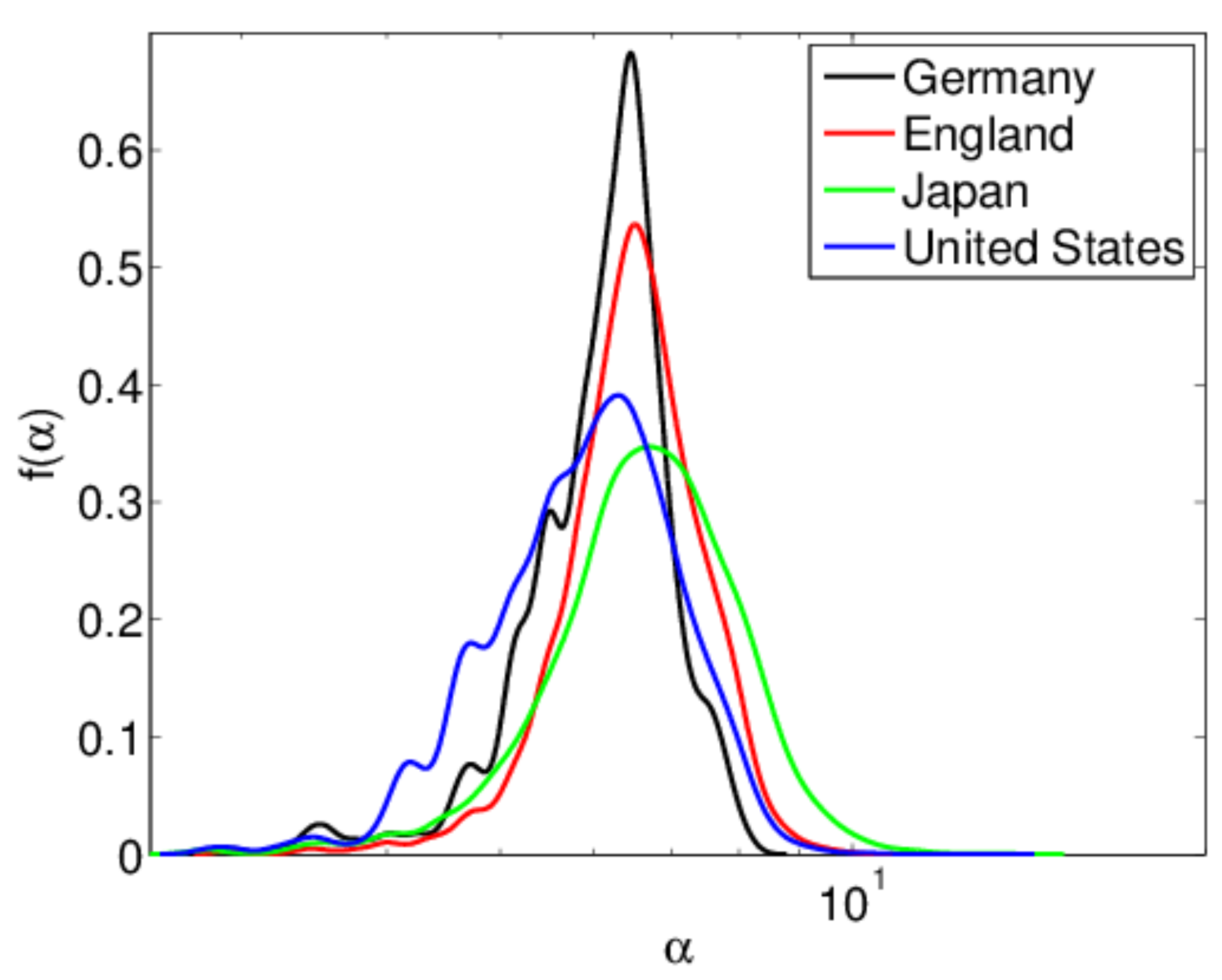}
\caption{Probability distribution of the accessibility in the road networks of Japan, Germany, US and England.}
\label{fig:distacc}
\end{figure}

Figure~\ref{fig:distacc} presents the probability distribution of the accessibility. For all cases, the distribution is asymmetric, presenting a long tail for higher values of accessibility, and centered at the same value. It is interesting to note that Germany and England has the smallest variation in the accessibility, whereas Japan has the highest one. This fact can be related to the rough of Japan, which influence directly how highways are distributed. 

\section{Non-spatial networks} \label{Sec:nonspatial}

We have also studied what happens for non-spatial networks using the same set of measurements considered in Secs.~\ref{Sec:spatial} and~\ref{Sec:cent}. Table~\ref{tab:structure} presents the average values of these measures calculated in the social networks and in synthetic BA networks. Table~\ref{tab:Corr_non_sp} presents the Spearman correlation coefficient calculated between the centrality metrics and final fraction of stiflers or recovered nodes in the epidemic and rumor processes, respectively. The results agree with the analysis of epidemic spreading presented in~\cite{Kitsak10:NP} and with the study of rumor diffusion in~\cite{Borge12}. In the case of the SIR model, the $k$-core and degree centralities are the most correlated with the final fraction of recovered nodes. Thus, the main hubs on the social networks are located in the center of the network, because they have the highest coreness, suggesting that such networks tend not to present peripheral hubs. Moreover, correlations are stronger when the parameter $\beta$ is decreased. On the contrary, the random walk accessibility yields the highest Spearman correlation for BA networks and for political blogs (for $\beta = 0.3$), although the correlation values are close to those obtained for the degree and $k$-core. All the remainder metrics exhibit smaller correlation coefficients than the $k$-core, $k$ and $\alpha$. 

\begin{figure}[t]
\includegraphics[width=0.9\linewidth]{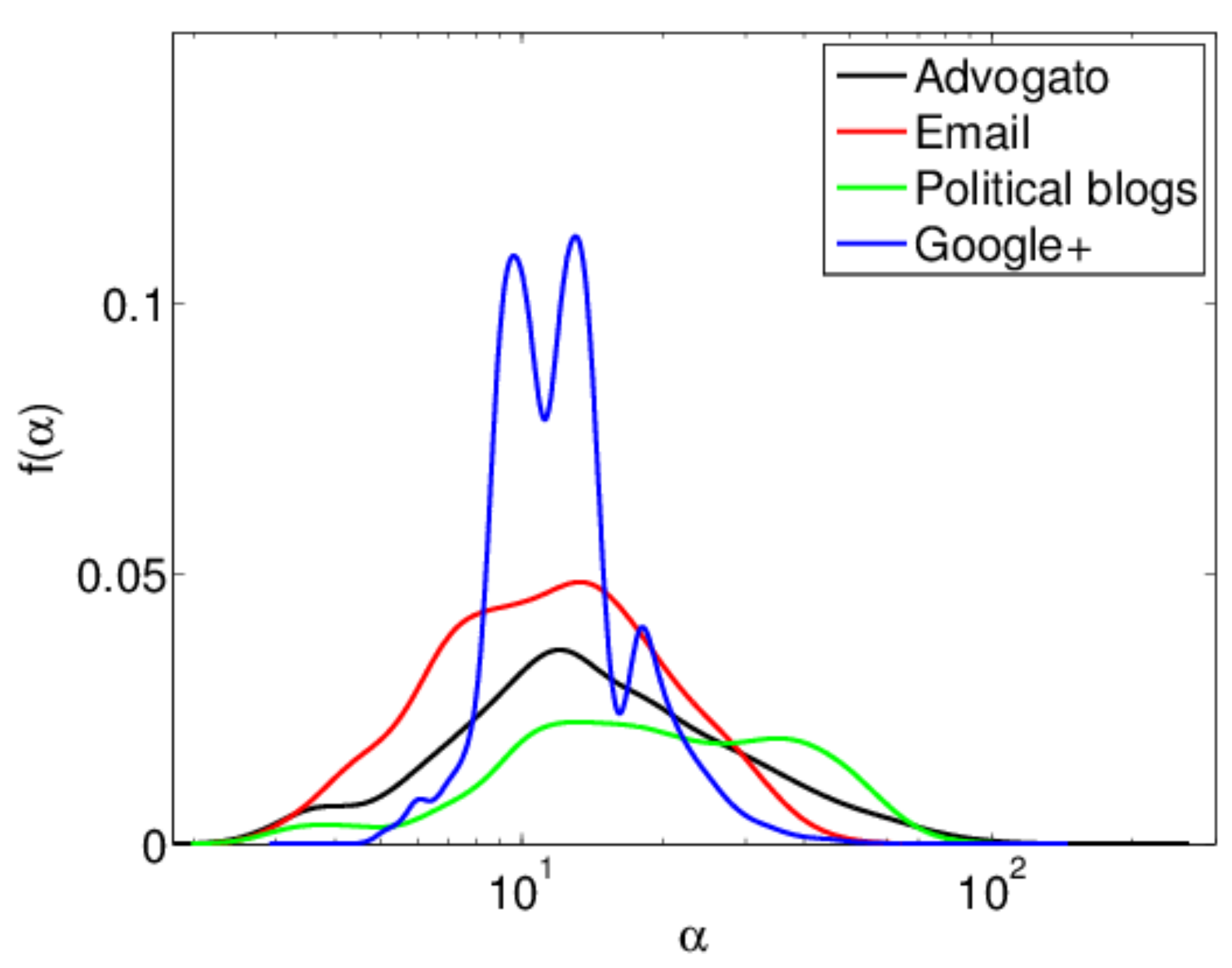}
\caption{Probability distribution of the accessibility in the social networks of advogato, email, political blogs and Google+.}
\label{fig:distacc_non_sp}
\end{figure}

With respect to the rumor dynamics, the CP and TP cases present different results. In the first case, the eigenvector and accessibility centralities are strongly correlated with the final fraction of stiflers, whereas, for the second case, closeness centrality and average neighborhood degree show the highest correlations. Considering the TP case with a stifling rate $\lambda = 1$, if the spreading rate is high, the average neighborhood degree is more related to the dynamics. However, for lower spreading rates the distance from one node to the rest of the network is more critical. This property is evinced in Table~\ref{tab:Corr_non_sp}. Note that $r$ presents higher correlations for higher spreading rates, whereas the closeness centrality is more correlated when spreading rates are smaller. Such analysis suggests that shortest paths get more important for information propagation proportionally to the inverse of the spreading rate. Furthermore, the $k$-core and degree centralities have not been found to exhibit strong correlations with the final fraction of  stiflers, supporting the results in~\cite{Borge12}. Finally, we note that at variance with previous cases, for the rumor dynamics on non-spatial networks, there is no single metric that has yielded the highest correlations for all the networks analyzed. In particular, the accessibility centrality does not seem to be in this case as distinct as before, likely because, as seen in Figure~\ref{fig:distacc_non_sp}, the distributions of accessibility in non spatial networks are asymmetric, with different mean values and characterized by a long tail distribution.

\begin{table*}[t] 
\begin{center}
\caption{Correlation between centrality measures and the final fraction of stiflers (MT model for the contact (CP) or truncated (TP) cases) or recovered individuals (SIR
model). The measures are the degree ($k$), clustering coefficient ($cc$), betweenness centrality ($B$, average neighborhood degree ($r$), PageRank ($\pi$), eigenvector
centrality ($x$), k-core index ($k_c$), closeness centrality ($C$) and accessibility ($\alpha$). The highest correlations are in bold. }
\begin{tabular}{|l|c|c|c|c|c|c|c|c|c|c|c|}
\hline
Process & Rates & Network & $k$ & $cc$ & $B$ & $r$ & $\pi$ & $x$ & $k_c$ & $C$ & $\alpha$ \\
\hline
\multirow{10}{*}{SIR} & \multirow{5}{*}{$\beta = 0.8. \mu = 1.0$} & advogato & \textbf{0.76} & 0.47 & 0.66 & 0.14 & 0.71 & 0.65 & \textbf{0.76} & 0.63 & 0.70 \\
 & & email & \textbf{0.67} & 0.41 & 0.57 & 0.09 & 0.64 & 0.58 & \textbf{0.67} & 0.57 & 0.63\\
 & & polblogs & \textbf{0.57} & 0.29 & 0.51 & 0.04 & 0.55 & 0.55 & \textbf{0.57} & 0.55 & \textbf{0.57}\\
 & & Google+ & \textbf{0.81} & 0.67 & 0.61 & 0.17 & 0.64 & 0.50 & \textbf{0.81} & 0.48 & 0.63 \\
 & & BA & 0.19 & 0.13 & 0.39 & 0.48 & -0.02 & 0.45 &  --  & 0.49 & \textbf{0.60} \\
\cline{2-12}
 & \multirow{5}{*}{$\beta = 0.3. \mu = 1.0$} & advogato & \textbf{0.97} & 0.40 & 0.85 & 0.19 & 0.92 & 0.88 & \textbf{0.97} & 0.84 & 0.92 \\
 & & email & \textbf{0.97} & 0.34 & 0.85 & 0.25 & 0.94 & 0.88 & \textbf{0.96} & 0.89 & 0.94\\
 & & polblogs & \textbf{0.89} & 0.25 & 0.78 & -0.08 & 0.86 & 0.85 & \textbf{0.89} & 0.82 & \textbf{0.88}\\
 & & Google+ & \textbf{0.81} & 0.67 & 0.61 & 0.23 & 0.60 & 0.56 & \textbf{0.81} & 0.54 & 0.68\\
 & & BA & 0.18 & 0.18 & 0.48 & 0.73 & -0.10 & 0.66 &  --  & 0.72 & \textbf{0.77}  \\
\hline
\multirow{20}{*}{MT TP} & \multirow{5}{*}{$\lambda = 0.8. \delta = 1.0$} & advogato & 0.23 & 0.17 & 0.26 & \textbf{0.55} & 0.17 & 0.38 & 0.22 & 0.43 & 0.35 \\
 & & email & 0.62 & 0.19 & 0.61 & 0.13 & 0.60 & 0.56 & 0.58 & 0.60 & \textbf{0.64}\\
 & & polblogs & -0.16 & -0.09 & -0.06 & \textbf{0.35} & -0.17 & -0.16 & -0.21 & -0.07 & -0.13\\
 & & Google+ & 0.04 & 0.03 & 0.00 & \textbf{0.78} & -0.07 & 0.27 & 0.04 & 0.25 & 0.40 \\
 & & BA & 0.16 & 0.19 & 0.48 & 0.76 & -0.12 & 0.70 &  --  & 0.75 & \textbf{0.78} \\
\cline{2-12}
 & \multirow{5}{*}{$\lambda = 0.8. \delta = 0.3$} & advogato & 0.05 & 0.15 & 0.10 & 0.51 & 0.02 & 0.20 & 0.05 & \textbf{0.27} & 0.16 \\
 & & email & 0.29 & 0.22 & \textbf{0.30} & 0.06 & 0.28 & 0.24 & 0.26 & 0.28 & \textbf{0.30}\\
 & & polblogs & -0.37 & -0.01 & -0.26 & \textbf{0.48} & -0.37 & -0.34 & -0.39 & -0.22 & -0.33\\
 & & Google+ & 0.00 & 0.03 & -0.05 & \textbf{0.69} & 0.01 & 0.11 & 0.004 & 0.09 & 0.27 \\
 & & BA & 0.14 & 0.19 & 0.47 & 0.77 & -0.11 & 0.79 &  --  & \textbf{0.82} & 0.72 \\
\cline{2-12}
 & \multirow{5}{*}{$\lambda = 0.3. \delta = 1.0$} & advogato & 0.54 & 0.11 & 0.47 & 0.64 & 0.45 & 0.74 & 0.55 & \textbf{0.76} & 0.73 \\
 & & email & 0.77 & 0.03 & 0.71 & 0.59 & 0.70 & 0.89 & 0.76 & \textbf{0.91} & 0.89\\
 & & polblogs & 0.19 & 0.05 & 0.19 & 0.41 & 0.17 & 0.19 & 0.16 & \textbf{0.31} & 0.26\\
 & & Google+ & 0.20 & 0.17 & 0.14 & \textbf{0.84} & -0.12 & 0.63 & 0.20 & 0.61 & 0.65 \\
 & & BA & 0.35 & 0.11 & 0.46 & 0.34 & 0.20 & 0.47 &  --  & 0.48 & \textbf{0.51} \\
\cline{2-12}
 & \multirow{5}{*}{$\lambda = 0.3. \delta = 0.3$} & advogato & 0.36 & 0.19 & 0.31 & 0.57 & 0.29 & 0.53 & 0.37 & \textbf{0.57} & 0.52 \\
 & & email & 0.70 & 0.19 & 0.62 & 0.35 & 0.65 & 0.71 & 0.70 & 0.74 & \textbf{0.77}\\
 & & polblogs & -0.16 & 0.09 & -0.12 & \textbf{0.52} & -0.18 & -0.13 & -0.18 & -0.02 & -0.10\\
 & & Google+ & 0.14 & 0.16 & 0.05 & \textbf{0.76} & -0.01 & 0.36 & 0.15 & 0.34 & 0.47 \\
 & & BA & 0.33 & 0.19 & 0.59 & 0.67 & 0.06 & 0.67 &  --  & 0.73 & \textbf{0.85} \\
\hline
\multirow{20}{*}{MT CP} & \multirow{3}{*}{$\lambda = 0.8. \delta = 1.0$} & advogato & 0.47 & 0.14 & 0.35 & 0.52 & 0.36 & 0.63 & 0.50 & 0.62 & \textbf{0.65} \\
 & & email & 0.69 & 0.19 & 0.56 & 0.57 & 0.61 & \textbf{0.81} & 0.73 & 0.79 & \textbf{0.81}\\
 & & polblogs & 0.29 & 0.14 & 0.21 & 0.26 & 0.25 & 0.28 & 0.29 & \textbf{0.34} & \textbf{0.34}\\
 & & Google+ & 0.40 & 0.32 & 0.31 & 0.55 & -0.12 & \textbf{0.84} & 0.40 & 0.76 & 0.75 \\
 & & BA & 0.56 & 0.19 & 0.78 & 0.63 & 0.30 & 0.74 &  --  & 0.80 & \textbf{0.94} \\
\cline{2-12}
 & \multirow{5}{*}{$\lambda = 0.8. \delta = 0.3$} & advogato & 0.29 & 0.14 & 0.21 & 0.35 & 0.21 & 0.38 & 0.31 & 0.39 & \textbf{0.41} \\
 & & email & 0.45 & 0.22 & 0.36 & 0.32 & 0.40 & 0.48 & 0.47 & 0.50 & \textbf{0.52}\\
 & & polblogs &0.01 & 0.14 & -0.01 & \textbf{0.26} & -0.03 & 0.02 & 0.00 & 0.07 & 0.05\\
 & & Google+ & 0.32 & 0.27 & 0.23 & 0.52 & -0.10 & \textbf{0.64} & 0.33 & 0.60 & 0.64 \\
 & & BA & 0.24 & 0.21 & 0.58 & 0.80 & -0.02 & 0.87 &  --  & \textbf{0.91} & 0.79 \\
\cline{2-12}
 & \multirow{5}{*}{$\lambda = 0.3. \delta = 1.0$} & advogato & 0.52 & 0.08 & 0.39 & 0.59 & 0.40 & 0.72 & 0.55 & 0.71 & \textbf{0.73} \\
 & & email & 0.64 & 0.05 & 0.55 & 0.68 & 0.55 & \textbf{0.83} & 0.67 & 0.82 & 0.80\\
 & & polblogs & 0.51 & 0.14 & 0.39 & 0.25 & 0.46 & 0.52 & 0.51 & 0.56 & \textbf{0.57}\\
 & & Google+ & 0.37 & 0.28 & 0.32 & 0.55 & -0.15 & \textbf{0.89} & 0.38 & 0.84 & 0.75 \\
 & & BA & \textbf{0.82} & 0.10 & 0.67 & -0.01 & 0.74 & 0.31 &  --  & 0.34 & 0.52 \\
\cline{2-12}
 & \multirow{5}{*}{$\lambda = 0.3. \delta = 0.3$} & advogato & 0.46 & 0.15 & 0.34 & 0.50 & 0.35 & 0.60 & 0.48 & 0.60 & \textbf{0.63} \\
 & & email & 0.68 & 0.23 & 0.55 & 0.50 & 0.60 & 0.76 & 0.71 & 0.76 & \textbf{0.79}\\
 & & polblogs & 0.19 & 0.14 & 0.13 & \textbf{0.27} & 0.15 & 0.18 & 0.18 & 0.24 & 0.23\\
 & & Google+ & 0.39 & 0.32 & 0.30 & 0.55 & -0.11 & \textbf{0.80} & 0.40 & 0.74 & 0.74 \\
 & & BA & 0.67 & 0.19 & 0.79 & 0.49 & 0.43 & 0.66 &  --  & 0.72 & \textbf{0.89} \\
\hline
\end{tabular}
\label{tab:Corr_non_sp}
\end{center}
\end{table*}

\begin{figure}[t]
\begin{center}
\subfigure[]{\includegraphics[width=0.45\columnwidth]{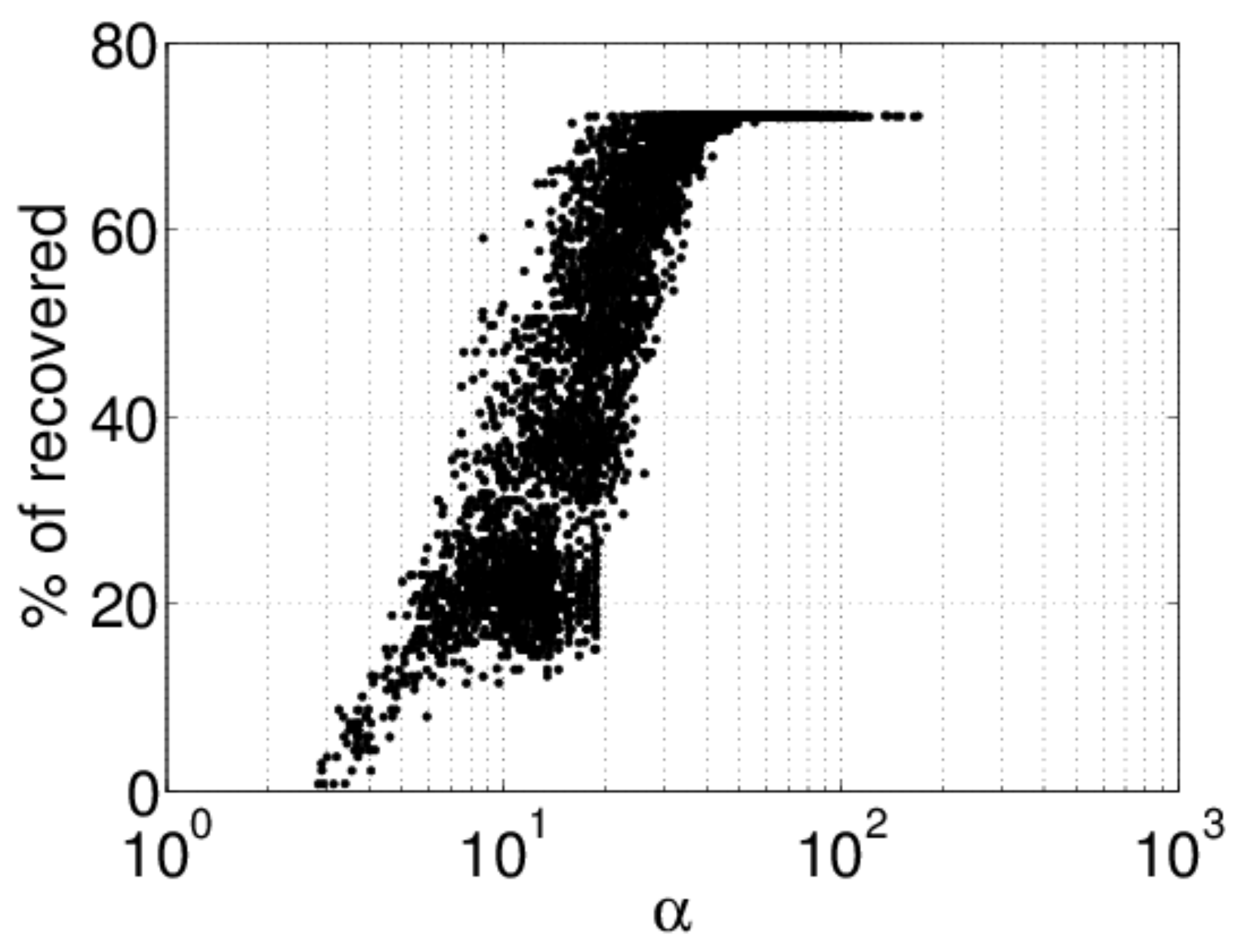}}
\subfigure[]{\includegraphics[width=0.45\columnwidth]{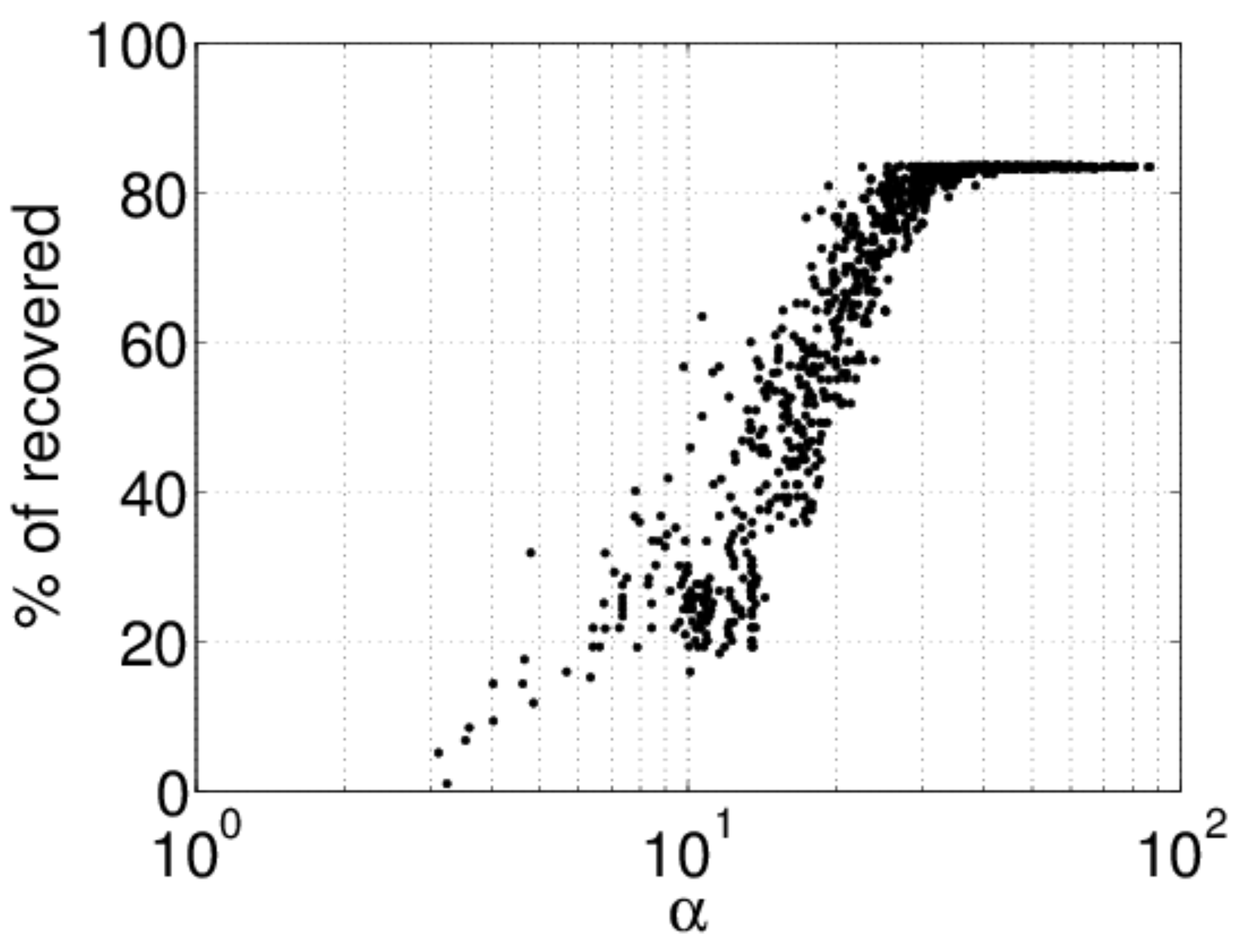}}
\subfigure[]{\includegraphics[width=0.45\columnwidth]{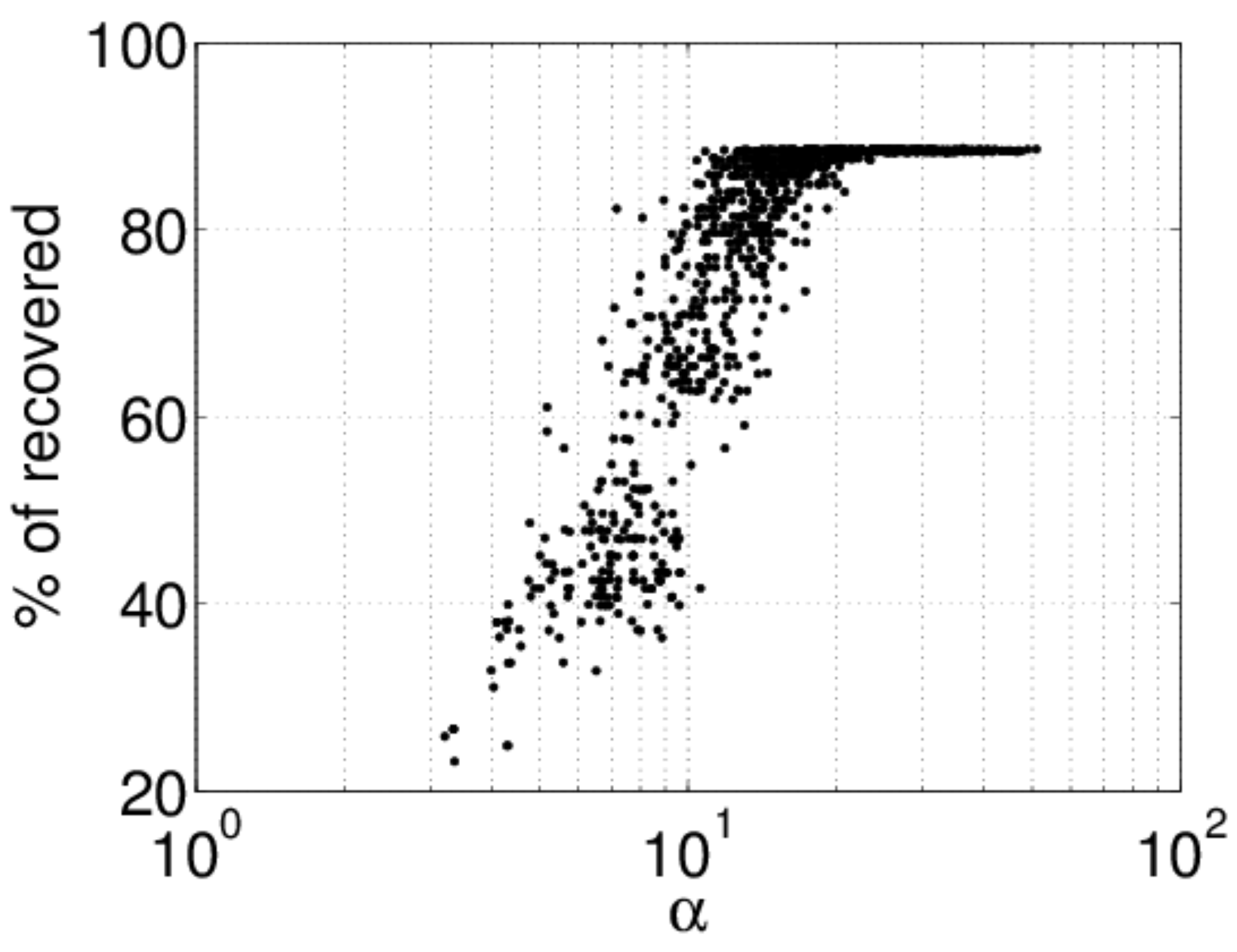}}
\subfigure[]{\includegraphics[width=0.45\columnwidth]{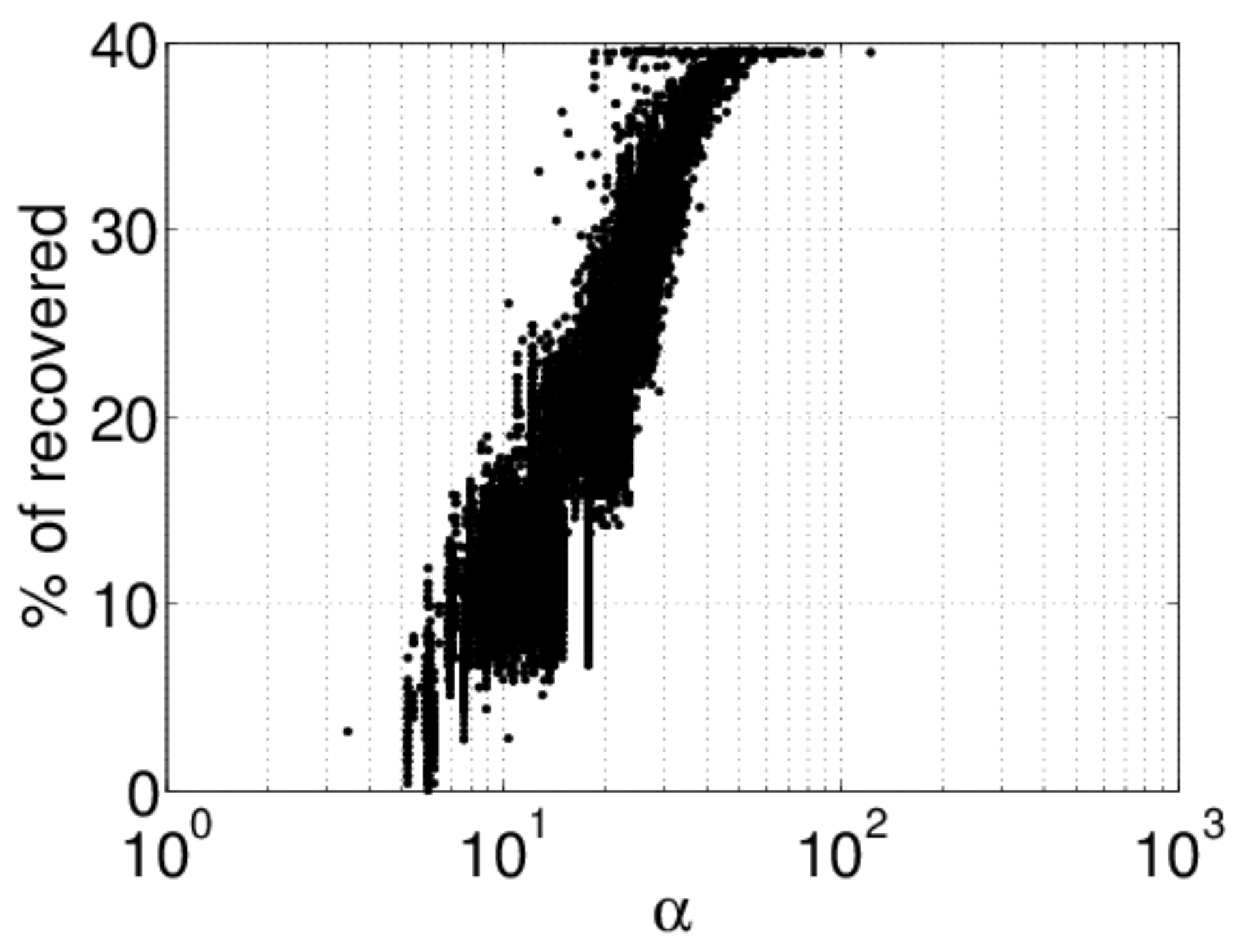}}
\caption{Scatter plots of the accessibility measure and the final percentage of recovered individuals on the SIR epidemic spreading model ($\beta = 0.3, \mu = 1.0$) for
the networks of (a) advogato, (b) political blogs, (c) e-mail, (d) Google+.}
\label{Fig:Corr_non_sp_alpha_SIR_RP}
\end{center}
\end{figure}

\begin{figure}[t]
\begin{center}
\subfigure[]{\includegraphics[width=0.45\columnwidth]{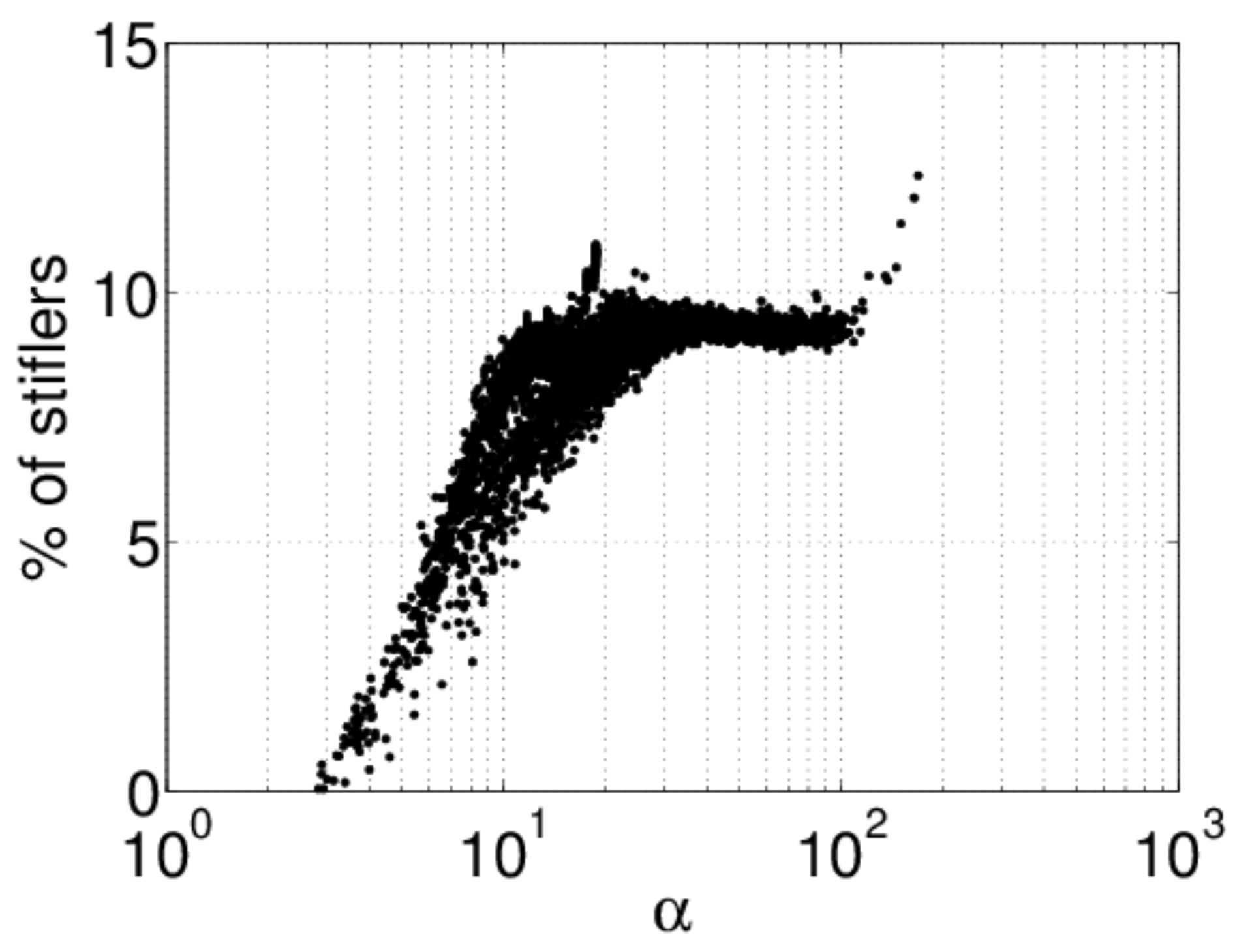}}
\subfigure[]{\includegraphics[width=0.45\columnwidth]{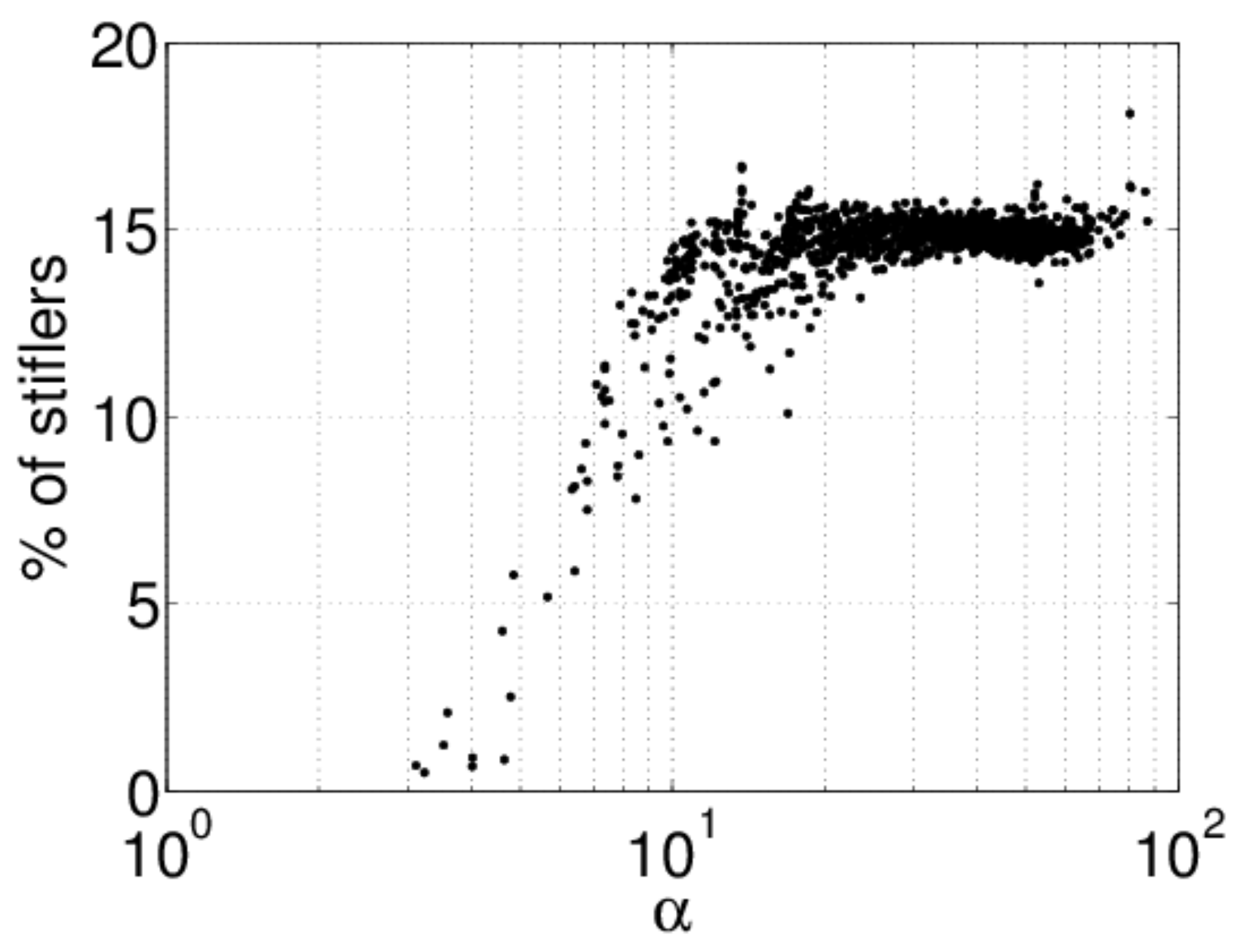}}
\subfigure[]{\includegraphics[width=0.45\columnwidth]{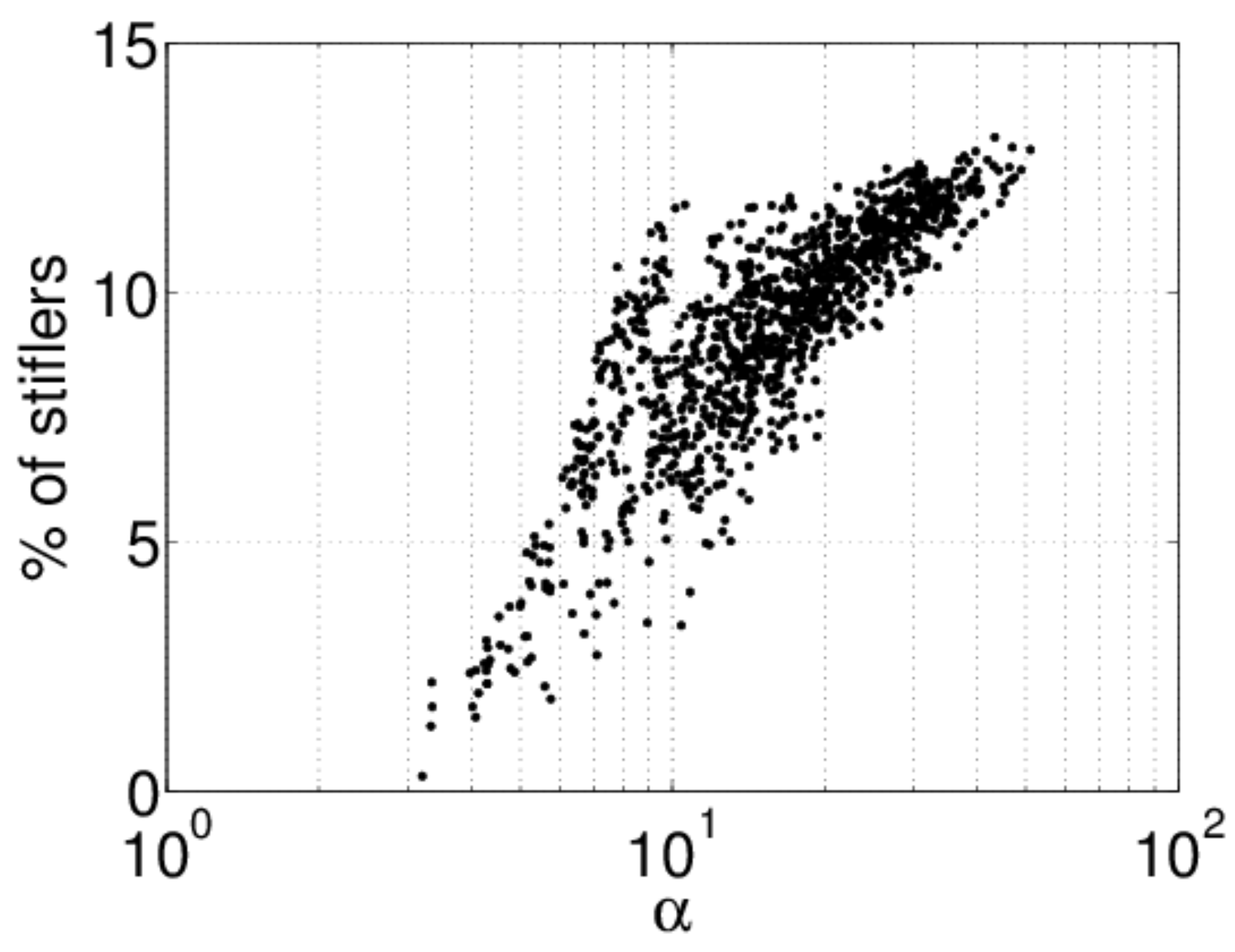}}
\subfigure[]{\includegraphics[width=0.45\columnwidth]{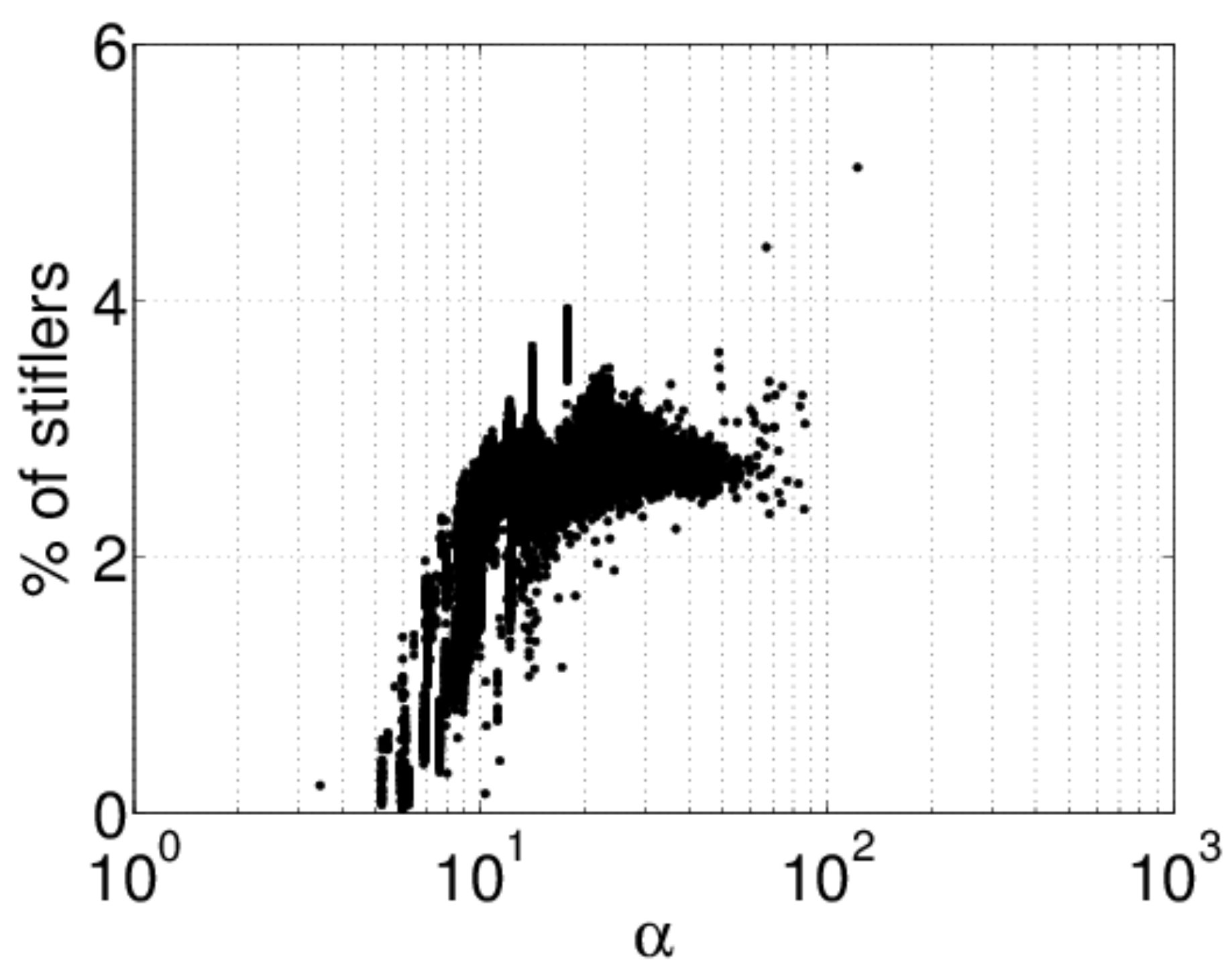}}
\caption{Scatter plots of the accessibility measure and the final percentage of stiflers on the MT (TP) rumor model ($\lambda = 0.3, \delta = 1.0$) for the networks of
(a) advogato, (b) political blogs, (c) e-mail, (d) Google+.}
\label{Fig:Corr_non_sp_alpha_MK_TP}
\end{center}
\end{figure}


\section{Conclusions} \label{Sec:conclusion}

In this paper we have studied the relation between the centrality of a node and the outcome of epidemic and rumor processes initiated in that node by means of extensive numerical simulations on top of several complex networks. We have considered eight network centrality metrics and two different kinds of networks: spatial and non-spatial ones. Networks generated by the Barab\'asi-Albert, Waxman and scale-free spatial models have also been considered. We have proposed a generalization of the accessibility measure introduced in~\cite{Travenccolo2008:PLA}, which allows the quantification of the potential of each node in accessing in a balanced and homogeneous manner other nodes. Such generalization takes into account walks of all lengths weighted by the inverse of the factorial of their lengths. 

Our results have shown that the generalised accessibility is the best metric to measure a node's spreading capacity in spatial networks. On the contrary, in non-spatial networks, the best correlations between a centrality metric and the dynamical outcome depends on the process. Thus, the degree and coreness (as given by the $k$-core) are the ones more suited when it comes to analyze epidemic spreading, confirming the results in~\cite{Kitsak10:NP}. However, these measures are not the best when a rumor model is considered. Indeed, for the latter case, the average neighborhood degree, the closeness centrality and accessibility gives higher correlations. 

We verified that the generalised accessibility is more related to spreading processes in spatial networks than in non-spatial networks. Indeed, Table~\ref{tab:Corr_non_sp} shows that this metric is the structural property that exhibits the highest correlation in most of the cases when the underlying network is spatial. Figs.~\ref{Fig:Corr_non_sp_alpha_SIR_RP} and~\ref{Fig:Corr_non_sp_alpha_MK_TP} show that the relationship between the accessibility and centrality measures are almost linear in spatial networks, whereas in non-spatial networks, such relationship is also almost linear, but only for values below a given threshold. Beyond that value, the fraction of stiflers and recovered nodes reaches a plateau, which is the maximum value of the dynamic measure in the networks. Such plateau reduces the Spearman correlation between the accessibility and the fraction of stiflers, since the relationship between these structural and dynamical measures is better defined for low values of accessibility. Therefore, due the higher distances in spatial networks, the value of accessibility does not saturate (i.e., there is no plateau), resulting in higher correlations.

The previous conclusions can be understood by looking with more care to the meaning of the new metric here discussed. The definition of the accessibility in terms of random walks is strictly related to the spreading processes \cite{Pinto12:PRL} and it is defined in terms of the diversity index of order one~\cite{Hill1973}. Thus, the higher the number of neighbors that a node can access with similar probability, the higher the expected number of infected nodes. In this way, the accessibility quantifies how many nodes can be effectively accessed during the spreading process. As reported in~\cite{Viana2012:PRE} this quantity is maximum whenever the exploration time is minimum. Thus, nodes presenting higher values of accessibility propagate viruses or rumors to the whole network faster than the nodes with smaller values, which results in a higher fraction of infected nodes before they become recovered. In summary, nodes with higher accessibility values should be 
the most influential spreaders.

The analysis presented here can be extended by considering other definitions of the accessibility in terms of other diversity indices~\cite{Jost06}. The role of the generalized random walk accessibility in other types of dynamical process, such as social dynamic models~\cite{Castellano09:RMP} and synchronization are also possible further researches. Ultimately, one important conclusion of our study, beyond the fact that the new metric appears to be the best way to detect influential spreaders in spatial networks, is that previous claims about whether a class of nodes are influential depends on both the metric used and most importantly, on the kind of network under study.

\section*{Acknowledgment}

FAR acknowledge CNPq (grant 305940/2010-4), Fapesp (grant 2011/50761-2 and 2013/26416-9) and NAP eScience - PRP - USP for financial support. LFC would like to acknowledge CNPq and Fapesp for the financial support. PMR is supported by Fapesp (grant 2013/03898-8) and CNPq (grant 479313/2012-1). GFA acknowledges Fapesp and ALB acknowledges CAPES for sponsorship provided. YM is partially supported by the EC FET-Proactive Project MULTIPLEX (grant 317532).

\bibliographystyle{apsrev}
\bibliography{paper}

\end{document}